%% file: main.tex
\pdfoutput=1
\documentclass[twocolumn]{aastex701}
\usepackage{subfig}
\usepackage{etoolbox}

\usepackage{amsmath}
\newcommand{\hi}{H\textsc{i}}
\newcommand{\hii}{H\textsc{i}\textsc{i}}
\newcommand{\htwo}{$\mathrm{H}_2$}

\newcommand{\changes}[1]{#1}
\newcommand{\changestwo}[1]{#1}
\defcitealias{lindberg2025}{L25}
\accepted{\today}

\submitjournal{ApJ}

\shorttitle{Scylla Dust Mapping}
\graphicspath{{./}{figures/}}

\begin{document}

\title{Scylla VI: Parsec-Scale Dust Extinction Maps in the SMC and LMC}

\author[0000-0003-0588-7360]{Christina W. Lindberg}
\altaffiliation{JC Ryan Post-Doctoral Fellow}
\affiliation{Center for Astrophysics $\vert$ Harvard \& Smithsonian, 60 Garden St., Cambridge, MA 02138, USA}
\affiliation{Space Telescope Science Institute, 
3700 San Martin Drive, 
Baltimore, MD 21218, USA}
\affiliation{The William H. Miller III Department of Physics \& Astronomy, Bloomberg Center for Physics and Astronomy, Johns Hopkins University, 3400 N. Charles Street, Baltimore, MD 21218, USA}

\email[show]{christina.lindberg@live.com}

\author[0000-0002-7743-8129]{Claire E. Murray}
\affiliation{Space Telescope Science Institute, 
3700 San Martin Drive, 
Baltimore, MD 21218, USA}
\affiliation{The William H. Miller III Department of Physics \& Astronomy, Bloomberg Center for Physics and Astronomy, Johns Hopkins University, 3400 N. Charles Street, Baltimore, MD 21218, USA}
\email[hide]{cmurray1@stsci.edu}  

\author[0000-0001-7959-4902]{Christopher J. R. Clark}
\affiliation{AURA for the European Space Agency, Space Telescope Science Institute, 3700 San Martin Drive, Baltimore, MD 21218, USA}
\email[hide]{cclark@stsci.edu}  


\author[0000-0001-6118-2985]{Caroline Bot}
\affiliation{Université de Strasbourg, CNRS, Observatoire astronomique de Strasbourg, UMR 7550, F-67000 Strasbourg, France}
\email[hide]{caroline.bot@astro.unistra.fr}

\author[0009-0005-0339-015X]{Clare Burhenne}
\affiliation{Department of Physics and Astronomy, Rutgers the State University of New Jersey, 136 Frelinghuysen Rd., Piscataway, NJ, 08854, USA}
\email[hide]{cdb201@physics.rutgers.edu}

\author[0000-0003-1680-1884]{Yumi Choi}
\affiliation{NSF National Optical-Infrared Astronomy Research Laboratory, 950 N. Cherry Avenue, Tucson, AZ 85719 USA}
\email[hide]{yumi.choi@noirlab.edu}  

\author[0000-0002-2970-7435]{Roger E. Cohen}
\affiliation{Department of Physics and Astronomy, Rutgers the State University of New Jersey, 136 Frelinghuysen Rd., Piscataway, NJ, 08854, USA}
\email[hide]{rc1273@physics.rutgers.edu}

\author[0000-0002-8937-3844]{Steven R. Goldman}
\affil{Space Telescope Science Institute, 3700 San Martin Drive, Baltimore, MD 21218, USA}
\email[hide]{sgoldman@stsci.edu}

\author[0000-0001-5340-6774]{Karl D.\ Gordon}
\affiliation{Space Telescope Science Institute, 3700 San Martin Drive, Baltimore, MD 21218, USA}
\email[hide]{kgordon@stsci.edu}

\author[0000-0001-5538-2614]{Kristen B. W. McQuinn}
\affiliation{Space Telescope Science Institute, 3700 San Martin Drive, 
Baltimore, MD 21218, USA}
\affiliation{Department of Physics and Astronomy, Rutgers the State University of New Jersey, 136 Frelinghuysen Rd., Piscataway, NJ, 08854, USA}
\email[hide]{kmcquinn@stsci.edu}

\author[0000-0001-6326-7069]{Julia Roman-Duval}
\email{duval@stsci.edu}
\affiliation{Space Telescope Science Institute, 3700 San Martin Drive, Baltimore, MD 21218, USA}
\email[hide]{duval@stsci.edu}

\author[0000-0002-4378-8534]{Karin M. Sandstrom}
\affil{Department of Astronomy \& Astrophysics, University of California San Diego, 9500 Gilman Drive, La Jolla, CA 92093, USA}
\email[hide]{kmsandstrom@ucsd.edu}

\author[0000-0002-3569-7421]{Edward F. Schlafly}
\affiliation{Space Telescope Science Institute, 
3700 San Martin Drive, 
Baltimore, MD 21218, USA}
\email[hide]{eschlafly@stsci.edu} 

\author[0000-0003-1356-1096]{Elizabeth Tarantino}
\affiliation{Space Telescope Science Institute, 3700 San Martin Drive, Baltimore, MD 21218, USA}
\email[hide]{etarantino@stsci.edu}  

\author[0000-0002-7502-0597]{Benjamin F. Williams}
\affiliation{Department of Astronomy, University of Washington, Box 351580, U. W., Seattle, WA 98195-1580, USA}
\email[hide]{benw1@uw.edu}  

\author[0000-0002-9912-6046]{Petia Yanchulova Merica-Jones}
\affiliation{University of Sofia, Faculty of Physics, 5 James Bourchier Blvd., 1164 Sofia, Bulgaria}
\affiliation{Institute of Astronomy and NAO, Bulgarian Academy of Sciences, 72 Tsarigradsko Chaussee Blvd., 1784 Sofia, Bulgaria}
\email[hide]{pyanchulova@astro.bas.bg}  

\author[0000-0002-2250-730X]{Catherine Zucker}
\affil{Center for Astrophysics $\vert$ Harvard \& Smithsonian, 60 Garden St., Cambridge, MA 02138, USA}
\affiliation{Space Telescope Science Institute, 
3700 San Martin Drive, 
Baltimore, MD 21218, USA}
\email[hide]{catherine.zucker@cfa.harvard.edu}
 
\begin{abstract}



We present a novel methodology for mapping dust extinction in nearby galaxies at parsec-scale resolution. We apply it to HST 68 fields within the Small and Large Magellanic Clouds (23 fields in the SMC and 45 fields in the LMC) using multi-band HST photometry from the Scylla and METAL surveys. Our technique leverages \textit{kriging}, a geostatistical interpolation method built on the principles of Gaussian Process regression, combined with Gaussian mixture modeling to statistically isolate background stellar sources and account for line-of-sight depth effects. 3D dust simulations demonstrate the method's capability to recover column densities to an accuracy of $A_V \approx 0.1$ mag in fields with at least 1000 sources. The resulting $4^{\prime\prime}$ resolution ($\sim1$-pc) dust maps reveal detailed structure and strong spatial correlation with ancillary ISM tracers, especially in star-forming regions like 30 Doradus. Global extinction of total column densities follows log-normal profiles in both galaxies, with the SMC exhibiting slightly higher mean extinction ($e^{\mu}=0.47$ mag) than the broader LMC ($e^{\mu}=0.43$ mag), likely due to significant line-of-sight depths. We find systematic offsets between dust mass surface densities ($\Sigma_{D}$) derived from extinction versus FIR emission in both galaxies, with $\Sigma_{D, FIR}/\Sigma_{D, A_V}$ ratios ranging from $0.6-1.8$. This work provides the highest-resolution dust extinction maps in SMC and LMC to date, which offer a vital independent benchmark for constraining dust emissivity, $\text{CO}$-dark gas fractions, and the multi-scale structure of the ISM in low-metallicity environments. 

\end{abstract}


\section{Introduction}
\label{sec:intro}


The interstellar medium (ISM) plays a critical role in the evolution of galaxies, facilitating the exchange of energy, momentum, mass, and metals between stars and the ISM. Although the majority of mass in the ISM is composed of gas, 
interstellar dust grains are well-mixed with both atomic and molecular gas \citep{Boulanger1996, Rachford2009}. 
Dust allows us to trace the combined gas column densities across gas phases -- especially cold molecular hydrogen (\htwo), which is otherwise difficult to observe directly \citep{draine2003} -- ultimately providing a more complete estimate of the total ISM mass budget of galaxies. 


One way to observe dust is through its extinction of light. Due to the distribution of dust grain sizes, extinction from dust is minimal in the near-IR but increases towards optical and ultraviolet (UV) wavelengths \citep[e.g.,][]{cardelli1989, fitzpatrick1999, draine2003, gordon2023}. This wavelength-dependent characteristic means that the column density of dust extinction can be constrained with multi-wavelength data (e.g., spectra or multi-band photometry) as long as the background source of light can be modeled adequately. 

For nearby galaxies with resolved stellar populations, multi-band photometry offers a unique opportunity to map the 2D distribution of the ISM via dust extinction \citep[e.g.,][]{zaritsky2002, zaritsky2004, Imara2007, Haschke2011, dalcanton2015, choi2018}. 
The observed photometry from a star is the result of four main processes: (1) a star emits photons in the form of blackbody radiation over a range of wavelengths, including absorption lines from the stellar photosphere; (2) intervening dust between Earth and the star scatters and absorbs some fraction of these photons; (3) photons with frequencies inside a filter wavelength range are collected by telescopes; and (4) point spread functions are used to isolated and distinguish flux from different sources. The presence of dust grains along a stellar sightline will not only dim the observed flux of a star, but will do so preferentially at shorter/bluer wavelengths, making a star appear dimmer and redder than it intrinsically is. This extinction from dust creates observational degeneracies in which a star can appear dim and red, either due to its intrinsic properties and distance, or due to the presence of dust along the line of sight.

With broad filter coverage, especially in the optical and near-UV, we can break degeneracies between stellar temperature and dust reddening for resolved stars and characterize their individual stellar properties, providing measurements of dust extinction towards millions of sources within galaxies \citep[e.g.,][]{Romaniello2002, gordon2016}. These dust extinction measurements can be used to estimate the escape fraction of ionizing photons from galaxies \citep[e.g.,][]{choi_mapping_2020, vandeputte2020} or study the small-scale structure of the ISM around individual sources of interest \citep[e.g., massive stars;][]{lindberg2024}. 


In this work, we demonstrate how extinction towards stellar sightlines can be modeled with kriging \citep{matheron1963, cressie1992} to construct 2D dust extinction maps at parsec scales in nearby galaxies. Kriging is a geostatistical interpolation method well-suited for data with spatial correlation, and generalizations of the method, like Gaussian Process regression, have been used extensively for mapping gas and dust within the Milky Way \citep[e.g.,][]{Rezaei2018, leike2020, thavisha2023, miller2022, soding2025}. Kriging models the correlation of data with a variogram, \changes{a geostatistical tool that measures dissimilarity between data as a function of distance separation. This setup} offers numerous benefits when modeling diffuse structures like the ISM, including: (1) explicitly incorporating the 2D/3D spatial correlation of measurements; and (2) providing predictive measurements for any location, along with the corresponding prediction variance. 


We perform dust extinction mapping on fields within the Small and Large Magellanic Clouds (SMC and LMC, respectively, MC for either), the most massive dwarf satellites of the Milky Way (MW). Likely on their first infall into the MW \citep{besla2007, bekki2009}, the SMC and LMC have been interacting for several Gyrs \citep{patel2020} and are thought to have collided as recently as 140-160 Myr ago \citep{zivick2018, choi2022, Cullinane2022}, the effects of which can still be observed through their gas kinematics \citep[e.g.,][]{murray2019, murray2024a}, morphology \citep[e.g.,][]{nidever2008, mastropietro2009, tsuge2019}, chemistry \citep[e.g.,][]{fox2018, romanduval2021}, and interaction-driven starbursts \citep[e.g.,][]{weisz2013, nidever2017, massana2022, burhenne2025}.
The close proximity of the LMC and SMC \citep[50 and 62 kpc, respectively;][]{degrijs2014, degrijs2015, pietrzynski2019, scowcroft2016} and low chemical enrichment \citep[$0.5\ Z_{\odot}$ and $0.2\ Z_{\odot}$, respectively;][]{russell1992} make them exceptional laboratories for studying the effects of low metallicity on small-scale ISM physics, including molecular gas formation, dust-to-gas ratios, and dust composition, all of which are known to vary significantly with enrichment. 

Both galaxies were recently observed as part of the Scylla survey \citep{murray2024b}. Scylla obtained multi-band photometry with the \emph{Hubble Space Telescope} (HST) Wide Field Camera 3 (WFC3) towards 96 fields in both galaxies. The survey imaged a diverse range of environments within each galaxy (e.g., gas column densities, radiation fields, etc.), resolving millions of stars down to sub-solar masses. For 1.5 million sources observed in three bands or more, we analyzed their spectral energy distribution (SED) to simultaneously characterize their stellar properties and line-of-sight dust extinction in \citet[][]{lindberg2025}, hereafter \citetalias[][]{lindberg2025}. 

In this paper, we leverage line-of-sight extinction measurements from Scylla in the SMC and LMC to map the 2D distribution of dust in fields across the SMC and LMC. Using kriging, we construct the highest-resolution dust extinction maps in the SMC and LMC to date, providing new insights into the small-scale structure of the ISM at low metallicities. We adopt an LMC distance of 50 kpc \citep{pietrzynski2019} and an SMC distance of 62 kpc \citep{scowcroft2016} throughout this paper, for which $4^{\prime\prime}$ is equivalent to 0.97 pc and 1.2 pc, respectively. The angular coverage of a singular HST WFC3 field spans $160\times160$ arcsec, roughly equivalent to 40 and 45~pc in physical scales in the LMC and SMC, respectively. 

In Section \ref{sec:data}, we recap the HST photometry and SED-fitting technique used to acquire the extinction measurements and establish the upper limits of detection for extinction. In Section \ref{sec:method}, we address line-of-sight depth effects and describe the kriging method in detail. In Section \ref{sec:mw_val}, we validate the methodology on 3D molecular clouds and test how the technique compares with alternative dust mapping methods. In Section \ref{sec:results}, we present examples of dust extinction maps in both low- and high-column density environments across the SMC and LMC and describe the global extinction distributions found within each galaxy. In Section \ref{sec:discussion}, we compare the maps with ancillary ISM tracers and FIR dust emission maps, discuss how these maps differ from previous extinction maps towards the SMC and LMC, and identify future applications for the dust maps and methodology. Finally, in Section \ref{sec:conclusion}, we recap any notable biases or limitations of the presented methodology and summarize our findings. 

\begin{figure*}[t]
 	\centering
\includegraphics[width=\textwidth]{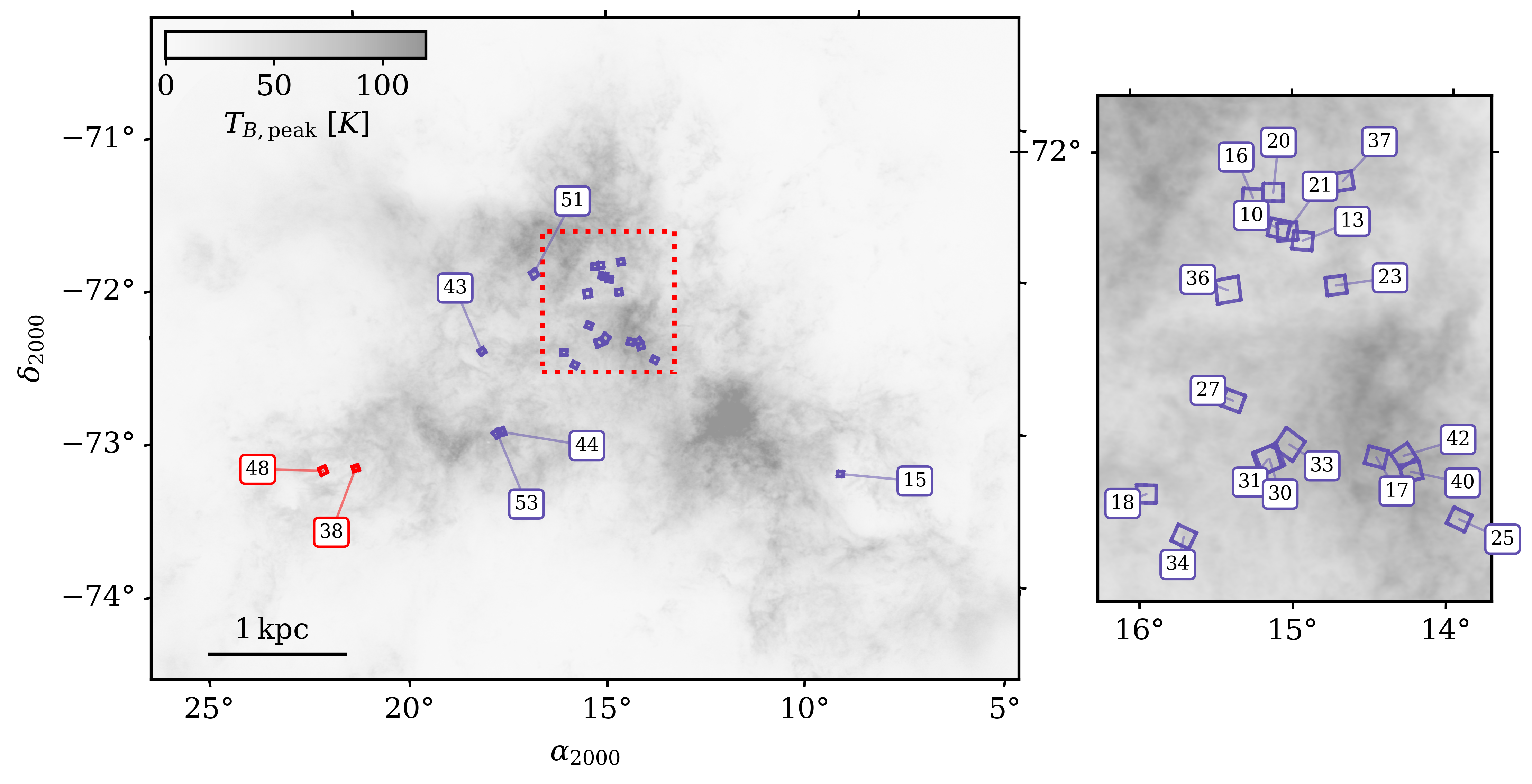}
\includegraphics[width=\textwidth]{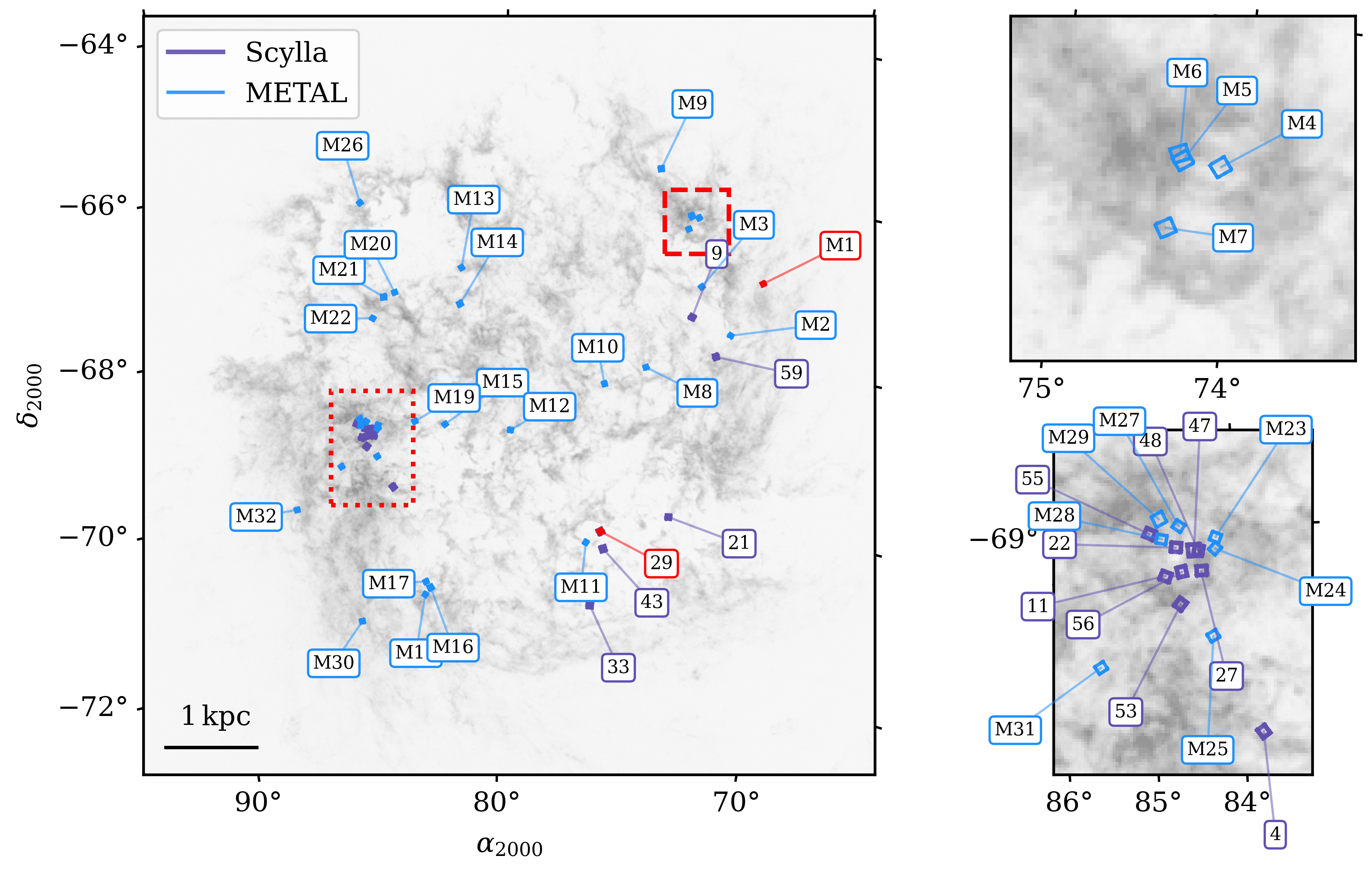}
	\caption{\changes{\textbf{Scylla and METAL dust map locations}: Maps of the peak brightness temperature of $21\rm\,cm$ emission in the SMC \citep[top;][]{pingel2022} and the LMC \citep[bottom;][]{kim1999}, overlaid with the footprints of the 25 Scylla fields in the SMC and 47 Scylla and METAL fields in the LMC that have $N\geq4$ filters needed to perform dust extinction mapping. In each panel, regions where fields overlap are identified by dashed red bounding boxes and expanded at right. Each field is tagged by its ``short" name number where the METAL fields start with ``M" \citetalias[][Table 1]{lindberg2025}. Four fields (red) have insufficient stellar densities to construct dust extinction maps (see details in Section \ref{sec:results}).}}
    \label{fig:targets}
\end{figure*}

\section{Observations and Data Preparation}
\label{sec:data}

In this section, we discuss the HST photometry used in the paper, followed by a description of the SED analysis performed in \citetalias{lindberg2025} to obtain individual dust extinction measurements for resolved stars. Finally, we discuss the upper detection limits on $A_V$ and the quality cuts we apply to avoid biasing resultant dust extinction maps.
 
\subsection{HST Photometry}

The Scylla survey is a multi-cycle, pure-parallel HST program (PIDs 15891, 16235, 16786) that used Wide Field Camera 3 (WFC3) to obtain multi-band imaging of 96 individual fields across the SMC and LMC \citep{murray2024b}. Depending on scheduling availability, fields were observed with anywhere from 2 to 7 bands, with a priority of obtaining coverage in optical bands (F475W, F814W), followed by near-UV (F336W, F275W), near-IR (F110W, F160W), and finally the additional UV filter F225W, if time permitted. 

To ensure quality SED fits, \citetalias{lindberg2025} performed SED fitting on fields observed in at least three bands. Additional validation simulations in \citetalias{lindberg2025} demonstrate that sources must be observed in at least four bands to recover dust extinction accurately; with only 3-band photometry, sources were systematically biased high by $\sim$0.1 mag of extinction, and had much greater scatter. To avoid biased dust extinction measurements, we opt to only analyze fields observed with at least 4 bands of photometry, limiting the Scylla survey to 25 fields in the SMC and 15 fields in the LMC. 

In addition to the Scylla survey, we also analyze multi-band photometry from the Metal Evolution Transport and Abundance in the LMC (METAL) survey \citep{romanduval2019}. 
The METAL survey obtained four-, six-, and seven-band parallel imaging with HST WFC3 towards 32 fields scattered across the LMC with the same filters as Scylla \citep{romanduval2019}. The SED fits for the METAL survey were also performed in \citetalias[][]{lindberg2025}. 

With the inclusion of the METAL survey, this paper analyzes a combined count of 47 fields with 229,005 stars in the LMC and 25 fields with 182,780 stars in the SMC, following high-reliability photometric quality cuts described in \citetalias{lindberg2025}. \changes{In Figure \ref{fig:targets}, we plot the locations of these fields, along with a summary table of their names, positions, and resultant extinction properties. An overview of all Scylla and METAL field locations, filters, and photometric reduction pipeline can be found in \citet[][Table 2]{murray2024b}.}

\begin{figure*}[t]
    \centering
    \includegraphics[width=\linewidth]{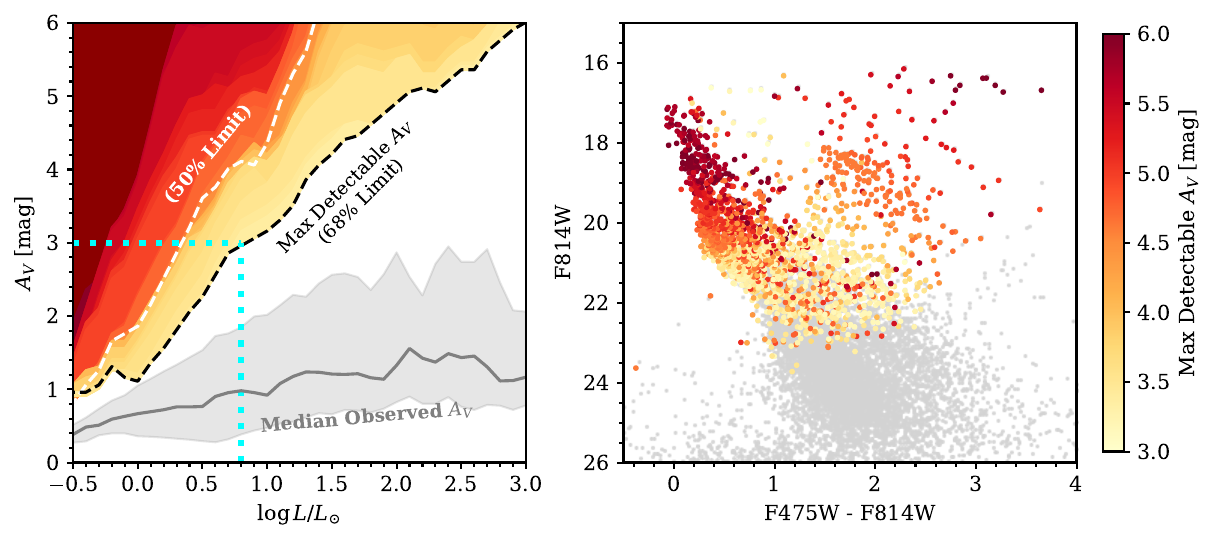}
    \caption{\textbf{Extinction detection limits and completeness}: (Left) Maximum detectable $A_{V}$ (black dashed line) as a function of intrinsic stellar luminosity, defined by the level where 68\% of artificial stars remain detectable in the F336W filter. The grey shaded region represents the median observed $A_{V}$ in a dusty 30 Doradus field, showing that most sources fall safely below the detection threshold. The cyan dashed line indicates the $\log(L/L_{\odot}) \ge 0.8$ quality cut applied to ensure sensitivity up to $A_{V} = 3.0$ mag. (Right) Optical CMD of observed stars (grey). Stars passing the luminosity cut are colored by their theoretical $A_{V}$ detection limit, demonstrating that upper main sequence and RGB stars are generally sensitive to $A_{V} \ge 4.0$ mag.}
    \label{fig:max_av}
\end{figure*}

\input{table.txt}

\subsection{Stellar SED Fitting}

In \citetalias{lindberg2025}, the multi-band SEDs of individual sources in the Scylla and METAL surveys were analyzed using the Bayesian Extinction And Stellar Tool \citep[BEAST;][]{gordon2016}, a probabilistic Bayesian tool designed to simultaneously fit stars and line-of-sight dust extinction. 
To characterize stellar sightlines, the BEAST compares the observed SED of individual dust reddened stars with a grid of model SEDs generated from PARSEC stellar evolutionary tracks \citep{bressan2012}. Models were constructed to span a wide range of initial stellar masses, ages, metallicities, and distances. To the intrinsic spectra of each stellar model, we apply increasing amounts of extinction ($A_V = 0.1 - 10.0$ mag), assuming different extinction curves ($R_V = 2.0 - 6.0$), ranging from MW-like extinction curves (i.e.~with a carbonaceous bump at 2175~\AA) to SMC-like extinction curves (i.e.~without a bump at 2175~\AA). These extinguished stellar spectra were then convolved with different HST filter bandpasses to simulate the expected photometric flux. To account for systematic effects in the observations (e.g., high background noise, crowding, etc.), we employed artificial star tests (ASTs) to quantify the average bias and uncertainty for a star in any observed flux range. 

We compared the grid of models to every observed star in a galaxy and evaluated the median -- plus 16th and 84th percentile -- parameter values that best reproduce the observed stellar flux. 
In addition to the main stellar and dust parameters, the BEAST also reports secondary stellar parameters like surface gravity, temperature, and intrinsic luminosity, the last of which we use in the following subsection to apply completeness cuts to our catalog of extinguished sources.
\subsection{Extinction Detection Limits and Source Completeness}
\label{sec:complet}

To ensure the accuracy of our dust maps, we must account for the observational selection effects that occur when stars are extinguished by dust. Because the distribution of dust grain sizes causes extinction to increase toward shorter ultraviolet (UV) and optical wavelengths, stars become progressively more difficult to detect in bluer filters. This creates a risk that stars in high-$A_V$ regions may fall below the detection limits of the HST WFC3. The absence of these extinguished sources -- coupled with the surviving abundance of unextinguished sources -- will bias the average observed extinction low within a field if not accounted for.

With the exception of dense star-forming regions, the SMC and LMC contain relatively little dust due to their low chemical enrichment \citep[e.g., $A_V = 0.2-1.0$ mag;]{skowron2021, Chen2022}, especially compared to larger Local Group galaxies \citep[e.g., M31 with $A_V = 1-4$ mag;][]{dalcanton2015}. However, the distribution is highly variable based on local environment; dense star-forming regions can harbor much higher column densities where extinction values frequently exceed $A_V >> 2.0$ mag. To capture this full dynamic range, our mapping methodology must be sensitive to both these modest diffuse levels and the high-extinction cores of molecular clouds.

We characterize the dynamic range of extinction we can probe by defining the maximum $A_V$ to which our stellar catalogs remain sensitive. This limit is intrinsically tied to a star's intrinsic luminosity; more luminous stars remain detectable behind significantly more dust than their dimmer counterparts.

To quantify $A_V$ detection limits, we employ artificial star tests (ASTs). ASTs are principally used to quantify how observing conditions (e.g., blending with emission, crowding from nearby stars, and background sky noise) can bias and scatter the observed magnitude of a star. We quantified these effects using 18,575 ASTs in a dusty 30 Doradus field (15891\textunderscore LMC-5421ne-12728). By injecting artificial stars with a range of extinctions ($A_{V} = 0–10$ mag) into the images and measuring their recovery rate, we can quantify how extinction impacts the recovery rate for stars with different intrinsic luminosities. 

Using these extinguished artificial stars, we define the maximum $A_V$ detection limit as the level of extinction at which 68\% of ASTs in a specific luminosity bin ($\log L/L_{\odot} = -1.0\mathrm{\ to\ }3.0, 0.1$ dex) \textit{remain} detectable in F336W. We chose a detection limit of 68\% to align with $1\sigma$ statistical confidence. We base our detection thresholds on the F336W filter because it is the bluest band available across all fields analyzed with the BEAST, and is therefore the most sensitive to dust extinction, as opposed to F475W and F814W. 
By adopting F336W as our maximally constraining filter, we account for the selective absence of reddened sources that would otherwise bias the average observed extinction low.

In Figure \ref{fig:max_av} (left), we plot the maximum detectable $A_V$ (based on the 68\% detection limit) as a function of the intrinsic luminosity ($\log L/L_{\odot}$) of the artificial stars (dashed black line). We compare this $A_V$ limit to the median $\pm 1\sigma$ observed $A_V$ (grey) measured in Scylla field 15891$\textunderscore$LMC-5421ne-12728. 
For most sources, we find that the average observed extinction remains far below their respective maximum $A_V$ limits, confirming that most sources are unlikely to suffer from completeness issues due to extinction. 

We define an intrinsic luminosity-based quality cut based on $A_V$ sensitivity in F336W using the relationship between luminosity and extinction shown in Figure \ref{fig:max_av} (left). To support a high level of source completeness, we applied a conservative quality cut of $\log L/L_{\odot} \ge 0.8$ (dotted cyan line). This ensures that even in active star-forming regions, stars are bright enough to be recovered with $A_V \geq 3$ mag. 

In Figure \ref{fig:max_av} (right), we plot the optical CMD of observed stars in the field (grey, $n=15,318$ stars). All stars that pass the luminosity cut ($\log L/L_{\odot}) \ge 0.8$; $n=3,028$ stars) are colored by their theoretical upper detection limit of $A_V$ based on complementary ASTs with similar luminosities. Stars in the upper main sequence and RGB are generally sensitive to $A_V = +3$ mag, while the most luminous sources ($\log L/L_{\odot} > 3$) remain detectable through $A_V > 6$ mag. 

Intrinsic luminosities for all observed sources are obtained from the BEAST SED fits, which derive theoretical intrinsic luminosities based on the median mass, age, metallicity, and distance derived for each star. With this luminosity cut, we retain 47,922 of the original 229,005 catalog stars in the LMC, with an average of 1,065 sources per field. In the SMC, we retain 36,988 of the 182,780 catalog stars, with an average of 1,608 sources per field.

\section{Method}
\label{sec:method}

To enable comparison with emission-based ISM tracers and constrain the total ISM mass budget within each region, we aim to measure the total column density of dust extinction. However, without precise distance measurements for individual stars, we rely on a statistical methodology to disentangle the line-of-sight distribution of dust from the distribution of stars sampling it. Our technique works by identifying the spatial correlation of extinction measurements while accounting for the fact that stars are distributed at various depths throughout the SMC and LMC.

In Section \ref{sec:los}, we discuss the line-of-sight decomposition technique we use to isolate background stars probing the entire ISM column. In Section \ref{sec:gp}, we describe the 2D spatial interpolation method used to construct the $A_V$ maps, followed by a discussion of the resolution, uncertainties, and MW foreground subtraction. \footnote{An example notebook of the dust extinction mapping methodology described in this section is publicly available at https://doi.org/10.5281/zenodo.18602607, including custom functions and an example data set.}

\subsection{Line-of-Sight Depth}
\label{sec:los}

To compensate for our lack of distance measurements for individual stars \changes{in the MCs, we separate these extragalactic stars into ``foreground" and ``background" populations} based on their relative extinction distributions. This methodology works by assuming that foreground stars (i.e., \changes{extragalactic} stars in front of the bulk ISM of the SMC or LMC) will only experience extinction from the MW ISM, while \changes{extragalactic} background sources will experience extinction from both the MW and any ISM within the SMC or LMC. Previous studies estimate that MW foreground extinction contributes $A_V = 0.1\mbox{--}0.4$ mag \citep{Subramaniam2010, Chen2022}, whereas contributions from the SMC and LMC depend on the location within each galaxy. 

We expect that the surface density of the ISM will follow a log-normal distribution due to injected turbulence \citep{kolmogorov, vaszquez-semadeni2001, federrath2010, dalcanton2015, burkhart2015}. The $A_V$ experienced by a population of foreground sources should thus follow a log-normal distribution due to the MW ISM. Meanwhile, the $A_V$ experienced by background sources should follow a log-normal distribution with larger amplitude based on the MC (LMC or SMC) ISM, plus the contribution from the MW ISM \citep{ymj2021}. Therefore, it should be possible to model the $\log(A_V)$ distribution of the two populations as a combination of two Gaussian distributions.

\begin{figure*}
    \centering
    \includegraphics[width=0.8\linewidth]{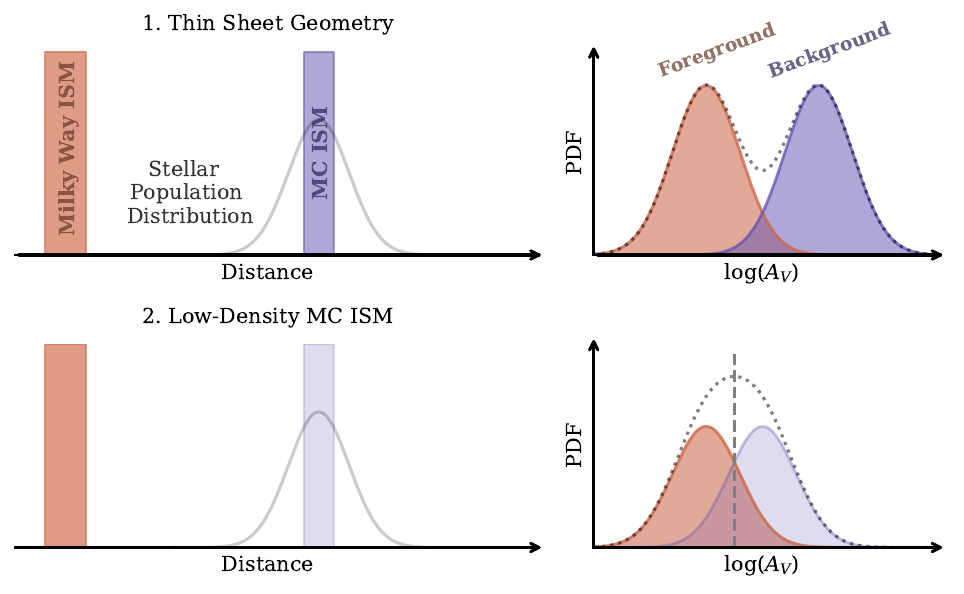}
    \caption{\textbf{Line-of-sight geometries:} 1. Thin Sheet Geometry: Assuming a Gaussian stellar population distribution where the stellar distribution is much broader than the galaxy ISM (left), one can assume that any foreground stars will only experience extinction from the MW, while background stars will experience extinction from both the MW and the galaxy, making it possible to distinguish the two populations by their $\log(A_V)$ distribution (right). However, if the dust contributions from the galaxy are minimal (2.~Low-Density MC ISM), it becomes harder to disentangle the foreground and background stellar populations with a simple two-component GMM, and we resort to a single Gaussian fit with a mean cut (dashed grey line) to isolate background sources. }
    \label{fig:LOS_geometry}
\end{figure*}

In the following section, we motivate different potential star-dust geometries (Figure \ref{fig:LOS_geometry}), and describe the resulting extinction distribution we anticipate for each scenario. Based on these possible geometries, we outline a multi-step workflow for disentangling foreground from background sources in each field.

\begin{enumerate}
    \item Thin Sheet Geometry (Fig.~\ref{fig:LOS_geometry}, top): If the distribution of sources is centered around the MC ISM (SMC or LMC), and the scale height of the ISM is much smaller than the scale height of the sources, then the MC ISM can be approximated as a thin sheet. Any sources in front of the galaxy will only experience extinction from the MW ISM (shown in red), while sources behind the galaxy will experience cumulative extinction from both the MW and the MC ISM (shown in blue).
    
    If the MC ISM contribution is on the order of -- or greater than -- the MW ISM, then the observed distribution of $\log(A_V)$ can be individually characterized using a Bayesian Gaussian mixture model (GMM). We perform our decomposition with $\log(A_V)$ values since GMMs are more mathematically tractable and robust to outliers than log-normal mixture models. In this scenario, we should be able to separate foreground sources from background sources using a simple two-component Gaussian Mixture model.

    \item Low-Density MC ISM (Fig.~\ref{fig:LOS_geometry}, bottom): 
    As extinction contributions from the MC ISM approach zero, the two Gaussian distributions of $\log(A_V)$ merge, and it becomes harder to cleanly separate the foreground and background stellar populations. In this scenario, we opt to fit the distribution with a single Gaussian and assume any sources with $A_V$ greater than the fitted mean $A_V$ to be background sources.

\end{enumerate}


Across our sample, we expect to observe a range of geometries spanning these examples, so we employ the following methodology to isolate background sources: 

\begin{enumerate}
    \item Within each field, split the field into $n\times m$ sub-regions and increase the sub-region sizes until there are \textit{at least} $N$ sources within each sub-region ($N=[10, 20, 30, .., 100]$). The range of $N$ was selected based on the general stellar density within the Scylla and METAL fields. This process results in maps with $3 - 6$ pixels per side, depending on the stellar density. 
    
    We break each Scylla field into multiple regions to account for variable dust structure within each field. For instance, if a field contained a molecular cloud in one quadrant, fitting the fields as a single region would bias all ``background sources" to be located in that same quadrant. By splitting the field up into multiple sub-regions, we avoid many of these issues and ensure homogeneous sampling across the field.
    \item For all sources within each sub-region, use the \texttt{scikit-learn} Bayesian Gaussian mixture model to see if the distribution of $\log(A_V)$ is better fit as a single- or two-component model.
    For a two-component model to be preferred, we require that: 
    \begin{enumerate}
        \item the weight of either mixture component is at least 10\%. This avoids scenarios where the second component only represents a small fraction of sources.
        \item the means of the two components are at least 0.2 log-mag apart. This avoids scenarios where two components are fit to represent a broader Gaussian or Student-T distribution.
    \end{enumerate}
    \item If a two-component fit is preferred, then tag any sources in the higher $A_V$ component as background sources. 
    \item If a single-component fit is preferred, then tag any sources with $A_V$ measurements greater than the Gaussian mean $A_V$ as background sources.
        
    \item To ensure each stellar tag is robust, bootstrap the catalog of sources within each sub-region 100 times and recompute the foreground and background component labels. Only sources tagged as background at least 90\% of the time are included in the catalogs for constructing the final dust maps.
\end{enumerate}

To recap, this methodology for decomposing sources along the line-of-sight relies on the following assumptions: (1) the column density of the ISM within a galaxy can be approximated as a log-normal distribution; (2) both the MW ISM and the MC ISM contribute to the observed extinction of background sources; and (3) the scale height of stars in the SMC and LMC is greater than the scale height of the ISM. With this methodology, we retain an average of 30\% of stars, $338^{+238}_{-167}$ stars, within each field as background stars.

Previous dust mapping of the SMC and LMC has largely relied on inferring extinction from stellar populations, as opposed to $A_V$ measurements from individual stars \citep{zaritsky2002, zaritsky2004, Haschke2011, skowron2021}. These studies typically leverage older stellar populations, such as red giant branch (RGB) or red clump stars, to separate foreground and background sources. This decomposition is possible because older stars are more dynamically heated than the star-forming thin disk, and therefore have a scale height greater than the ISM within a galaxy \citep{dalcanton2015, hawcroft2024}. 
While this methodology is valid for the SMC and LMC, it limits the sample size of sources considerably, as there are only an average of $160^{+175}_{-80}$ RGB stars (defined as $\log T/T_{\odot} < 3.72$ or $5250$ K) present within a single Scylla or METAL field. Given that only half of these sources are expected to be located behind the bulk ISM content of the SMC or LMC, this reduction in source density severely limits the spatial resolution of the resultant dust maps. 

Ultimately, the notion of tagging stars as foreground or background sources will always result in some amount of contamination from foreground sources or over-restriction on background sources. In the future, if improved distance estimates towards individual stars become available, this methodology can be updated to incorporate distance measurements.

\subsection{2D Dust Mapping}
\label{sec:gp}

Having isolated background sources within each field, we now interpolate between these sources to infer the 2D distribution of extinction. We implement kriging, an advanced geostatistical method used to estimate continuous surfaces from scattered sets of measured data points \citep{matheron1963, cressie1992}. Similar to Gaussian processes, kriging uses statistical models to account for spatial relationships between measurements, providing both predictions and measures of prediction uncertainty. With the package \texttt{pykrige} \citep{murphy2021}, we employ an ordinary kriging model, which assumes a constant mean and no underlying trends across the region, combined with a spherical variogram model (similar to a Matern kernel) to model the 2D distribution of $A_V$. We sample the model across the spatial extent of the original $HST$ field to obtain a 2D map of the predicted $A_V$ and standard deviation across the field. 

We perform this fitting using $\log(A_V)$ measurements to ensure positivity. To convert the map values and uncertainties back to regular quantities, we implement the following conversions:

$${A_V} =\exp{( \mu_Y + \sigma^2_Y/2})$$

and 
    
$$\sigma^2_{A_V}= (\exp{(\sigma^2_Y)}-1) * \exp{(2 \mu_Y + \sigma^2_Y)}$$

where $\mu_Y$ is the mean kriging prediction from the log-transformed data and $\sigma^2_{Y}$ is the variance from the log-transformed data. 

We compute ten individual dust maps for each field, varying the minimum required sources needed within each sub-region ($N=[10, 20, 30, ... , 100]$) for constructing the sample of background stars. The final dust map for each field is the median of all ten individual dust maps, weighted equally, allowing us to avoid edge effects from sub-regions. To calculate the total uncertainty of each map ($\sigma_{A_V,\ total}$), we compute the median variance across all map instances and combine this with the variance between map instances to present the average uncertainty. 

$$\sigma_{A_V,\ total} = \sqrt{\textrm{median}(\sigma^2_{A_V}) + \textrm{Var}(\sigma^2_{A_V})}$$

While the first term broadly measures the inherent increase in variance at greater $A_V$, the secondary term captures a significant portion of the stochastic variations that occur in the variance maps due to diffraction spikes or other sources of anomalously high extinction.

\subsubsection{Resolution}

We construct dust maps with $4^{\prime \prime}$ resolution, equivalent to 0.96 parsec and 1.20 parsec in the LMC and SMC, respectively. This resolution is based on the length scales of correlation between $A_V$ measurements captured in the variograms. While most variograms typically peak at larger scales (e.g., 20-pc or half an $HST$ field), some also exhibit smaller peaks indicative of multi-scale structure persistent down to sub-parsec scales. A 1-pc resolution allows us to capture both broad atomic structures and small-scale molecular structures identified with other ISM tracers \citep{wong2022}.



\subsubsection{Uncertainty}\label{sec:uncert}

There are several sources of uncertainty within the extinction maps: (1) uncertainty of individual $A_V$ measurements for each star; (2) assignment of foreground/background sources; (3) contamination by anomalously high-$A_V$ sources; and (4) variance within the maps themselves. In this section, we discuss the relative contributions of each of these uncertainty sources.

The $A_V$ uncertainty of individual stellar SED fits varies between sources. On average, though, $\sigma_{A_V}< 0.1$ mag, and therefore does not constitute a significant source of uncertainty in the final dust extinction maps. 

The assignment of foreground and background provides another source of uncertainty. To limit variance from this source, we bootstrap the assignment of sources, only keeping sources considered "background" at least 90\% of the time. This technique reduced the variance of the background catalogs considerably in each field.


Despite previous high-reliability quality cuts imposed on the SED fits in \citetalias{lindberg2025}, we still find occasional contamination in the SED catalogs from diffraction spikes, background galaxies, and diffuse gas emission, in about $\sim35\%$ of fields. Due to their peculiar SEDs, these contaminants are often characterized by high values of extinction (see Figure \ref{fig:artifacts}b for an example). Unfortunately, because of the variable filter coverage, exposure times, and CCD locations, no singular quality cut can be applied across all fields to remove all contaminant sources without risking the removal of genuine high $A_V$ sources, which would introduce significant biases in the extinction maps.

We test the influence of these anomalous high-$A_V$ sources by randomly inserting sources with high extinction ($A_V=4.0 \mbox{--}10.0$ mag) into the background source catalog of a low extinction field ($A_V<1.0$ mag) and reconstructing the predicted extinction maps. We find that contaminant sources will generally create a spike with a radius of 1-2 pc within the $A_V$ maps; however, the median reported variance ($\textrm{median}(\sigma^2_{A_V})$) from kriging will also increase at the locations of these spikes, such that the signal-to-noise ratio remains proportional. Moreover, the average magnitudes of the spikes remain between $A_V = 1.5\mbox{--}3.5$ mags, regardless of the simulated extinction of the source ($A_V=4.0\mbox{--}10.0$ mag). Unfortunately, these anomalous high-$A_V$ sources are consistently selected as background sources, meaning that the secondary variance term ($\textrm{Var}(\sigma^2_{A_V})$) fails to capture their presence. 

Anomalously high-Av sources provide a source of contamination in the extinction maps, however, their impact is generally localized and does not impact the overall $A_V$ measured within a field. The one exception to this is 16786$\textunderscore$SMC-4451ne-16362 (see Figure \ref{fig:artifacts}c), which is dominated by anomalous high-$A_V$ sources. In this scenario, $A_V$ is overestimated across

\begin{figure*}
    \centering
    \includegraphics[width=\linewidth]{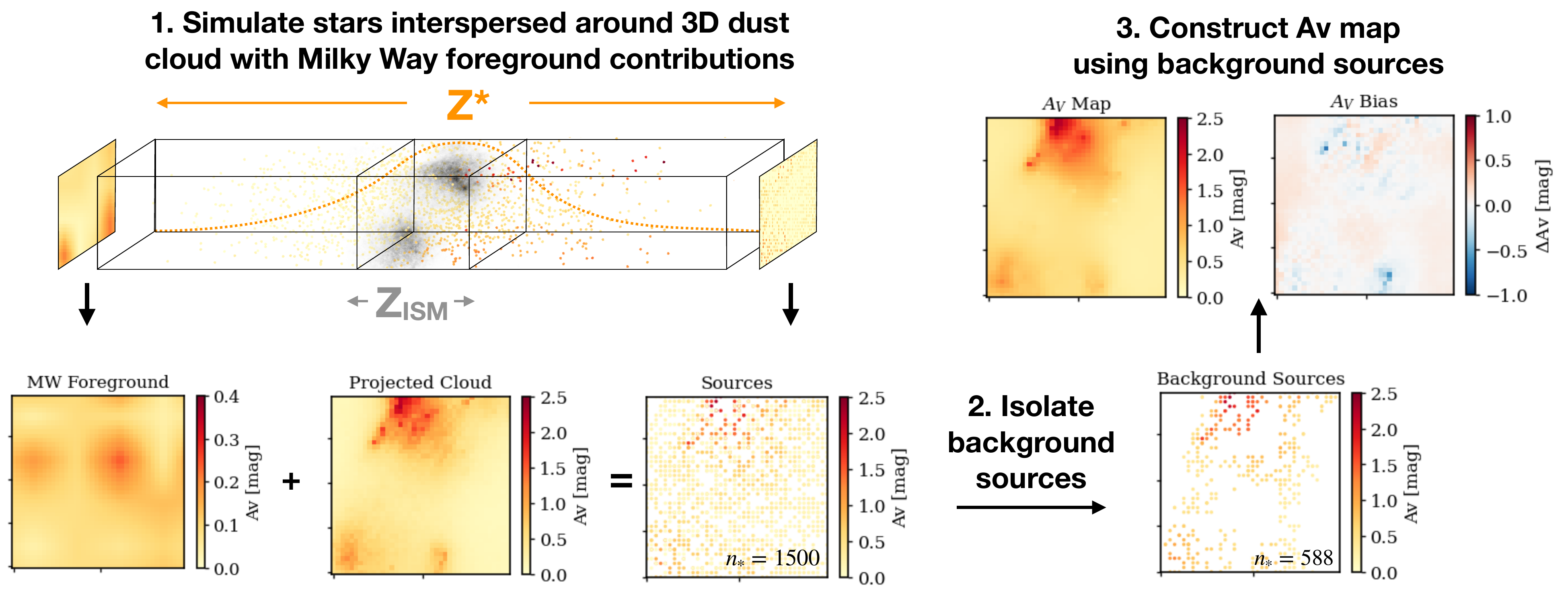}
    \caption{\textbf{Schematic of validation tests with 3D dust maps}: We use MW 3D dust clouds from \citet{zucker_three-dimensional_2021} to test how different star-dust geometries hinder our ability to recover total column densities of dust extinction. \changestwo{In a $40\times40$~pc face-on box}, $n_{\star}$ stars (e.g., 1500) are simulated and interspersed along the line-of-sight ($Z_{\star} = f \times Z_{ISM}$, where $f=1\mbox{--}10$), following a Gaussian distribution. A modest amount of extinction from the MW foreground is added to the integrated dust column density towards each source. We isolate background sources, using the methodology outlined in Section \ref{sec:los}, and construct 2D dust maps using the background sources to compare with the true column densities of the 3D dust cloud.}
    \label{fig:mw_sim}
\end{figure*}

The dominant source of uncertainty stems from the intrinsic scatter in $A_V$ between spatially neighboring sources. This uncertainty is directly captured by the kriging method in the variogram uncertainty, defined as the data-driven ``nugget'' of variability that exists even at infinitesimally small distance separations. This uncertainty metric encompasses both the random error inherent in the measurement process (e.g., $\sigma_{A_V}$) and the spatial variation that occurs at scales finer than the sampling resolution. We report this variogram uncertainty for each extinction map. We generally find $A_V$ signal-to-noise ratios (SNR) of $2.5\mbox{--}3.0$ within each field.

\subsubsection{Milky Way Foreground Subtraction}

To compare dust maps with other ISM tracers, we need to remove dust contributions from the MW. Previous studies towards the LMC and SMC have measured MW foreground extinction around $A_V = 0.2$ mag towards the LMC and $A_V = 0.1$ mag towards the SMC \citep{Chen2022}
To estimate the contribution of the MW foreground, we use observations of neutral hydrogen (\hi) emission at $21\rm\,cm$ from the Galactic All Sky Survey \citep[GASS;][]{mccluregriffiths2009}. We identify MW gas in the direction of the LMC and SMC as emission with velocities in the local standard of rest between $-100$ and $60\rm\,km\,s^{-1}$, and compute the foreground \hi\ column density by integrating over this range. We then convert \hi\ column to $A_V$ using the conversion factor calibrated for the diffuse, high-latitude ISM \citep{liszt2014}, and assuming a MW value of $R_V=3.1$. The MW foreground extinction measurements towards Scylla and METAL fields range from $A_V = 0.10\mbox{--}0.12$ mag towards the SMC, and $A_V = 0.24\mbox{--}0.28$ towards the LMC, in agreement with the existing literature.

The GASS survey has an angular resolution of $23^{\prime}$, meaning the pixel resolution of MW foreground extinction map is much larger than the size of the individual $HST$ fields in the Scylla and METAL survey. To remove the MW contribution, we measure the MW foreground values at the center of each field location and subtract this value from the entire $A_V$ map within each field. \changes{The foreground values removed from each field can be found in Table \ref{tab:fields}.}

\begin{figure*}%
    \centering
    \subfloat[\centering Extinction Maps]{{\includegraphics[width=0.46\textwidth]{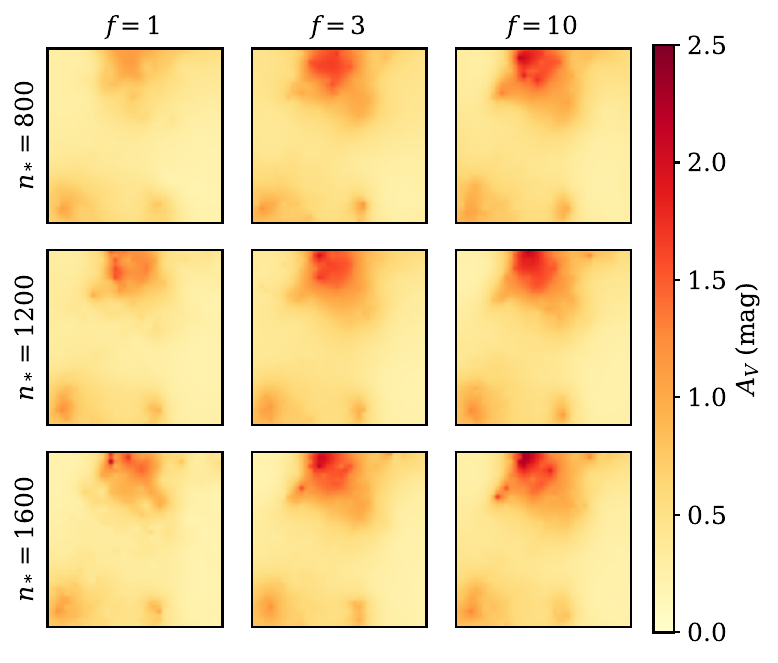} }}%
    \qquad
    \subfloat[\centering Residual Bias Maps]{{\includegraphics[width=0.48\textwidth]{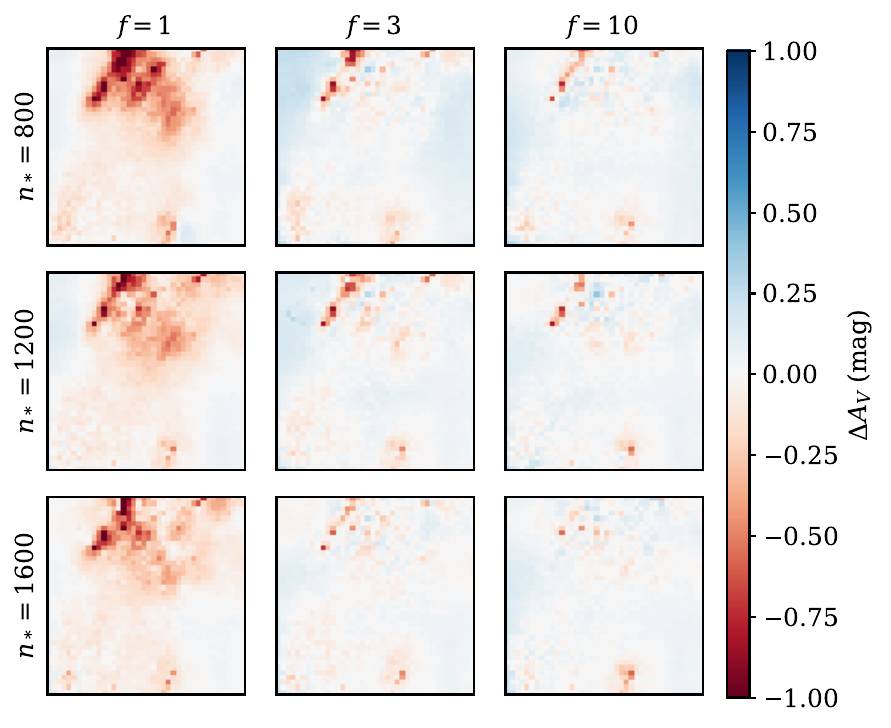}}}%
    \caption{\textbf{Simulation results}: Resultant $A_V$ maps (left) and biases (right) simulated with different stellar densities ($n_{\star}$) and star-dust geometries ($f$, where $Z_\star = f \times Z_{ISM}$) to mimic the potential observational conditions within the Scylla and METAL fields ($\bar n_\star = 1000-1600$). As the scale height of stars increases ($f\gg1$), the method can more accurately recover high $A_V$ structures, since a larger fraction of sources sample the total column density of the ISM. However, even under the most ideal of conditions, the method can still fail to capture compact, high $A_V$ clouds simply because there are no sources that probe these structures. }%
    \label{fig:grid_oriona}%
\end{figure*}

\section{Validation with 3D Dust Maps}\label{sec:mw_val}

To validate the methodology, we perform a series of tests using 3D dust maps of molecular clouds located in the Milky Way. These clouds were originally modeled in the \citet{leike2020} dust map, which characterized the 3D structure of the ISM within $\sim$400 pc of the Sun to 1 parsec spatial-resolution using Gaia extinction and parallax measurements \citep{gaiacollaboration2018}. This map was used by \citet{zucker_three-dimensional_2021} to isolate and study the 3D structure of ten molecular clouds and complexes (e.g., Orion A, Perseus, Taurus, Chameleon, etc.), producing volume density cubes for each cloud\footnote{Available online: https://doi.org/10.7910/DVN/IADP7W}. Using these data cubes, we can simulate different star-dust geometries we might observe in the SMC and LMC (Figure \ref{fig:LOS_geometry}), and test how robust the method is at recovering the total integrated dust extinction of each cloud.

In Figure \ref{fig:mw_sim}, we show a diagram of the setup described below. The average molecular cloud volume cube spans $70\mbox{--}150$ pc along any axis. Rather than performing tests on the whole cloud, we randomly restrict the face-on xy size of the cloud to an HST WFC3 field-of-view in the SMC/LMC (e.g., $\sim40\times40$ pc). This allows for a more realistic comparison between the MW maps and the observations in the SMC and LMC.

We include the effects of foreground dust from the MW by adding a diffuse screen of extinction to the front of the cube. We would expect any MW ISM to have angular length scales larger than any ISM in the SMC or LMC (e.g., 10 times larger if assuming an average distance of 5 kpc). To simulate these increased length scales, we randomly sample $4\times4$ values from a log-normal distribution ($\mu=-2.5$, $\sigma=0.7$) with a peak of $\sim0.08$ mag \citep{Subramaniam2010, Chen2022}. We then resize these pixel values to $10\times10$ parsec pixels each and smooth between the edges using anti-aliasing, creating a smooth map of simulated MW foreground extinction with $A_V = 0.02 \mbox{--} 0.2$ mag.

To simulate different scale height ratios between the ISM and stars, we expand the z-axis of the cube (i.e.~distance) by a factor $f=1\mbox{--}10$, such that $Z_\star = f \times Z_{ISM}$. This allows us to place stars both in front of, inside of, and behind the molecular cloud. We place a range of stars ($n_{*} = 800\mbox{--}1600$) to mimic the stellar densities observed in the HST fields after applying luminosity-based quality cuts ($\bar n_\star = 1000-1600$). Stars are placed randomly on the x- and y-axis. Along the z-axis, stars are placed in a Gaussian distribution, centered on the middle of the cube, with $\sigma=1/6$ of the full z-axis. For simplicity, we assume all simulated sources are complete, since source completeness is already addressed for our actual observations (Section \ref{sec:complet}).

Finally, we measure the total column density of extinction towards each star by integrating all pixels along the line-of-sight ($Z$) towards the front of the cube, adding any contributions from the simulated MW foreground. We then isolate background stars with the selection process described in Section \ref{sec:los}, and interpolate between stars with kriging to produce 2D extinction maps, as described in Section \ref{sec:gp}. We compare these simulated extinction maps with the ``true" projected cloud column densities to determine the accuracy of the methodology in isolating background sources and recovering the total integrated dust extinction of clouds.

\begin{figure*}
    \centering
    \includegraphics[width=\linewidth]{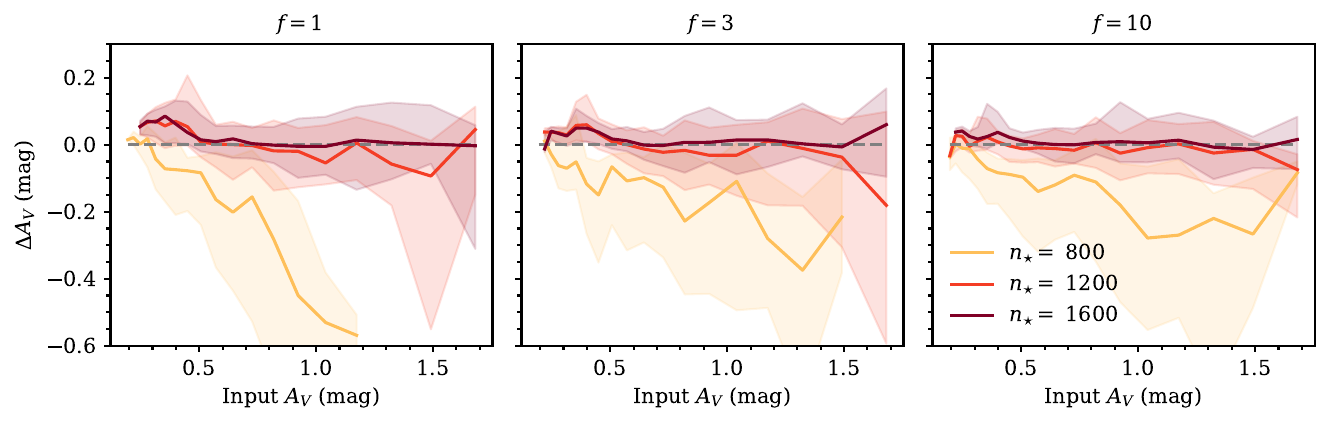}
    \caption{\textbf{Simulation biases}: Median extinction bias ($\Delta A_V = A_{V,\,sim} - A_{V,\,true}$) and the $\pm 1 \sigma$ spread (shaded regions) as a function of the true $A_V$ from the simulations presented in Figure \ref{fig:grid_oriona}b. Each panel represents a different star-dust geometry, $f$, where $f$ is the ratio of scale heights between the stars and the ISM ($Z_\star = f \times Z_{ISM}$). These simulations consistently show that increasing the number of stars ($n_{\star}$)  leads to a decrease in the scatter (shaded regions) and an improvement in the accuracy (colored lines) of the recovered extinction values. }
    \label{fig:sim_nstars}
\end{figure*}

\begin{figure*}
    \centering
    \includegraphics[width=0.8\linewidth]{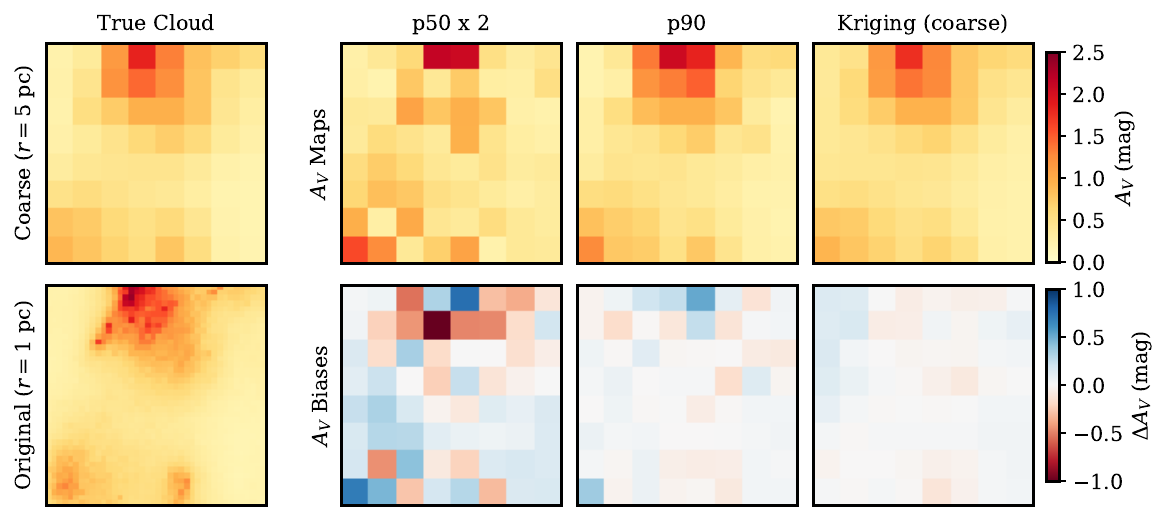}
    \caption{\textbf{Comparison with alternative methods for dust mapping}: We compare $A_V$ maps (top) and their resulting biases (bottom) generated by different dust mapping techniques. These include a ``true'' $A_V$ map (left), along with maps derived from two times the 50th percentile of the total $A_V$ distribution, the 90th percentile of the distribution, and our kriging method. Maps are based on the optimal simulation using data from the Orion A molecular cloud with $n_\star=1200$ sources and a stellar scale height factor $f=3$, reprojected to 5-pc pixel scales (Figure \ref{fig:grid_oriona}a). Even with a fraction of the original stars, the kriging method effectively captures the structure and $A_V$ range of the molecular cloud and can be sampled at higher resolutions.}
    \label{fig:alt}
\end{figure*}

To assess the impact of source density and star-dust geometry on the extinction mapping methodology, we perform a series of simulations using data from the 3D dust map of the Orion A molecular cloud. 
Implementing the extinction mapping methodology, Figure \ref{fig:grid_oriona}a presents nine $A_V$ maps constructed using different simulation parameters, i.e.~varying the number of stars ($n_{
\star}$) and the ratio of scale heights between the ISM and the stars ($f = Z_{\star}/Z_{ISM}$). Figure \ref{fig:grid_oriona}b displays the corresponding residual bias maps ($\Delta A_V = A_{V,\,sim} - A_{V,\,true}$) for each simulation, defined as the constructed $A_V$ maps minus the true $A_V$ maps. The true $A_V$ map can be viewed for comparison in Figure \ref{fig:alt}.

In general, the predicted extinction in low $A_V$ areas tends to be overestimated, while high $A_V$ areas are generally underestimated, resulting in a flattening of the dynamic range of extinctions measured. These trends are especially severe in fields where stars are only embedded within the ISM ($f=1$). However, as soon as the scale height of the stars increases ($f\geq3$), the methodology is much better at recovering structures with high $A_V$. For the average simulation with $n_{\star}=1200$ sources and $f=3$ (center), we find a median bias of $\mu=0.02$ mag with a standard deviation of $\sigma=0.1$ mag. 
We repeat these simulations for all ten 3D molecular clouds available, and find similar modest trends in over-/underestimation for low and high $A_V$ regimes, respectively.

While Figure \ref{fig:grid_oriona} provides a sense of the spatial biases present in the simulations, we also want to quantify how the accuracy of the extinction mapping methodology varies across different extinction regimes. In Figure \ref{fig:sim_nstars}, we calculate the median extinction bias (colored line) and $\pm1\sigma$ (shaded regions) as a function of input $A_V$ for the nine simulations shown in Figure \ref{fig:grid_oriona}. For visual clarity, we present the three star-dust geometries ($f=1,3,10$) in separate panels, with each panel displaying three lines representing different stellar densities: $n_{\star}=800$ (yellow), $n_{\star}=1200$ (orange), $n_{\star}=1600$ (red). These simulations consistently demonstrate that increasing the number of stars ($n_{\star}$) significantly decreases the scatter (shaded regions) and improves the accuracy (colored lines) of the recovered extinction values. Again, we find that low extinction regions ($A_V < 0.3$ mag) are slightly overestimated by $A_V\approx0.05$ mag, especially when $f=1$. Meanwhile, high extinction regions ($+0.4$ mag) are generally underestimated; however, with increased source density, this underestimation is reduced to $A_V=0.1$ mag.


The scale height ratio between stars and the ISM (quantified as $f$ in these simulations) has a measurable impact on our ability to accurately measure the total column density of extinction, especially for fields with low source density (Figure \ref{fig:sim_nstars}). However, this scale height ratio is difficult to measure in the LMC and SMC due to their irregular stellar and gas morphologies, which likely stem from recent dynamical interactions \citep[$<250$ Myrs ago;][]{donghia2016, zivick2018, murray2019, choi2022}. 
While the scale height of \hi\ and dust in the LMC has been measured to be on the order of 100-200 pc \citep{Padoan2001, elmegreen2001, Block2010}, the flared disk structure of the LMC causes the scale height of stars to vary radially, from 270 pc in the center to as much as 1.5 kpc at a radius of 5.5 kpc, resulting in the scale height ratio varying from $2\mbox{--}10$. However, in central fields near active star formation, we expect the majority of dust to stem from denser molecular gas, which should have a much smaller scale height than the diffuse atomic gas, increasing the expected scale height ratio between stars and the ISM. Similar principles apply to the SMC, with the exception that the SMC is much more irregular and is thought to have a stellar line-of-sight depth on the order of $12-14$ kpc \citep{subramanian2012, ymj2021}, resulting in large scale height ratios, even when the ISM is dominated by atomic gas.

With applied completeness cuts (Section \ref{sec:complet}), the number of sources in the HST fields ranges from $n_{\star} = 337 \mbox{--} 2790$. Even assuming a baseline scale height ratio of $f=10$, fields with source densities of $n_*<800$ will likely be underestimated in their total column density. From Figure \ref{fig:sim_nstars}, for simulations with $n_*=800$ and $f=3$ and $f=10$, we see that $\Delta A_V$ is correlated with $A_V$ and generally increases at a rate of $-0.3$ mag per magnitude of actual extinction, meaning that while denser structures with $A_V>3$ mag might be underestimated by up to $1$ mag, any lower density structures should still have $\Delta A_V < 0.5$ mag. 

We conclude that, for the average Scylla or METAL field with $n_*\geq 1200$, the extinction mapping methodology presented here should be capable of reliably recovering the total column density distribution of dust extinction within the Scylla fields to an accuracy of $A_V=0.1$ mag, with the exception of the dustiest compact molecular regions. However, for fields with $n_*<1000$, we urge users to treat the resultant $A_V$ maps as a lower limit, especially if these fields show evidence of dense gas structures with $A_V > 1$ mag.


 

\subsection{Alternative Dust Mapping Methods}
\label{sec:alt}

The dust extinction methodology presented above offers a remarkable improvement compared to other common dust mapping techniques, such as characterizing dust extinction through stellar population averages. For example, one basic technique is to bin stars into angular bins with some minimum number of sources, and then characterize the total column density of dust extinction as either double the median ($2\times p_{50}$) or the 90th percentile ($p_{90}$) of the $A_V$ distribution within each bin.

To illustrate this improvement, we divide all the sources in the average Orion A simulation ($f=3$, $n_\star=1600$, Figure \ref{fig:grid_oriona}) into 5-pc bins and compute the $2\times p_{50}$ and $p_{90}$ $A_V$ values for each pixel. To compare with the original dust extinction mapping method, we resample the kriging model with a resolution of 5 pc, instead of 1 pc. In Figure \ref{fig:alt}, we compare these maps (top) and their resulting biases (bottom) with the ``true" $A_V$ map (left), reprojected to 5-pc pixel scales.


While all three techniques broadly capture the existence of a molecular cloud, there are notable differences in the structure and column density ranges they recover. The $2\times p_{50}$ method grossly underestimates and overestimates $A_V$ across the entire map and does a poor job of recovering the structure of the cloud. By comparison, the $p_{90}$ method does a much better job of capturing the structure of the molecular cloud, but overestimates the extinction of the cloud across most of the region. Meanwhile, the original kriging methodology, which uses only a fraction of the original number of stars ($\sim 30\%$), can capture both the structure and $A_V$ range of the molecular cloud, and is capable of producing maps with higher resolutions than the other two methods.

\begin{figure*}[ht!]
\includegraphics[width=\linewidth]{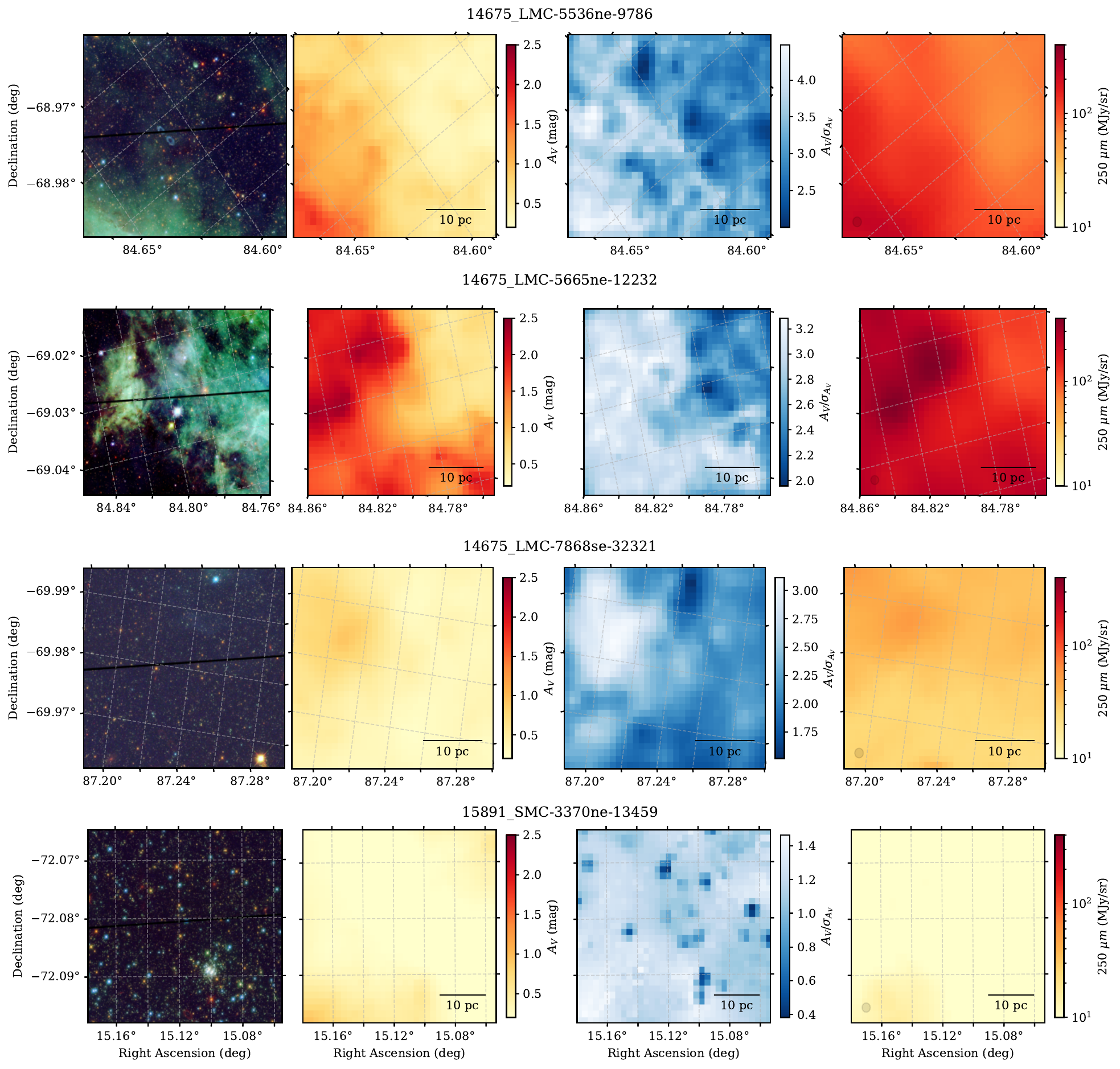} 
\caption{\textbf{Dust extinction and emission maps}: (Left) Three-color images of four Scylla and METAL fields, constructed with HST WFC3 filters F336W (blue), F475W (green), and F814W (red); (middle panels) resultant dust extinction maps and uncertainties derived from stellar SED fits (LMC: $4.1^{\prime\prime}$; SMC: $3.5^{\prime\prime}$); (right) Herschel $250\, \mu m$ dust emission \citep[$6.0^{\prime\prime}$;][]{clark2021}. Dust extinction spans over an order of magnitude ($A_V=0.2\mbox{--}2.5$ mag) and shows spatial correlation between areas of elevated $250\, \mu m$ dust emission. These are representative images from the complete figure set (68 fields total) available in the online journal. The full set includes dust extinction maps, uncertainties, and FIR emission for all Scylla and METAL fields processed in this paper. \label{fig:results}}
\end{figure*}

\begin{figure}
    \centering
    \includegraphics[width=\linewidth]{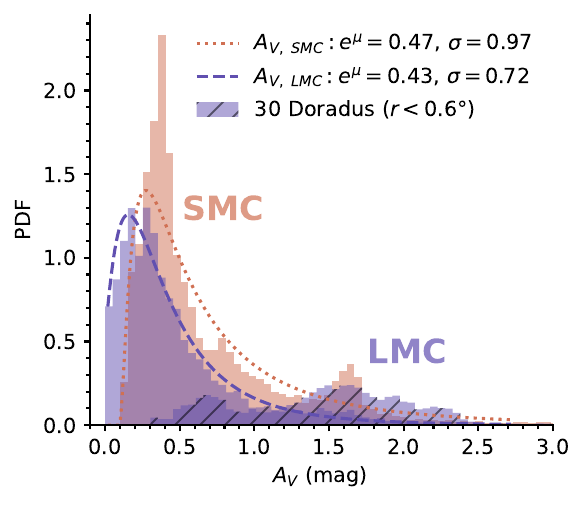}
    \caption{$A_V$ distribution for all fields in the SMC (red) and LMC (blue) based on total column density measurements from extinction maps (dashed and dotted lines). These $A_V$ distributions can be characterized as log-normal distributions, especially when excluding fields within the 30 Doradus region (hatched), \changes{and peak at $A_{V,\ SMC}= 0.38$ mag and $A_{V,\ LMC} = 0.28$ mag}.}
    \label{fig:av_distr}
\end{figure}


\begin{figure*}
    \centering
    \includegraphics[width=0.9\linewidth]{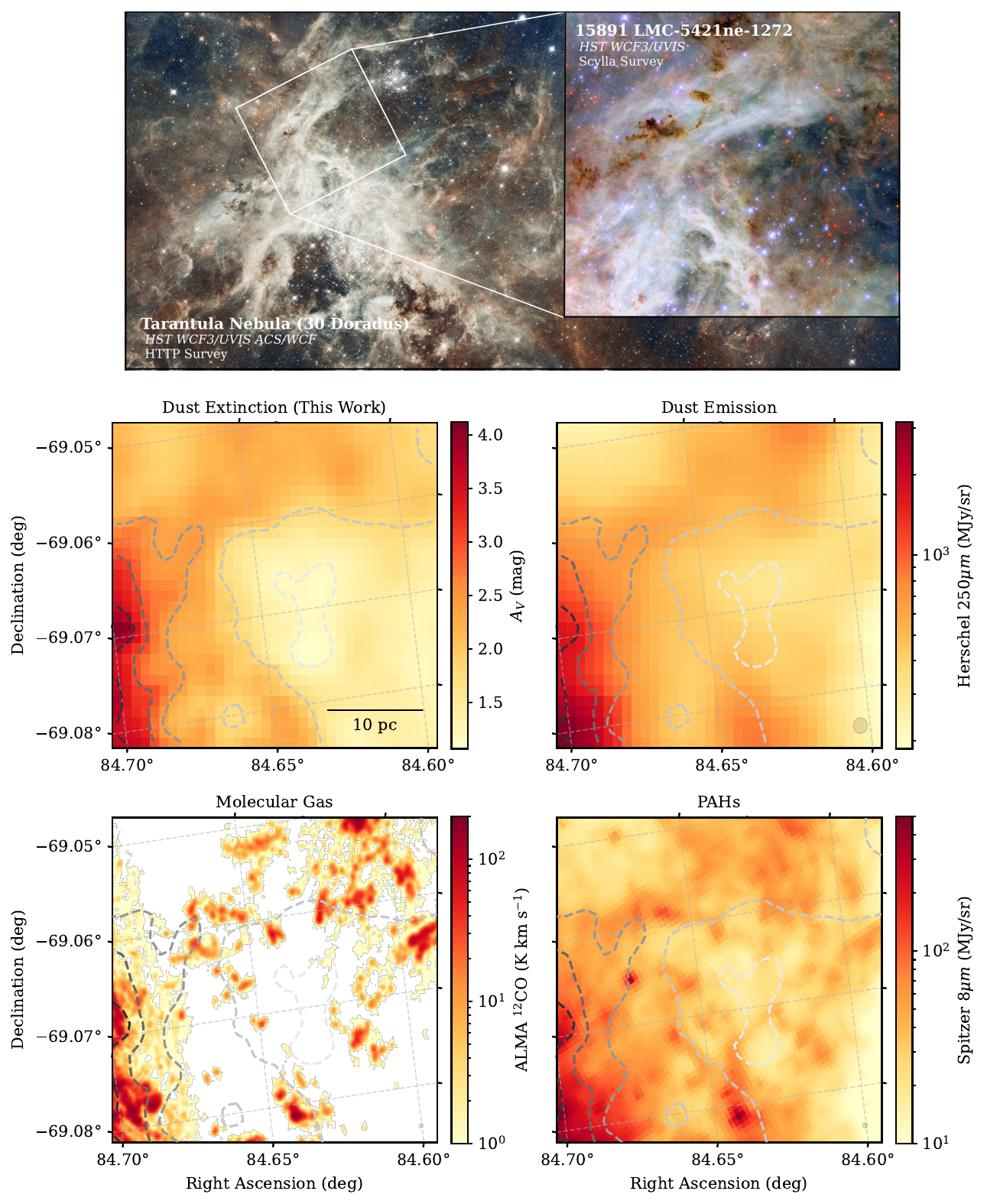}
    \caption{\textbf{Scylla field in 30 Doradus region}: (Top) Three-color image of the broader 30 Dor region (Image credit: NASA, ESA, Lennon; GO-12499) with a zoom-in of the Scylla field (15891$\textunderscore$LMC-5421ne-12728; Image credit: ESA/Hubble \& NASA, C. Murray; GO-15891). 4-panel, clockwise from top left: (1) dust extinction constructed from SED fitting and mapping in this work ($4^{\prime\prime}$/0.97 pc); (2) Herschel $250\, \mu m$ tracing emission from heated dust grains \citep[$6^{\prime\prime}$/1.45 pc;][]{clark2021}; (3) Spitzer IRAC $8\, \mu m$ emission tracing emission from polycyclic aromatic hydrocarbons (PAHs), stochastically heated dust grains, and stellar continuum \citep[$2^{\prime\prime}$/0.5 pc;][]{meixner2006}; (4) ALMA $^{12}\mathrm{CO}(2-1)$ emission tracing dense molecular gas \citep[$1^{\prime\prime}$/0.4 pc;][]{wong2022}. Contours of dust extinction are overlaid on all subplots to aid in visual comparison. The dust extinction map shows high agreement with ancillary ISM tracers, and is particularly effective at capturing broad molecular structures ($5\mbox{--}10$ pc).} 
    \label{fig:30dor}
\end{figure*}


\section{Results}
\label{sec:results}

We apply our methodology to 47 fields in the LMC and 25 fields in the SMC to construct dust extinction maps with 1-pc resolution. There are four fields \changes{(Figure \ref{fig:targets}, red)} which contain too few background stars ($N<200$ sources) to perform kriging (16786$\textunderscore$SMC-8904se-15007, 16235$\textunderscore$SMC-8151se-32530, 16235$\textunderscore$LMC-5812sw-7744, and 14675$\textunderscore$LMC-15978nw-34999). We omit these fields from our analysis and subsequent discussion, leaving us with a total of 68 fields. \changes{We report the median and standard deviation of extinction for each field in Table \ref{tab:fields}.}

In the following section, we highlight a subset of these fields as examples and discuss the global distributions of extinction. A complete set of all fields is shown in Figure Set 1 in the online journal. Extinction maps, including individual instances and associated uncertainties, are available as high-level science data products on MAST at https://archive.stsci.edu/hlsp/scylla.


\subsection{Example Maps}

In Figure \ref{fig:results}, we present dust extinction maps for a subset of fields in the Scylla and METAL surveys. The fields span a range of SFR environments to highlight the diversity of fields that exist within the clouds. The left panels show the corresponding three-color images for each field, constructed using HST WFC3 filters F336W (blue), F475W (green), and F814W (red). The middle panels show the derived dust extinction maps ($A_V$) and their uncertainties ($\sigma_{A_V}$) within each field at $4^{\prime\prime}$ resolution (0.97 pc and 1.2 pc in the LMC and SMC, respectively). Finally, the right panels show Herschel $250\, \mu m$ emission \citep[originally $6.0^{\prime\prime}$;][]{clark2021}, tracing thermal emission from heated dust grains. For comparison, we adopt the same color scale across all four example fields.

Both the first and second fields in Figure \ref{fig:results} are located within the 30 Doradus (30 Dor) star-forming region \citep{sabbi2013}, and contain high amounts of extinction. In the first field (14675\textunderscore LMC-5536ne-9786), dust extinction and emission spatially coincide with nebular emission in the three-color image \citep{romanduval2019}, tracing the surface of a molecular cloud.

In the second field (14675\textunderscore LMC-5665ne-12232), both dust extinction and emission are inversely correlated with optical emission features, likely because the field is near an \hii\ region. In this field, emission and extinction are offset from each other, potentially due to drastic fluctuations in the local radiative field intensity, illuminating dust grains outside the molecular clouds while dust within stays shielded. 

The third field (14675\textunderscore LMC-7868se--9786) lies outside the 30 Dor region and contains modest amounts of dust ($A_V \sim 1$ mag) traced in both extinction and emission. Interestingly, emission from a distant background galaxy ($z=0.026$) is clearly visible in the $250\, \mu m$ dust emission map, but is notably absent in the dust extinction map. Since extinction is measured using stars within the galaxies, extinction maps are impervious to contamination from intermediate-redshift galaxies \citep{casey2012}.

The fourth field lies within the SMC near the N66 star-forming region and star cluster NGC 346 \citep{henize1956}. However, despite an abundance of ionizing sources in the region -- an indication of nearby and recent star formation -- the field contains relatively little dust, especially compared to the other star-forming regions within the LMC shown above. Potential drivers for these lower dust values could be clearing from stellar feedback or low dust-to-gas ratios within the SMC \citep{Clark2023}. 

These examples demonstrate a key advantage of dust extinction maps over emission maps, namely, the lack of contamination from background IR sources, mapping resolution that is dependent on source density, and the ability to disentangle ISM structures from their radiative environments. 

\subsection{Global Extinction Distributions}\label{sec:global}

The distributions of dust extinction across the SMC and LMC provide new insights into the typical density of the ISM in low-metallicity galaxies. In Figure \ref{fig:av_distr}, we plot the distribution of $A_{V}$ across all fields for both the SMC (red) and LMC (blue). The extinction distributions for both the SMC and LMC are well-characterized by log-normal profiles, a characteristic feature of turbulent ISM density structures \citep{kolmogorov, dalcanton2015}. The broader tail towards higher $A_{V}$ values in the LMC reflects the presence of dustier fields near the 30 Doradus star-forming region (hatched). In these active environments, extinction values frequently exceed $A_{V} > 1.0$ mag, underscoring the high column densities associated with these massive star-forming complexes.

The SMC exhibits a narrower distribution with a mean extinction of $e^{\mu} = 0.47$ mag and a dispersion of $\sigma = 0.97$. Meanwhile, the LMC has a mean extinction of $e^{\mu} = 0.43$ mag and a dispersion of $\sigma = 0.72$, when excluding fields within 30 Dor (hatched). Due to the parallel observing strategy with ULYSSES and METAL, which targeted massive stars \citep{romanduval2019, romanduval2025}, a large number of fields are located within the 30 Dor region ($n=13/45$ LMC fields; loosely defined as any fields within $0.6$ deg of RA$=84.6$ deg, DEC$=-68.8$ deg). We opt to exclude these 30 Dor fields to better characterize the global distribution of $A_V$ across the LMC. 

Although the LMC reaches greater column densities, especially within the 30 Dor region, it is perhaps surprising to see that $A_V$ within the SMC is, on average, greater than that of the LMC. The SMC is a lower metallicity system, with $Z= 20\%\ Z_{\odot}$ (compared to 50\% in the LMC), which results in lower dust-to-gas ratios \citep{Clark2023} and consequently should translate to lower volume densities of dust extinction. However, geometry plays a major role: while the LMC is thought to be a face-on dwarf galaxy, the SMC exhibits complex 3D morphology, both in terms of its stellar and gas distributions \citep{zivick2018, pingel2022}. Most notably, the SMC is thought to be extremely elongated along the line-of-sight \citep[$62\pm10$ kpc][]{subramanian2012, Jacyszyn2016, murray2024a, lindberg2025}, consisting of multiple stellar and ISM structures along the line-of-sight. This geometry means that any column densities of $A_V$ will be the integration of the total ISM content along the line-of-sight, and cannot directly be interpreted as a reflection of the volume densities present within the galaxy. 



\section{Discussion}
\label{sec:discussion}


In the following sections, we provide further validation of the extinction maps by comparing a Scylla field within 30 Doradus to ancillary ISM tracers and alternative $A_V$ measurements (Section \ref{sec:30dor}). Then, we discuss how Scylla and METAL dust extinction maps compare with previous extinction maps (Section \ref{sec:previous}) and FIR dust emission maps (Section \ref{sec:dust_mass}), and conclude with a discussion of planned future science and methodological developments.

\begin{figure*}
    \centering
    \includegraphics[width=0.7\linewidth]{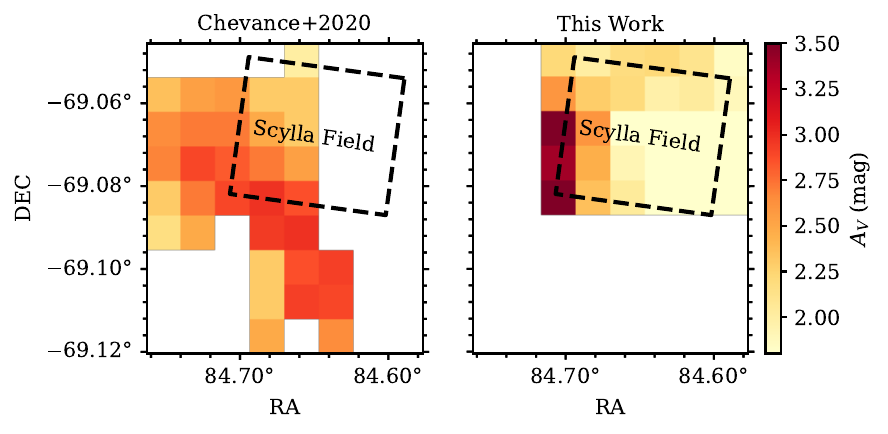}
    \caption{\textbf{$A_V$ comparison with previous extinction map in 30 Doradus region}: (Left) Reproduction $A_V$ map, shown in Figure 4 from \citet{Chevance2020} \changes{with the footprint of Scylla field presented in Figure \ref{fig:30dor} overlaid in black. (Right) Scylla field reprojected and regridded to the same resolution as \citet[][$43^{\prime\prime}$ or 10.5 pc]{Chevance2020}.} Despite differences in the dynamic range, the median $A_V$ for all overlapping pixels is $A_V=2.7$ mag and $A_V = 2.6$ mag for \citet{Chevance2020} and the Scylla field, respectively, indicating that the average recovered extinction from SED fitting is comparable to extinction maps derived from emission-line modeling.}
    \label{fig:chevance}
\end{figure*}

\subsection{Validation in 30 Doradus Region}
\label{sec:30dor}

With pc-scale dust extinction maps of star-forming regions in both the SMC and LMC, we unlock a new opportunity to anchor emission-based SFR tracers (e.g., dust emission, molecular gas emission, etc.) to radiation-independent measurements of the ISM. Here, we analyze one of the Scylla fields located within the 30 Dor star-forming region of the LMC (15891$\textunderscore$LMC-5421ne-12728). This field lies just outside the R136 star cluster, the highest mass cluster in the Local Group, powering the radiative energy in the vicinity \citep{hunter1995, crowther2016, bestenlehner2020}. We chose to analyze this field because it is the only Scylla or METAL field that spatially coincides with sub-parsec resolution CO observations from ALMA \citep{wong2022}, providing a valuable opportunity to directly observe how the dust extinction map compares with molecular gas. 

In Figure \ref{fig:30dor}, we compare the extinction map within the 30 Dor region to three other ancillary ISM tracers. In the top panel, we show a three-color image of the broader 30 Dor region with a footprint and inset of the Scylla field being analyzed. 
In the upper left panel, we plot the dust extinction map constructed from stellar SEDs. The map has a resolution of $4^{\prime\prime}$ or 0.97 pc. Of the original 1989 stars within the field that passed the luminosity threshold cuts, only $686_{-18}^{+32}$ stars were identified as background stars and were used to construct the resultant map. Extinction ranges from $A_V = 1.4 \mbox{--} 4.2$ mag, far above the transition from atomic to molecular gas \citep[e.g., $A_V = 0.3-0.5$ in the MW, $A_V = 0.7-1.2$ in the LMC;][]{liszt2014, wolfire2010}, and is concentrated in structures spanning $3 \mbox{--} 10$ pc. For all ancillary tracers, we overlay contours of the dust extinction map to aid in visual comparison.

In the upper right panel, we plot $250\, \mu m$ emission from Herschel at a resolution of $6.0^{\prime\prime}$ or 1.45 pc, tracing dust emission from heated grains \citep{clark2021}. In the bottom left panel, we plot $^{12}\mathrm{CO(2-1)}$ emission from ALMA with a resolution of $1^{\prime\prime}$ or 0.4 pc, tracing dense molecular gas \citep[][]{wong2022}. Finally, in the bottom right panel, we plot $8\, \mu m$ emission from Spitzer IRAC at a resolution of $2^{\prime\prime}$ or 0.5 pc, broadly tracing emission from polycyclic aromatic hydrocarbons (PAHs), heated dust, and stellar continuum \citep{meixner2006}.

The dust extinction map does an excellent job of capturing broad ISM structures ($5\mbox{--}10$ pc), such as the molecular cloud in the southeast corner and the molecular core in the northwest. Perhaps unsurprisingly, it struggles to capture concentrated clumps ($<2$ pc) due to the low density of sources (average of 686 stars in this field). However, this limitation is not unique to the LMC and SMC, and applies to any methodology that relies on point sources to probe diffuse structure, including 3D dust extinction maps within the MW \citep{edenhofer2024}. 


While Figure \ref{fig:30dor} illustrates good spatial agreement between $A_V$ and ancillary ISM tracers, we do not expect a one-to-one correlation since emission-based tracers vary greatly as a function of radiative environment. 
Dust extinction maps of 30 Dor exist \citep[e.g.,][discussed further in Section \ref{sec:previous}]{skowron2021, Chen2022}, however, these maps are of much coarser resolution (e.g., $\sim 0.5^{\prime} \mbox{--}1^{\prime}$).
To date, the only $A_V$ map of 30 Dor with comparable angular resolution comes from \citet{Chevance2020}, who used SOFIA/FIFI-LS observations of ionized and neutral gas lines to constrain the fraction of CO-dark gas. Using a Meudon PDR model and emission line ratios, they modeled the spatial distributions of physical properties like pressure, radiation field, and $A_V$ at a resolution of $43^{\prime\prime}$ or 10.5 pc \citep[see Figure 4 in][]{Chevance2020}.

We reproduce the $A_V$ map from \citet{Chevance2020} to compare with the Scylla $A_V$ map (Figure \ref{fig:chevance}, left). The $A_V$ map from \citet{Chevance2020} does not cover the exact spatial extent of the Scylla field (black box), so we reproject the Scylla $A_V$ map to the same resolution and gridding as the map from \citet{Chevance2020}\changes{, shown in Figure \ref{fig:chevance}, right}.
We find broad agreement between the two maps, with median $A_V=2.7$ mag and $A_V=2.6$ mag in the \citet{Chevance2020} map and the Scylla map, respectively, for overlapping pixels. However, the Scylla map has nearly double the dynamic range in extinction ($A_V = 1.8 \mbox{--}3.6$ mag) compared to the \citet{Chevance2020} map ($A_V = 2.0 \mbox{--}3.0$ mag). 

While our methodology integrates the total column density of dust along the entire line-of-sight using background stellar probes, the $A_{V}$ values in \citet{Chevance2020} are defined by the physical depth of the cloud as constrained by PDR modeling. Specifically, their extinction measurements are derived by comparing PDR-tracing fine-structure lines, which originate from distinct layers within the PDR structure. Consequently, their technique may be inherently limited to probing the $A_{V}$ accessible to these specific gas tracers rather than capturing the cumulative dust column, biasing the $A_V$ dynamic range low. Alternatively, emission line ratios might capture both extinction and attenuation, as radiation is absorbed and scattered along different lines of sight. The combination of these processes could smooth and average out the underlying extinction structures, resulting in the reduced dynamic range observed in their $A_V$ map.

This comparison reveals a notable paradox: despite validation simulations in Section \ref{sec:mw_val} showing a tendency for kriging to flatten the dynamic range, the Scylla maps still recover nearly double the $A_{V}$ range of the \citet{Chevance2020} map. This suggests that the inherent smoothing effects of modeling extinction via PDR emission-line ratios are even more restrictive than the sampling limitations of our stellar-probe methodology. Despite disagreements between dynamic ranges of $A_V$, similar spatial structures and comparable median extinction corroborate the pc-scale dust extinction map presented in this subsection, validating that the methodology presented in this paper can characterize dust structures with $A_V > 3$ mag.

\subsection{Dust Mass Surface Density}
\label{sec:dust_mass}

One of the major benefits of measuring dust content through extinction is its ability to constrain the amount of dust mass available in a region, independent of the ISM temperature or radiation field, which can provide valuable constraints on the dust-to-gas ratio within a galaxy. In this section, we compare estimates of the dust mass surface density ($\Sigma_D$) derived from both extinction and emission and discuss the potential sources of discrepancies in the measurements.

Dust grain models provide theoretical estimates for how dust extinction scales as a function of $\Sigma_D$ \citep{draine2007}. While there are no relations to directly convert from dust extinction ($A_V$) to dust mass surface density ($\Sigma_{D,\,A_V}$) when assuming a modified blackbody, we can approximate the dust mass surface density by applying the following relation, as originally defined in \citet{draine2014}:

\begin{equation}
A_V = 0.74 \left( \frac{\Sigma_{D,\,A_V}}{10^5\ M_{\odot}\ \mathrm{kpc\ pc}^{-2} } \right) \mathrm{mag} 
\end{equation}

\begin{figure}
    \centering
    \includegraphics[width=\linewidth]{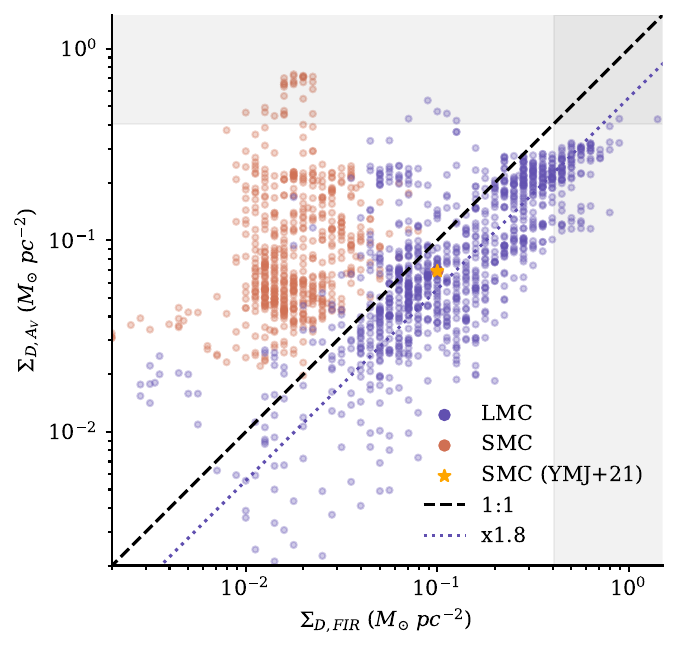}
    \caption{\textbf{Dust mass surface density}: Comparison of dust mass surface densities at $36^{\prime\prime}$ derived from FIR dust emission ($\Sigma_{D,\,FIR}$) and dust extinction ($\Sigma_{D,\,A_V}$) for fields in the SMC (red) and LMC (blue). Our minimum extinction threshold of $A_V=3$ mag is shaded in, denoting the range above which completeness might become an issue. Both galaxies show systematic offsets between the two measurements.}
    \label{fig:dust_mass_ratio}
\end{figure}

Dust mass surface density can also be measured via emission by modeling the observed FIR emission across wavelengths as a broken-emissivity modified blackbody model \citep{gordon2014}. By combining FIR emission maps from Herschel HERITAGE data \citep[][]{meixner2013}, Planck \citep{planck2011}, IRAS \citep[][]{neugebauer1984}, and COBE \citep[][]{boggess1992, silverberg1993}, \citet{Clark2023} constructed new $\Sigma_{D,\,FIR}$ maps at $36^{\prime\prime}$ resolution spanning the SMC, LMC, M31, and M33.  

We compare our extinction-derived dust mass surface densities ($\Sigma_{D,\,A_V}$) to FIR emission-based estimates ($\Sigma_{D,\,FIR}$) by convolving $\Sigma_{D,\,A_V}$ to $36^{\prime\prime}$ resolution with a Gaussian kernel and re-projecting to the same spatial grid. This allows us to achieve a pixel-by-pixel comparison between $\Sigma_{D,\,A_V}$ and $\Sigma_{D,\,FIR}$, as opposed to the field-average comparison that was performed in \citetalias{lindberg2025}. For both sets of measurements, MW foreground contamination was removed beforehand.

In Figure \ref{fig:dust_mass_ratio}, we plot the extinction-derived dust mass surface densities ($\Sigma_{D,\ A_V}$) as a function of the FIR emission-based estimates ($\Sigma_{D,\ FIR}$) for each Scylla and METAL field in both the SMC (red) and LMC (blue). We include a one-to-one guideline (black dashed) and optimized correction for the LMC (purple dotted). Based on the quality cut defined in Section \ref{sec:complet}, our sources should be complete up to an $A_V=3$ mag. We denote this limit with a shaded region in the plot, marking the range above which completeness might become an issue,\changes{ however, only 2\% of the entire survey surface area has $A_V > 3$ mag.}
We find that $\Sigma_D$ spans a considerably larger range in the LMC ($\Sigma_{D,\,FIR} = 10^{-2}  - 10^0\ M_{\odot}\ \mathrm{pc}$), as opposed to the SMC ($\Sigma_{D,\,FIR} = 10^{-2} - 5\times10^{-2}\ M_{\odot}\ \mathrm{pc}$). Additionally, both galaxies show offsets between their extinction- and emission-derived $\Sigma_D$, however, in the SMC, extinction overpredicts $\Sigma_D$ compared to emission, whereas in the LMC, extinction generally underpredicts $\Sigma_D$ compared to emission.

The $\Sigma_D$ discrepancy between dust emission and extinction in the LMC has been discussed previously. In M31, \citet{dalcanton2015} found a factor of $\sim2.5$ difference (meaning $\Sigma_{D,\,FIR}$ is $X$ times larger than $\Sigma_{D,\,A_V}$) between their extinction-derived $\Sigma_{D,\,A_V}$ inferred from CMD fitting versus $\Sigma_{D,\,FIR}$ measured by \citet{draine2014}. In the MW, \citet{planck2016} also found a factor of $\sim2$ factor difference between FIR-derived extinction estimates using models from \citet{draine2007} versus $A_V$ measurements from background quasars. In the SMC, \citet{ymj2021} found a 1.8 factor offset between $\Sigma_{D,\,A_V}$ versus $\Sigma_{D,\,FIR}$ (marked as a yellow star in Figure \ref{fig:dust_mass_ratio}). 

We observe a similar general offset of 1.8 in the LMC (blue dotted line), whereby dust mass surface density is greater than inferred from dust emission ($\Sigma_{D,\,FIR}$) than extinction($\Sigma_{D,\,A_V}$). However, in the SMC, the correlation between dust mass surface density measurements is less linear, with extinction greatly overpredicting the dust mass surface density relative to extinction. 

One potential explanation for the discrepancy in the SMC is its extended geometry. As briefly discussed in Section \ref{sec:global}, the stellar content of the SMC spans several kpc along the line-of-sight \citep[$62\pm 10$ kpc;][]{Subramanian2017, murray2024a}. If the diffuse ISM spans comparable ranges, then the measured $A_V$ will reflect the integrative quantity of dust along the entire line-of-sight, and the total column densities of extinction will not reflect the true volume densities present within the galaxy. While dust emission might span similar extents, the bulk of emission will be driven by the interstellar radiation field, which we cannot assume to stay constant over similar ranges. 

The one exception to this trend is the result from \citet{ymj2021} (Figure \ref{fig:dust_mass_ratio}; yellow star), which probes a denser star-forming region (N13). In this region, we expect any dense atomic or molecular gas to dominate the contribution of total extinction, resulting in much better agreement with the observed dust emission. 

Theories predict that dust opacity (and thereby the ratio of $\Sigma_{D,\,FIR}$ to $\Sigma_{D,\,A_V}$ when assuming a constant dust opacity) should increase in denser ISM environments as grains grow to larger sizes \citep{kohler2015}. Future work will investigate whether variations between $\Sigma_{D,\,FIR}$ and $\Sigma_{D,\,A_V}$ are correlated with gas densities, indicative of dust opacity evolution as a function of ISM environment.

\subsection{Comparison with Other Extinction Maps}
\label{sec:previous}


\begin{figure}
    \centering
    \includegraphics[width=\linewidth]{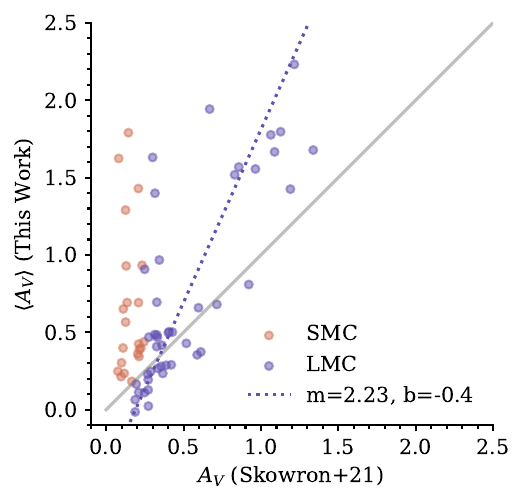}
    \caption{\changes{\textbf{Other Dust Maps}: Median dust extinction within each Scylla and METAL field compared to reddening maps from \citet{skowron2021}. \changestwo{We convert to $E(V-I)$ to $A_V$, we adopt the relation $E(B-V) = 0.808 \times E(V-I)$} and conservatively assume $R_V=3.4$ for the LMC and $R_V=2.7$ for the SMC for both reddening maps \citep{gordon2014}. We find that the \citet{skowron2021} reddening map in the LMC underestimates $A_V$ by a factor of $2.23$, which is consistent if we assume that the average extinction experienced by stars is half the total column density measured in this work. In the SMC, line-of-sight effects obscure any correlation.}}
    \label{fig:previous}
\end{figure}

\changes{To validate the Scylla and METAL extinction maps, we compare these maps with existing dust maps of the SMC and LMC. However, while several reddening maps exist already \citep[e.g., ][]{choi2018, gorski2020, skowron2021, Chen2022}, there are key differences in their methodology and intent of use that make them difficult to perform an apples-to-apples comparison with the Scylla and METAL extinction maps.}

\changes{Most reddening maps towards the SMC and LMC were constructed to measure the average extinction experienced by the stellar population within the SMC and LMC}. These maps leverage red clump (RC) stars with optical and/or NIR photometry to either statistically model population reddening using a CMD \citep[e.g.,][]{choi2018, gorski2020, skowron2021}, or combine several surveys to perform individual stellar SED modeling, from which a population average can be computed \citep[e.g.,][]{Chen2022}. \changes{Meanwhile, the Scylla and METAL maps attempt to measure the total column density of dust extinction, independent of any geometric effects between the stars and dust (e.g., star-dust geometries in Section \ref{sec:mw_val}).}


\changestwo{To compare with reddening maps from \citet{skowron2021}, we adopt the relation $E(B-V) = 0.808 \times E(V-I)$, just as in \citet{Chen2022}. We then conservatively assume $R_V=3.4$ in the LMC and $R_V=2.7$ \citep{gordon2003} in the SMC to convert to $A_V$.}
Indeed, when we convolve the Scylla and METAL maps to the same central $1^{\prime}7^{\prime\prime}$ resolution from \citet{skowron2021}, \changes{shown in Figure \ref{fig:previous}}, we find that Scylla and METAL maps report an average of 2.23 times more total extinction than inferred from the reddening map in the LMC. In the SMC, we struggle to fit any correlation between the two maps. \changes{For a face-on galaxy like the LMC,} this disagreement is largely resolved if one assumes the reddening map from \citet{skowron2021} captures half of the total column density of the Scylla and METAL maps \changes{since the reddening maps are ultimately a product of the geometry between the stars and the ISM. However, in the violently disrupted SMC \citep{choi2018}, offsets between ISM structures \citep{murray2024a} and stellar distributions \citep{ymj2021} obscure any clear relation between average stellar reddening and total column density. 
In summation, we cannot assume that the correlation between total column density extinction maps and population average extinction maps is inherently linear, as their relation is strongly dependent on the star-dust geometry.}

\subsection{Future Work}

Dust extinction maps offer a new independent tracer of the ISM, providing a more comprehensive understanding of how small-scale physics within the ISM varies as a function of metallicity and environment. We identify several key opportunities for future scientific and methodological developments.

\changes{In this publication, we focus on dust extinction mapping to leverage as much of the existing Scylla and METAL coverage as possible. However, a similar framework could be applied to fields with adequate filter coverage to map $R_V$ across fields. BEAST validation simulations in \citetalias[][]{lindberg2025} (Figure 8) found that 6-7 filters were needed to accurately recover $R_V$.}
Upcoming Scylla publications will investigate whether variations between $\Sigma_{D,\,FIR}$ and $\Sigma_{D,\,A_V}$ are correlated with gas densities, indicative of dust opacity evolution as a function of ISM environment. Furthermore, we plan to leverage the improved spatial resolution of the extinction maps to constrain the fraction of CO-dark molecular gas (i.e., \htwo\ not traced by CO emission) within regions, like 30 Doradus \citep[e.g.,][]{Chevance2020}. By comparing dust extinction with $^{12}$CO and HI emission, we can map how the fraction of CO-dark gas varies with decreasing metallicity, a factor that remains critical for understanding star formation efficiency, especially in dwarf galaxies.

While this paper demonstrates that the current methodology works for nearby galaxies like the SMC and LMC, we believe that a few key modifications could improve the application of this methodology to more distant galaxies. Rather than relying on kriging as a form of 2D interpolation using only background sources, we intend to develop an integrative 3D mapping, similar to Milky Way dust maps \citep{edenhofer2024}. Such a technique would allow us to utilize all stellar extinction measurements along a line-of-sight, rather than only background stars, eliminating the need for any Gaussian mixture model decomposition. However, careful consideration would be needed for modeling stellar completeness, especially due to selection effects from dust extinction. 
\changes{Additionally, by implementing a hierarchical Bayesian framework, we could leverage population-level statistics from initial BEAST fits (e.g., mass/age distribution, age/metallicity correlations, extinction maps, etc.) to update prior models (e.g., IMF, SFH, Av, etc.) for individual stellar BEAST fits, allowing us to obtain more realistic posteriors for each parameter.}

Expanding this high-resolution methodology to other galaxies with multi-band photometry like M31, M33, and other nearby dwarf galaxies within the Local Group \citep{dalcanton2012, williams2021, gilbert2024} would offer a vital independent benchmark for constraining dust emissivity and abundances across a wider range of galactic environments. Additionally, these methodological improvements will complement upcoming optical and near-IR observations of nearby galaxies with the Vera C. Rubin Observatory \citep{ivezic2019} and Nancy Grace Roman Space Telescope \citep{akeson2019}.

\section{Conclusion}
\label{sec:conclusion}



The methodology introduced in this work leverages kriging to map dust extinction at parsec scales in the SMC and LMC. Before summarizing our primary scientific findings, we first address the systematic constraints and uncertainties inherent in this approach to provide a clear framework for interpreting the resulting maps.

\subsection{Methodological Biases and Limitations}

While this methodology offers a significant increase in resolution and completeness over previous dust mapping techniques, several inherent biases and limitations remain:

\begin{enumerate}

    \item \textit{Resolution and Source Density}: The 1-pc resolution of the maps is ultimately constrained by the spatial density of background stars. As such, the method will struggle to capture compact molecular structures smaller than a few pc if no background sources are available, or structures with $A_V \geq 6$ mag. (\S\ref{sec:complet} and \S\ref{sec:mw_val}; Figures \ref{fig:max_av} and \ref{fig:grid_oriona})

    \item \textit{Flattening of Dynamic Range}: Simulation results indicate that the method tends to overestimate extinction in low-$A_{V}$ regions and underestimate it in high-$A_{V}$ regions. This leads to a potential flattening of the recovered dynamic range, especially in fields with low stellar densities (\S\ref{sec:mw_val}; Figure \ref{fig:sim_nstars}). However, when we compare with other extinction maps from the literature, we find that maps constructed with this methodology have nearly twice the dynamic range in observed extinction. (\S\ref{sec:30dor}; Figure \ref{fig:chevance})    
    
    \item \textit{Observational Artifacts}: Occasional contamination from diffraction spikes, background galaxies, or diffuse gas emission can result in $A_{V}$ spikes with a radius of 1–2 pc. These impacts are localized and generally do not bias field-wide measurements. Additionally, in fields with low intrinsic dust ($A_{V} < 0.5$ mag), the methodology can occasionally exhibit linear artifacts stemming from over-corrections in charge-transfer efficiency (CTE) loss along the HST WFC3 chip gaps. These are generally mitigated by luminosity quality cuts but remain present in a small subset of fields. (\S\ref{sec:uncert} and Appendix \ref{sec:artifacts}, Figure \ref{fig:artifacts}).

\end{enumerate}

\subsection{Summary of Findings}

The dust extinction maps produced in this paper reveal detailed ISM structures and show strong spatial correlation with ancillary ISM tracers, offering a significant increase in resolution and completeness over previous dust mapping techniques in the SMC and LMC. Additionally, these maps provide a critical independent tracer of the total ISM, allowing us to probe the multi-scale structure of the ISM across different metallicity regimes. Our main findings are as follows:

\begin{enumerate}
    \item \textit{Validation Simulations} - With 3D dust maps simulations, we find that the dust mapping methodology reliably recovers the total column density distribution of dust extinction to an accuracy of $A_V=0.1$ mag in fields with more than 1200 sources, with the exception of the dustiest compact regions. (\S\ref{sec:mw_val}; Figure \ref{fig:grid_oriona})    

    \item \textit{Global Distribution} - The global extinction distributions of total column densities follow log-normal profiles in both galaxies, with the SMC exhibiting a slightly higher mean extinction ($e^{\mu}=0.47$ mag) than the broader LMC ($e^{\mu}=0.43$ mag), likely due to the significant line-of-sight depth in the SMC. (\S\ref{sec:global}; Figure \ref{fig:av_distr})
    
    \item \textit{30 Doradus} - In high-density environments like 30 Dor, we find high spatial agreement between the $A_V$ maps and ancillary ISM tracers like Herschel dust emission, ALMA $^{12}$CO, and Spitzer $8\, \mu m$ emission, validating the method ability to characterize structures over broad scales ($5-10$ pc). Moreover, extinction maps from \citet{Chevance2020} constructed from emission line ratios find similar median $A_V$ values within the field. (\S\ref{sec:30dor}; Figures \ref{fig:30dor} and \ref{fig:chevance})
    
    \item \textit{Dust Mass Surface Density} - Comparisons between FIR emission- and extinction-derived dust mass surface densities ($\Sigma_D$) show offsets in both the SMC and LMC, suggesting variations in dust opacity or dust-to-gas ratios linked to the surrounding ISM environment. (\S\ref{sec:dust_mass}; Figure \ref{fig:dust_mass_ratio})
    
\end{enumerate}

\facilities{HST (WFC3/IR), HST (WFC3/UVIS), XSEDE \citep{xsede2014}}

\software{DOLPHOT \citep{dolphin2002}, BEAST \citep{gordon2016}, astropy \citep{2013A&A...558A..33A, astropy}, numpy \citep{harris2020array}, scipy \citep{Virtanen2020}, matplotlib \citep{Hunter2007}, PyKrige \citep{murphy2021}}

\begin{acknowledgments}

This research is based on observations made with the NASA/ESA Hubble Space Telescope obtained from the Space Telescope Science Institute, which is operated by the Association of Universities for Research in Astronomy, Inc., under NASA contract NAS 5–26555. These observations are associated with programs 15891, 16235, and 16786. This research has made use of NASA’s Astrophysics Data System. All of the data presented in this paper were obtained from MAST at the Space Telescope Science Institute. The specific observations analyzed can be accessed via \dataset[https://doi.org/10.17909/8ads-wn75]{https://doi.org/10.17909/8ads-wn75}. Support to MAST for these data is provided by the NASA Office of Space Science via grant NAG5–7584 and by other grants and contracts.

The authors acknowledge Interstellar Institute's program "II7" and the Paris-Saclay University's Institut Pascal for hosting discussions that nourished the development of the ideas behind this work.

CWL would like to thank Jeremy Chastenet for insightful comments on dust mapping in the early stages of this project, as well as David Nataf for his suggestion to compare with previous reddening maps towards the SMC and LMC, and Erik Tollerud for invaluable guidance on handling reprojections. 

\end{acknowledgments}

\bibliography{main}{}
\bibliographystyle{aasjournalv7}

\appendix


\begin{figure}
    \centering
    \includegraphics[width=0.9\linewidth]{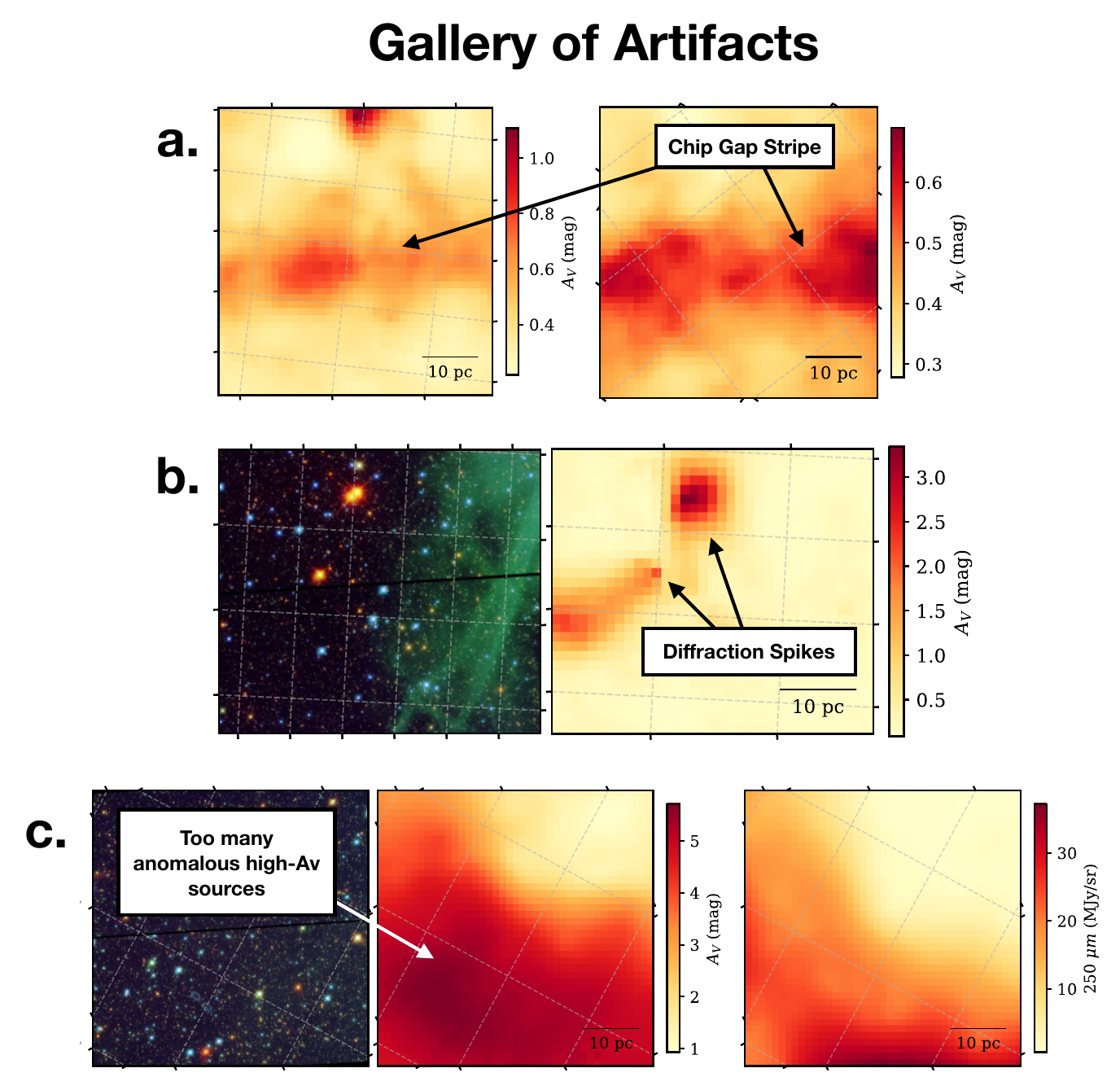}
    \caption{Gallery of known artifacts in the Scylla and METAL dust extinction maps. Each row highlights a different artifact. See Appendix \ref{sec:artifacts} for a description of each artifact.
    }
    \label{fig:artifacts}
\end{figure}

\section{Artifacts}
\label{sec:artifacts}

We highlight several artifacts present in the Scylla and METAL extinction maps. While the reported uncertainty should capture and propagate the increased uncertainties surrounding these structures, we urge users to carefully consider the inclusion of any maps with prominent artifacts in their scientific investigations. In Figure \ref{fig:artifacts}, we show a gallery of known artifacts, where each row illustrates a different artifact.

In Row A, we show abnormal linear structures present in four fields within low column density environments. 
These dust structures likely stem from over-corrections in charge-transfer efficiency (CTE) loss along the WFC3 UVIS chip gap. Space-based CCD detectors, like WFC3 UVIS, suffer from CTE degradation when exposed to cosmic rays, the effects of which cumulatively impact pixels further from the readout amplifiers. These effects are usually remedied by applying CTE corrections as a part of the photometric reduction pipeline. However, without perfect calibrations, these corrections can produce minor over- or under-estimations of the true fluxes \citep{williams2014}, the impact of which will vary spatially based on proximity from the CCD chip gap \citep[Appendix C;][]{murray2024b}.



The effects of these CTE over- and under-corrections are minor within each filter; however, a slight over-correction in optical flux bands, relative to UV bands, could propagate into the SED fits, resulting in an overestimation of the inferred dust extinction. 
Sources with over-predicted dust extinction from CTE miscorrections are generally low-mass ($M_{ini} < 1\ M_{\odot}$) and dim, and the majority of these sources are removed with quality cuts imposed in Section \ref{sec:complet}. However, these CTE artifacts are still present in four fields with low intrinsic dust ($A_V < 0.5$ mag). We confirm that these structures are artifacts by comparing with FIR emission maps from \citet{Clark2023}.   



We note that these CTE artifacts should not be an issue for other WFC3 photometric surveys with dithered observations since dithering will shift the location of the chip gap. Unfortunately, since Scylla was designed as pure-parallel programs, the opportunity to dither observations was not available. 

In Row B, we highlight localized, high-$A_V$ spikes that primarily originate from observational contamination by bright, saturated stars. These artifacts occur when photons from a bright source are scattered along the telescope’s secondary mirror support vanes, creating prominent "diffraction spikes" in the imaging. Due to their peculiar and non-stellar spectral energy distributions (SEDs), these features can be erroneously characterized by the BEAST as sources with high levels of extinction, often reaching $A_V \approx 4.0-10.0$ mag. These contaminants typically manifest in the final dust maps as circular or linear spikes with a radius of 1–2 pc. While we find that $\sim35\%$ of fields in the survey exhibit some evidence of these diffraction spike "bumps," they are generally localized and do not bias field-wide measurements

In Row C, we identify a much rarer artifact resulting from an extreme density of anomalous high-$A_V$ sources within a single field. While localized contamination is common, we identify 2-3 fields in our entire sample that are truly dominated by these features. In these rare instances, the concentration of contaminants is sufficient to bias the kriging interpolation across large spatial scales. As shown in Figure \ref{fig:artifacts}c, the density of these contaminants can overwhelm the background source catalog, leading to a regional overestimation of $A_V$ that does not reflect the true ISM column density. In these instances, we rely on ancillary FIR emission maps to help distinguish these rare, dominated fields from genuine high-column density molecular structures.

\end{document}

%% file: table.txt
\begin{deluxetable*}{llcccccc}
\tablecaption{\label{tab:fields} Dust Extinction Field Locations}
\tablehead{
\colhead{Field Name} &  \colhead{Field Name} & \colhead{Field RA} & \colhead{Field Dec} & \colhead{$N_{\rm filters}$} & \colhead{Median $A_V$} & \colhead{StD $A_V$} & \colhead{Foreground $A_V$}\\
\colhead{(short)} & \colhead{(long)} &  \colhead{($^{\circ}$)} & \colhead{($^{\circ}$)} & \colhead{} & \colhead{(mag)} & \colhead{(mag)} & \colhead{(mag)} \\
\colhead{(1)} & \colhead{(2)} & \colhead{(3)} & \colhead{(4)} & \colhead{(5)} & \colhead{(6)} & \colhead{(7)} & \colhead{(8)} 
}
\startdata
SMC\_10 & 15891\_SMC-3149ne-12269  & 15.0793 & -72.1515 & 7 & 0.303   & 0.524   & 0.107  \\
SMC\_13 & 15891\_SMC-2983ne-12972  & 14.9313 & -72.1748 & 5 & 0.651   & 0.201   & 0.103  \\
SMC\_15 & 15891\_SMC-4292sw-13841  & 9.562   & -73.4013 & 7 & 0.248   & 0.037   & 0.083  \\
SMC\_16 & 15891\_SMC-3435ne-13258  & 15.2425 & -72.0912 & 4 & 1.624   & 0.061   & 0.107  \\
SMC\_17 & 15891\_SMC-1588ne-12105  & 14.4594 & -72.5908 & 6 & 0.359   & 0.329   & 0.106  \\
SMC\_18 & 15891\_SMC-3029ne-13288  & 15.9413 & -72.6589 & 6 & 0.182   & 0.597   & 0.104  \\
SMC\_20 & 15891\_SMC-3370ne-13459  & 15.1156 & -72.082  & 6 & 0.213   & 0.126   & 0.107  \\
SMC\_21 & 15891\_SMC-3104ne-13781  & 15.0284 & -72.1569 & 7 & 0.234   & 0.505   & 0.107  \\
SMC\_23 & 15891\_SMC-2584ne-14274  & 14.7218 & -72.2607 & 5 & 0.692   & 0.299   & 0.103  \\
SMC\_25 & 16235\_SMC-879ne-11082   & 13.9209 & -72.7082 & 4 & 0.929   & 0.396   & 0.110  \\
SMC\_27 & 16235\_SMC-2668ne-11415  & 15.379  & -72.4818 & 6 & 1.790   & 0.971   & 0.104  \\
\enddata
\tablecomments{The complete table is available in machine-readable format in the online journal.}

\end{deluxetable*}

%% file: main.bib
@ARTICLE{Romaniello2002,
       author = {{Romaniello}, Martino and {Panagia}, Nino and {Scuderi}, Salvatore and {Kirshner}, Robert P.},
        title = "{Accurate Stellar Population Studies from Multiband Photometric Observations}",
      journal = {\aj},
         year = 2002,
        month = feb,
       volume = {123},
       number = {2},
        pages = {915-940},
          doi = {10.1086/338430},
archivePrefix = {arXiv},
       eprint = {astro-ph/0111399},
 primaryClass = {astro-ph},
       adsurl = {https://ui.adsabs.harvard.edu/abs/2002AJ....123..915R}
}

@ARTICLE{vaszquez-semadeni2001,
       author = {{V{\'a}zquez-Semadeni}, Enrique and {Garc{\'\i}a}, Nieves},
        title = "{The Probability Distribution Function of Column Density in Molecular Clouds}",
      journal = {\apj},
         year = 2001,
        month = aug,
       volume = {557},
       number = {2},
        pages = {727-735},
          doi = {10.1086/321688},
archivePrefix = {arXiv},
       eprint = {astro-ph/0103199},
 primaryClass = {astro-ph},
       adsurl = {https://ui.adsabs.harvard.edu/abs/2001ApJ...557..727V}
}

@ARTICLE{bestenlehner2020,
       author = {{Bestenlehner}, Joachim M. and {Crowther}, Paul A. and {Caballero-Nieves}, Saida M. and {Schneider}, Fabian R.~N. and {Sim{\'o}n-D{\'\i}az}, Sergio and {Brands}, Sarah A. and {de Koter}, Alex and {Gr{\"a}fener}, G{\"o}tz and {Herrero}, Artemio and {Langer}, Norbert and {Lennon}, Daniel J. and {Ma{\'\i}z Apell{\'a}niz}, Jesus and {Puls}, Joachim and {Vink}, Jorick S.},
        title = "{The R136 star cluster dissected with Hubble Space Telescope/STIS - II. Physical properties of the most massive stars in R136}",
      journal = {\mnras},
         year = 2020,
        month = dec,
       volume = {499},
       number = {2},
        pages = {1918-1936},
          doi = {10.1093/mnras/staa2801},
archivePrefix = {arXiv},
       eprint = {2009.05136},
 primaryClass = {astro-ph.SR},
       adsurl = {https://ui.adsabs.harvard.edu/abs/2020MNRAS.499.1918B}
}

@ARTICLE{crowther2016,
       author = {{Crowther}, Paul A. and {Caballero-Nieves}, S.~M. and {Bostroem}, K.~A. and {Ma{\'\i}z Apell{\'a}niz}, J. and {Schneider}, F.~R.~N. and {Walborn}, N.~R. and {Angus}, C.~R. and {Brott}, I. and {Bonanos}, A. and {de Koter}, A. and {de Mink}, S.~E. and {Evans}, C.~J. and {Gr{\"a}fener}, G. and {Herrero}, A. and {Howarth}, I.~D. and {Langer}, N. and {Lennon}, D.~J. and {Puls}, J. and {Sana}, H. and {Vink}, J.~S.},
        title = "{The R136 star cluster dissected with Hubble Space Telescope/STIS. I. Far-ultraviolet spectroscopic census and the origin of He II {\ensuremath{\lambda}}1640 in young star clusters}",
      journal = {\mnras},
         year = 2016,
        month = may,
       volume = {458},
       number = {1},
        pages = {624-659},
          doi = {10.1093/mnras/stw273},
archivePrefix = {arXiv},
       eprint = {1603.04994},
 primaryClass = {astro-ph.SR},
       adsurl = {https://ui.adsabs.harvard.edu/abs/2016MNRAS.458..624C}
}

@ARTICLE{romanduval2025,
       author = {{Roman-Duval}, Julia and {Fischer}, William J. and {Fullerton}, Alexander W. and {Taylor}, Jo and {Plesha}, Rachel and {Proffitt}, Charles and {Monroe}, TalaWanda and {Fischer}, Travis C. and {Aloisi}, Alessandra and {Bouret}, Jean-Claude and {Britt}, Christopher and {Calvet}, Nuria and {Carlberg}, Joleen K. and {Crowther}, Paul A. and {De Rosa}, Gisella and {Dixon}, William V. and {Espaillat}, Catherine C. and {Evans}, Christopher J. and {Fox}, Andrew J. and {France}, Kevin and {Garcia}, Miriam and {Fleming}, Scott W. and {Frazer}, Elaine M. and {G{\'o}mez de Castro}, Ana I. and {Herczeg}, Gregory J. and {Hernandez}, Svea and {Hirschauer}, Alec S. and {James}, Bethan L. and {Johns-Krull}, Christopher M. and {Leitherer}, Claus and {Lockwood}, Sean and {Najita}, Joan and {Oey}, M.~S. and {Oliveira}, Cristina and {Pauly}, Tyler and {Reid}, I. Neill and {Riedel}, Adric and {Rodriguez}, David R. and {Sahnow}, David and {Sankrit}, Ravi and {Sembach}, Kenneth R. and {Shaw}, Richard and {Smith}, Linda J. and {Sohn}, S. Tony and {Som}, Debopam and {{\'U}beda}, Leonardo and {Welty}, Daniel E.},
        title = "{The UV Legacy Library of Young Stars as Essential Standards (ULLYSES) Large Director's Discretionary Program with Hubble. I. Goals, Design, and Initial Results}",
      journal = {\apj},
         year = 2025,
        month = may,
       volume = {985},
       number = {1},
          eid = {109},
        pages = {109},
          doi = {10.3847/1538-4357/adc45b},
archivePrefix = {arXiv},
       eprint = {2504.05446},
 primaryClass = {astro-ph.SR},
       adsurl = {https://ui.adsabs.harvard.edu/abs/2025ApJ...985..109R}
}

@ARTICLE{hunter1995,
       author = {{Hunter}, Deidre A. and {Shaya}, Edward J. and {Holtzman}, Jon A. and {Light}, Robert M. and {O'Neil}, Jr., Earl J. and {Lynds}, Roger},
        title = "{The Intermediate Stellar Mass Population in R136 Determined from Hubble Space Telescope Planetary Camera 2 Images}",
      journal = {\apj},
         year = 1995,
        month = jul,
       volume = {448},
        pages = {179},
          doi = {10.1086/175950},
       adsurl = {https://ui.adsabs.harvard.edu/abs/1995ApJ...448..179H}
}

@misc{murphy2021,
       author = {{Murphy}, Benjamin and {M{\"u}ller}, Sebastian and {Yurchak}, Roman},
        title = "{GeoStat-Framework/PyKrige: v1.6.0}",
         year = 2021,
        month = apr,
          eid = {10.5281/zenodo.4661732},
          doi = {10.5281/zenodo.4661732},
      version = {v1.6.0},
    publisher = {Zenodo},
       adsurl = {https://ui.adsabs.harvard.edu/abs/2021zndo...4661732M}
}

@ARTICLE{burkhart2015,
       author = {{Burkhart}, Blakesley and {Collins}, David C. and {Lazarian}, Alex},
        title = "{Observational Diagnostics of Self-gravitating MHD Turbulence in Giant Molecular Clouds}",
      journal = {\apj},
         year = 2015,
        month = jul,
       volume = {808},
       number = {1},
          eid = {48},
        pages = {48},
          doi = {10.1088/0004-637X/808/1/48},
archivePrefix = {arXiv},
       eprint = {1505.03855},
 primaryClass = {astro-ph.SR},
       adsurl = {https://ui.adsabs.harvard.edu/abs/2015ApJ...808...48B}
}

@ARTICLE{federrath2010,
       author = {{Federrath}, C. and {Roman-Duval}, J. and {Klessen}, R.~S. and {Schmidt}, W. and {Mac Low}, M.-M.},
        title = "{Comparing the statistics of interstellar turbulence in simulations and observations. Solenoidal versus compressive turbulence forcing}",
      journal = {\aap},
         year = 2010,
        month = mar,
       volume = {512},
          eid = {A81},
        pages = {A81},
          doi = {10.1051/0004-6361/200912437},
archivePrefix = {arXiv},
       eprint = {0905.1060},
 primaryClass = {astro-ph.SR},
       adsurl = {https://ui.adsabs.harvard.edu/abs/2010A&A...512A..81F}
}

@ARTICLE{Block2010,
       author = {{Block}, David L. and {Puerari}, Iv{\^a}nio and {Elmegreen}, Bruce G. and {Bournaud}, Fr{\'e}d{\'e}ric},
        title = "{A Two-component Power Law Covering Nearly Four Orders of Magnitude in the Power Spectrum of Spitzer Far-infrared Emission from the Large Magellanic Cloud}",
      journal = {\apjl},
         year = 2010,
        month = jul,
       volume = {718},
       number = {1},
        pages = {L1-L6},
          doi = {10.1088/2041-8205/718/1/L1},
archivePrefix = {arXiv},
       eprint = {1006.2080},
 primaryClass = {astro-ph.CO},
       adsurl = {https://ui.adsabs.harvard.edu/abs/2010ApJ...718L...1B}
}

@ARTICLE{elmegreen2001,
       author = {{Elmegreen}, Bruce G. and {Kim}, Sungeun and {Staveley-Smith}, Lister},
        title = "{A Fractal Analysis of the H I Emission from the Large Magellanic Cloud}",
      journal = {\apj},
         year = 2001,
        month = feb,
       volume = {548},
       number = {2},
        pages = {749-769},
          doi = {10.1086/319021},
archivePrefix = {arXiv},
       eprint = {astro-ph/0010578},
 primaryClass = {astro-ph},
       adsurl = {https://ui.adsabs.harvard.edu/abs/2001ApJ...548..749E}
}

@ARTICLE{nidever2017,
       author = {{Nidever}, David L. and {Olsen}, Knut and {Walker}, Alistair R. and {Vivas}, A. Katherina and {Blum}, Robert D. and {Kaleida}, Catherine and {Choi}, Yumi and {Conn}, Blair C. and {Gruendl}, Robert A. and {Bell}, Eric F. and {Besla}, Gurtina and {Mu{\~n}oz}, Ricardo R. and {Gallart}, Carme and {Martin}, Nicolas F. and {Olszewski}, Edward W. and {Saha}, Abhijit and {Monachesi}, Antonela and {Monelli}, Matteo and {de Boer}, Thomas J.~L. and {Johnson}, L. Clifton and {Zaritsky}, Dennis and {Stringfellow}, Guy S. and {van der Marel}, Roeland P. and {Cioni}, Maria-Rosa L. and {Jin}, Shoko and {Majewski}, Steven R. and {Martinez-Delgado}, David and {Monteagudo}, Lara and {No{\"e}l}, Noelia E.~D. and {Bernard}, Edouard J. and {Kunder}, Andrea and {Chu}, You-Hua and {Bell}, Cameron P.~M. and {Santana}, Felipe and {Frechem}, Joshua and {Medina}, Gustavo E. and {Parkash}, Vaishali and {Navarrete}, J.~C. Ser{\'o}n and {Hayes}, Christian},
        title = "{SMASH: Survey of the MAgellanic Stellar History}",
      journal = {\aj},
         year = 2017,
        month = nov,
       volume = {154},
       number = {5},
          eid = {199},
        pages = {199},
          doi = {10.3847/1538-3881/aa8d1c},
archivePrefix = {arXiv},
       eprint = {1701.00502},
 primaryClass = {astro-ph.GA},
       adsurl = {https://ui.adsabs.harvard.edu/abs/2017AJ....154..199N}
}

@ARTICLE{burhenne2025,
       author = {{Burhenne}, Clare and {McQuinn}, Kristen B.~W. and {Cohen}, Roger E. and {Murray}, Claire E. and {Patel}, Ekta and {Williams}, Benjamin F. and {Lindberg}, Christina W. and {Yanchulova Merica-Jones}, Petia and {Gordon}, Karl D. and {Choi}, Yumi and {Dolphin}, Andrew E. and {Roman-Duval}, Julia C.},
        title = "{Scylla V: Constraints on the spatial and temporal distribution of bursts and the interaction history of the Magellanic Clouds from their resolved stellar populations}",
      journal = {arXiv e-prints},
         year = 2025,
        month = nov,
          eid = {arXiv:2511.02947},
        pages = {arXiv:2511.02947},
          doi = {10.48550/arXiv.2511.02947},
archivePrefix = {arXiv},
       eprint = {2511.02947},
 primaryClass = {astro-ph.GA},
       adsurl = {https://ui.adsabs.harvard.edu/abs/2025arXiv251102947B}
}

@ARTICLE{Haschke2011,
       author = {{Haschke}, Raoul and {Grebel}, Eva K. and {Duffau}, Sonia},
        title = "{New Optical Reddening Maps of the Large and Small Magellanic Clouds}",
      journal = {\aj},
         year = 2011,
        month = may,
       volume = {141},
       number = {5},
          eid = {158},
        pages = {158},
          doi = {10.1088/0004-6256/141/5/158},
archivePrefix = {arXiv},
       eprint = {1104.2325},
 primaryClass = {astro-ph.GA},
       adsurl = {https://ui.adsabs.harvard.edu/abs/2011AJ....141..158H}
}

@ARTICLE{Imara2007,
       author = {{Imara}, Nia and {Blitz}, Leo},
        title = "{Extinction in the Large Magellanic Cloud}",
      journal = {\apj},
         year = 2007,
        month = jun,
       volume = {662},
       number = {2},
        pages = {969-979},
          doi = {10.1086/517911},
archivePrefix = {arXiv},
       eprint = {astro-ph/0703421},
 primaryClass = {astro-ph},
       adsurl = {https://ui.adsabs.harvard.edu/abs/2007ApJ...662..969I}
}

@ARTICLE{gorski2020,
       author = {{G{\'o}rski}, Marek and {Zgirski}, Bart{\l}omiej and {Pietrzy{\'n}ski}, Grzegorz and {Gieren}, Wolfgang and {Wielg{\'o}rski}, Piotr and {Graczyk}, Dariusz and {Kudritzki}, Rolf-Peter and {Pilecki}, Bogumi{\l} and {Narloch}, Weronika and {Karczmarek}, Paulina and {Suchomska}, Ksenia and {Taormina}, M{\'o}nica},
        title = "{Empirical Calibration of the Reddening Maps in the Magellanic Clouds}",
      journal = {\apj},
         year = 2020,
        month = feb,
       volume = {889},
       number = {2},
          eid = {179},
        pages = {179},
          doi = {10.3847/1538-4357/ab65ed},
archivePrefix = {arXiv},
       eprint = {2001.08242},
 primaryClass = {astro-ph.GA},
       adsurl = {https://ui.adsabs.harvard.edu/abs/2020ApJ...889..179G}
}

@ARTICLE{kohler2015,
       author = {{K{\"o}hler}, M. and {Ysard}, N. and {Jones}, A.~P.},
        title = "{Dust evolution in the transition towards the denser ISM: impact on dust temperature, opacity, and spectral index}",
      journal = {\aap},
         year = 2015,
        month = jul,
       volume = {579},
          eid = {A15},
        pages = {A15},
          doi = {10.1051/0004-6361/201525646},
archivePrefix = {arXiv},
       eprint = {1506.01533},
 primaryClass = {astro-ph.GA},
       adsurl = {https://ui.adsabs.harvard.edu/abs/2015A&A...579A..15K}
}

@ARTICLE{edenhofer2024,
       author = {{Edenhofer}, Gordian and {Zucker}, Catherine and {Frank}, Philipp and {Saydjari}, Andrew K. and {Speagle}, Joshua S. and {Finkbeiner}, Douglas and {En{\ss}lin}, Torsten A.},
        title = "{A parsec-scale Galactic 3D dust map out to 1.25 kpc from the Sun}",
      journal = {\aap},
         year = 2024,
        month = may,
       volume = {685},
          eid = {A82},
        pages = {A82},
          doi = {10.1051/0004-6361/202347628},
archivePrefix = {arXiv},
       eprint = {2308.01295},
 primaryClass = {astro-ph.GA},
       adsurl = {https://ui.adsabs.harvard.edu/abs/2024A&A...685A..82E}
}

@ARTICLE{williams2021,
       author = {{Williams}, Benjamin F. and {Durbin}, Meredith J. and {Dalcanton}, Julianne J. and {Lang}, Dustin and {Girardi}, Leo and {Smercina}, Adam and {Dolphin}, Andrew and {Weisz}, Daniel R. and {Choi}, Yumi and {Bell}, Eric F. and {Rosolowsky}, Erik and {Skillman}, Evan and {Koch}, Eric W. and {Lindberg}, Christina W. and {Hagen}, Lea and {Gordon}, Karl D. and {Seth}, Anil and {Gilbert}, Karoline and {Guhathakurta}, Puragra and {Lauer}, Tod and {Bianchi}, Luciana},
        title = "{The Panchromatic Hubble Andromeda Treasury: Triangulum Extended Region (PHATTER). I. Ultraviolet to Infrared Photometry of 22 Million Stars in M33}",
      journal = {\apjs},
         year = 2021,
        month = apr,
       volume = {253},
       number = {2},
          eid = {53},
        pages = {53},
          doi = {10.3847/1538-4365/abdf4e},
archivePrefix = {arXiv},
       eprint = {2101.01293},
 primaryClass = {astro-ph.GA},
       adsurl = {https://ui.adsabs.harvard.edu/abs/2021ApJS..253...53W}
}

@ARTICLE{2013A&A...558A..33A,
       author = {{Astropy Collaboration} and {Robitaille}, Thomas P. and
         {Tollerud}, Erik J. and {Greenfield}, Perry and {Droettboom}, Michael and
         {Bray}, Erik and {Aldcroft}, Tom and {Davis}, Matt and
         {Ginsburg}, Adam and {Price-Whelan}, Adrian M. and
         {Kerzendorf}, Wolfgang E. and {Conley}, Alexander and {Crighton}, Neil and
         {Barbary}, Kyle and {Muna}, Demitri and {Ferguson}, Henry and
         {Grollier}, Fr{\'e}d{\'e}ric and {Parikh}, Madhura M. and
         {Nair}, Prasanth H. and {Unther}, Hans M. and {Deil}, Christoph and
         {Woillez}, Julien and {Conseil}, Simon and {Kramer}, Roban and
         {Turner}, James E.~H. and {Singer}, Leo and {Fox}, Ryan and
         {Weaver}, Benjamin A. and {Zabalza}, Victor and {Edwards}, Zachary I. and
         {Azalee Bostroem}, K. and {Burke}, D.~J. and {Casey}, Andrew R. and
         {Crawford}, Steven M. and {Dencheva}, Nadia and {Ely}, Justin and
         {Jenness}, Tim and {Labrie}, Kathleen and {Lim}, Pey Lian and
         {Pierfederici}, Francesco and {Pontzen}, Andrew and {Ptak}, Andy and
         {Refsdal}, Brian and {Servillat}, Mathieu and {Streicher}, Ole},
        title = "{Astropy: A community Python package for astronomy}",
      journal = {\aap},
         year = "2013",
        month = "Oct",
       volume = {558},
          eid = {A33},
        pages = {A33},
          doi = {10.1051/0004-6361/201322068},
archivePrefix = {arXiv},
       eprint = {1307.6212},
 primaryClass = {astro-ph.IM},
       adsurl = {https://ui.adsabs.harvard.edu/abs/2013A&A...558A..33A}
}

@ARTICLE{pingel2022,
       author = {{Pingel}, N.~M. and {Dempsey}, J. and {McClure-Griffiths}, N.~M. and {Dickey}, J.~M. and {Jameson}, K.~E. and {Arce}, H. and {Anglada}, G. and {Bland-Hawthorn}, J. and {Breen}, S.~L. and {Buckland-Willis}, F. and {Clark}, S.~E. and {Dawson}, J.~R. and {D{\'e}nes}, H. and {Di Teodoro}, E.~M. and {For}, B. -Q. and {Foster}, Tyler J. and {G{\'o}mez}, J.~F. and {Imai}, H. and {Joncas}, G. and {Kim}, C. -G. and {Lee}, M. -Y. and {Lynn}, C. and {Leahy}, D. and {Ma}, Y.~K. and {Marchal}, A. and {McConnell}, D. and {Miville-Desch{\`e}nes}, M. -A. and {Moss}, V.~A. and {Murray}, C.~E. and {Nidever}, D. and {Peek}, J. and {Stanimirovi{\'c}}, S. and {Staveley-Smith}, L. and {Tepper-Garcia}, T. and {Tremblay}, C.~D. and {Uscanga}, L. and {van Loon}, J. Th. and {V{\'a}zquez-Semadeni}, E. and {Allison}, J.~R. and {Anderson}, C.~S. and {Ball}, Lewis and {Bell}, M. and {Bock}, D.~C. -J. and {Bunton}, J. and {Cooray}, F.~R. and {Cornwell}, T. and {Koribalski}, B.~S. and {Gupta}, N. and {Hayman}, D.~B. and {Harvey-Smith}, L. and {Lee-Waddell}, K. and {Ng}, A. and {Phillips}, C.~J. and {Voronkov}, M. and {Westmeier}, T. and {Whiting}, M.~T.},
        title = "{GASKAP-HI pilot survey science I: ASKAP zoom observations of HI emission in the Small Magellanic Cloud}",
      journal = {\pasa},
         year = 2022,
        month = feb,
       volume = {39},
          eid = {e005},
        pages = {e005},
          doi = {10.1017/pasa.2021.59},
archivePrefix = {arXiv},
       eprint = {2111.05339},
 primaryClass = {astro-ph.GA},
       adsurl = {https://ui.adsabs.harvard.edu/abs/2022PASA...39....5P}
}

@ARTICLE{weisz2013,
   author = {{Weisz}, D.~R. and {Dolphin}, A.~E. and {Skillman}, E.~D. and 
	{Holtzman}, J. and {Dalcanton}, J.~J. and {Cole}, A.~A. and 
	{Neary}, K.},
    title = "{Comparing the ancient star formation histories of the Magellanic Clouds}",
  journal = {\mnras},
archivePrefix = "arXiv",
   eprint = {1301.7422},
     year = 2013,
    month = may,
   volume = 431,
    pages = {364-371},
      doi = {10.1093/mnras/stt165},
   adsurl = {http://adsabs.harvard.edu/abs/2013MNRAS.431..364W}
}

@ARTICLE{besla2007,
   author = {{Besla}, G. and {Kallivayalil}, N. and {Hernquist}, L. and {Robertson}, B. and 
	{Cox}, T.~J. and {van der Marel}, R.~P. and {Alcock}, C.},
    title = "{Are the Magellanic Clouds on Their First Passage about the Milky Way?}",
  journal = {\apj},
   eprint = {astro-ph/0703196},
     year = 2007,
    month = oct,
   volume = 668,
    pages = {949-967},
      doi = {10.1086/521385},
   adsurl = {http://adsabs.harvard.edu/abs/2007ApJ...668..949B}
}

@ARTICLE{cardelli1989,
   author = {{Cardelli}, J.~A. and {Clayton}, G.~C. and {Mathis}, J.~S.},
    title = "{The relationship between infrared, optical, and ultraviolet extinction}",
  journal = {\apj},
     year = 1989,
    month = oct,
   volume = 345,
    pages = {245-256},
      doi = {10.1086/167900},
   adsurl = {http://adsabs.harvard.edu/abs/1989ApJ...345..245C}
}

@ARTICLE{kim1999,
   author = {{Kim}, S. and {Dopita}, M.~A. and {Staveley-Smith}, L. and {Bessell}, M.~S.
	},
    title = "{H I Shells in the Large Magellanic Cloud}",
  journal = {\aj},
     year = 1999,
    month = dec,
   volume = 118,
    pages = {2797-2823},
      doi = {10.1086/301116},
   adsurl = {http://adsabs.harvard.edu/abs/1999AJ....118.2797K}
}

@ARTICLE{sabbi2013,
   author = {{Sabbi}, E. and {Anderson}, J. and {Lennon}, D.~J. and {van der Marel}, R.~P. and 
	{Aloisi}, A. and {Boyer}, M.~L. and {Cignoni}, M. and {de Marchi}, G. and 
	{de Mink}, S.~E. and {Evans}, C.~J. and {Gallagher}, III, J.~S. and 
	{Gordon}, K. and {Gouliermis}, D.~A. and {Grebel}, E.~K. and 
	{Koekemoer}, A.~M. and {Larsen}, S.~S. and {Panagia}, N. and 
	{Ryon}, J.~E. and {Smith}, L.~J. and {Tosi}, M. and {Zaritsky}, D.
	},
    title = "{Hubble Tarantula Treasury Project: Unraveling Tarantula's Web. I. Observational Overview and First Results}",
  journal = {\aj},
archivePrefix = "arXiv",
   eprint = {1304.6747},
     year = 2013,
    month = sep,
   volume = 146,
      eid = {53},
    pages = {53},
      doi = {10.1088/0004-6256/146/3/53},
   adsurl = {http://adsabs.harvard.edu/abs/2013AJ....146...53S}
}

@ARTICLE{pietrzynski2019,
   author = {{Pietrzynski}, G. and {Graczyk}, D. and {Gallenne}, A. and {Gieren}, W. and 
	{Thompson}, I.~B. and {Pilecki}, B. and {Karczmarek}, P. and 
	{Gorski}, M. and {Suchomska}, K. and {Taormina}, M. and {Zgirski}, B. and 
	{Wielgorski}, P. and {Kolaczkowski}, Z. and {Konorski}, P. and 
	{Villanova}, S. and {Nardetto}, N. and {Kervella}, P. and {Bresolin}, F. and 
	{Kudritzki}, R.~P. and {Storm}, J. and {Smolec}, R. and {Narloch}, W.
	},
    title = "{A distance to the Large Magellanic Cloud that is precise to one per cent}",
  journal = {arXiv e-prints},
archivePrefix = "arXiv",
   eprint = {1903.08096},
     year = 2019,
    month = mar,
   adsurl = {http://adsabs.harvard.edu/abs/2019arXiv190308096P}
}

@ARTICLE{gordon2014,
   author = {{Gordon}, K.~D. and {Roman-Duval}, J. and {Bot}, C. and {Meixner}, M. and 
	{Babler}, B. and {Bernard}, J.-P. and {Bolatto}, A. and {Boyer}, M.~L. and 
	{Clayton}, G.~C. and {Engelbracht}, C. and {Fukui}, Y. and {Galametz}, M. and 
	{Galliano}, F. and {Hony}, S. and {Hughes}, A. and {Indebetouw}, R. and 
	{Israel}, F.~P. and {Jameson}, K. and {Kawamura}, A. and {Lebouteiller}, V. and 
	{Li}, A. and {Madden}, S.~C. and {Matsuura}, M. and {Misselt}, K. and 
	{Montiel}, E. and {Okumura}, K. and {Onishi}, T. and {Panuzzo}, P. and 
	{Paradis}, D. and {Rubio}, M. and {Sandstrom}, K. and {Sauvage}, M. and 
	{Seale}, J. and {Sewi{\l}o}, M. and {Tchernyshyov}, K. and {Skibba}, R.
	},
    title = "{Dust and Gas in the Magellanic Clouds from the HERITAGE Herschel Key Project. I. Dust Properties and Insights into the Origin of the Submillimeter Excess Emission}",
  journal = {\apj},
archivePrefix = "arXiv",
   eprint = {1406.6066},
     year = 2014,
    month = dec,
   volume = 797,
      eid = {85},
    pages = {85},
      doi = {10.1088/0004-637X/797/2/85},
   adsurl = {http://adsabs.harvard.edu/abs/2014ApJ...797...85G}
}

@ARTICLE{romanduval2019,
   author = {{Roman-Duval}, J. and {Jenkins}, E.~B. and {Williams}, B. and 
	{Tchernyshyov}, K. and {Gordon}, K. and {Meixner}, M. and {Hagen}, L. and 
	{Peek}, J. and {Sandstrom}, K. and {Werk}, J. and {Yanchulova Merica-Jones}, P.
	},
    title = "{METAL: The Metal Evolution, Transport, and Abundance in the Large Magellanic Cloud Hubble Program. I. Overview and Initial Results}",
  journal = {\apj},
archivePrefix = "arXiv",
   eprint = {1901.06027},
     year = 2019,
    month = feb,
   volume = 871,
      eid = {151},
    pages = {151},
      doi = {10.3847/1538-4357/aaf8bb},
   adsurl = {http://adsabs.harvard.edu/abs/2019ApJ...871..151R}
}

@ARTICLE{zivick2018,
   author = {{Zivick}, P. and {Kallivayalil}, N. and {van der Marel}, R.~P. and 
	{Besla}, G. and {Linden}, S.~T. and {Koz{\l}owski}, S. and {Fritz}, T.~K. and 
	{Kochanek}, C.~S. and {Anderson}, J. and {Sohn}, S.~T. and {Geha}, M.~C. and 
	{Alcock}, C.~R.},
    title = "{The Proper Motion Field of the Small Magellanic Cloud: Kinematic Evidence for Its Tidal Disruption}",
  journal = {\apj},
archivePrefix = "arXiv",
   eprint = {1804.04110},
     year = 2018,
    month = sep,
   volume = 864,
      eid = {55},
    pages = {55},
      doi = {10.3847/1538-4357/aad4b0},
   adsurl = {http://adsabs.harvard.edu/abs/2018ApJ...864...55Z}
}

@ARTICLE{ymj2021,
       author = {{Yanchulova Merica-Jones}, Petia and {Sandstrom}, Karin M. and {Johnson}, L. Clifton and {Dolphin}, Andrew E. and {Dalcanton}, Julianne J. and {Gordon}, Karl and {Roman-Duval}, Julia and {Weisz}, Daniel R. and {Williams}, Benjamin F.},
        title = "{Three-dimensional Structure and Dust Extinction in the Small Magellanic Cloud}",
      journal = {\apj},
         year = 2021,
        month = jan,
       volume = {907},
       number = {1},
          eid = {50},
        pages = {50},
          doi = {10.3847/1538-4357/abc48b},
archivePrefix = {arXiv},
       eprint = {2010.11181},
 primaryClass = {astro-ph.GA},
       adsurl = {https://ui.adsabs.harvard.edu/abs/2021ApJ...907...50Y}
}

@ARTICLE{gordon2003,
   author = {{Gordon}, K.~D. and {Clayton}, G.~C. and {Misselt}, K.~A. and 
	{Landolt}, A.~U. and {Wolff}, M.~J.},
    title = "{A Quantitative Comparison of the Small Magellanic Cloud, Large Magellanic Cloud, and Milky Way Ultraviolet to Near-Infrared Extinction Curves}",
  journal = {\apj},
   eprint = {astro-ph/0305257},
     year = 2003,
    month = sep,
   volume = 594,
    pages = {279-293},
      doi = {10.1086/376774},
   adsurl = {http://adsabs.harvard.edu/abs/2003ApJ...594..279G}
}

@ARTICLE{bressan2012,
   author = {{Bressan}, A. and {Marigo}, P. and {Girardi}, L. and {Salasnich}, B. and 
	{Dal Cero}, C. and {Rubele}, S. and {Nanni}, A.},
    title = "{PARSEC: stellar tracks and isochrones with the PAdova and TRieste Stellar Evolution Code}",
  journal = {\mnras},
archivePrefix = "arXiv",
   eprint = {1208.4498},
 primaryClass = "astro-ph.SR",
     year = 2012,
    month = nov,
   volume = 427,
    pages = {127-145},
      doi = {10.1111/j.1365-2966.2012.21948.x},
   adsurl = {http://adsabs.harvard.edu/abs/2012MNRAS.427..127B}
}

@ARTICLE{gaiacollaboration2018,
       author = {{Gaia Collaboration} and {Helmi}, A. and {van Leeuwen}, F. and
         {McMillan}, P.~J. and {Massari}, D. and {Antoja}, T. and
         {Robin}, A.~C. and {Lindegren}, L. and {Bastian}, U. and {Arenou}, F. and
         {Babusiaux}, C. and {Biermann}, M. and {Breddels}, M.~A. and
         {Hobbs}, D. and {Jordi}, C. and {Pancino}, E. and {Reyl{\'e}}, C. and
         {Veljanoski}, J. and {Brown}, A.~G.~A. and {Vallenari}, A. and
         {Prusti}, T. and {de Bruijne}, J.~H.~J. and {Bailer-Jones}, C.~A.~L. and
         {Evans}, D.~W. and {Eyer}, L. and {Jansen}, F. and {Klioner}, S.~A. and
         {Lammers}, U. and {Luri}, X. and {Mignard}, F. and {Panem}, C. and
         {Pourbaix}, D. and {Randich}, S. and {Sartoretti}, P. and
         {Siddiqui}, H.~I. and {Soubiran}, C. and {Walton}, N.~A. and
         {Cropper}, M. and {Drimmel}, R. and {Katz}, D. and {Lattanzi}, M.~G. and
         {Bakker}, J. and {Cacciari}, C. and {Casta{\~n}eda}, J. and
         {Chaoul}, L. and {Cheek}, N. and {De Angeli}, F. and {Fabricius}, C. and
         {Guerra}, R. and {Holl}, B. and {Masana}, E. and {Messineo}, R. and
         {Mowlavi}, N. and {Nienartowicz}, K. and {Panuzzo}, P. and
         {Portell}, J. and {Riello}, M. and {Seabroke}, G.~M. and {Tanga}, P. and
         {Th{\'e}venin}, F. and {Gracia-Abril}, G. and {Comoretto}, G. and
         {Garcia-Reinaldos}, M. and {Teyssier}, D. and {Altmann}, M. and
         {Andrae}, R. and {Audard}, M. and {Bellas-Velidis}, I. and
         {Benson}, K. and {Berthier}, J. and {Blomme}, R. and {Burgess}, P. and
         {Busso}, G. and {Carry}, B. and {Cellino}, A. and {Clementini}, G. and
         {Clotet}, M. and {Creevey}, O. and {Davidson}, M. and {De Ridder}, J. and
         {Delchambre}, L. and {Dell'Oro}, A. and {Ducourant}, C. and
         {Fern{\'a}ndez-Hern{\'a}ndez}, J. and {Fouesneau}, M. and
         {Fr{\'e}mat}, Y. and {Galluccio}, L. and {Garc{\'\i}a-Torres}, M. and
         {Gonz{\'a}lez-N{\'u}{\~n}ez}, J. and {Gonz{\'a}lez-Vidal}, J.~J. and
         {Gosset}, E. and {Guy}, L.~P. and {Halbwachs}, J. -L. and
         {Hambly}, N.~C. and {Harrison}, D.~L. and {Hern{\'a}ndez}, J. and
         {Hestroffer}, D. and {Hodgkin}, S.~T. and {Hutton}, A. and
         {Jasniewicz}, G. and {Jean-Antoine-Piccolo}, A. and {Jordan}, S. and
         {Korn}, A.~J. and {Krone-Martins}, A. and {Lanzafame}, A.~C. and
         {Lebzelter}, T. and {L{\"o}ffler}, W. and {Manteiga}, M. and
         {Marrese}, P.~M. and {Mart{\'\i}n-Fleitas}, J.~M. and {Moitinho}, A. and
         {Mora}, A. and {Muinonen}, K. and {Osinde}, J. and {Pauwels}, T. and
         {Petit}, J. -M. and {Recio-Blanco}, A. and {Richards}, P.~J. and
         {Rimoldini}, L. and {Sarro}, L.~M. and {Siopis}, C. and {Smith}, M. and
         {Sozzetti}, A. and {S{\"u}veges}, M. and {Torra}, J. and
         {van Reeven}, W. and {Abbas}, U. and {Abreu Aramburu}, A. and
         {Accart}, S. and {Aerts}, C. and {Altavilla}, G. and
         {{\'A}lvarez}, M.~A. and {Alvarez}, R. and {Alves}, J. and
         {Anderson}, R.~I. and {Andrei}, A.~H. and {Anglada Varela}, E. and
         {Antiche}, E. and {Arcay}, B. and {Astraatmadja}, T.~L. and {Bach}, N. and
         {Baker}, S.~G. and {Balaguer-N{\'u}{\~n}ez}, L. and {Balm}, P. and
         {Barache}, C. and {Barata}, C. and {Barbato}, D. and {Barblan}, F. and
         {Barklem}, P.~S. and {Barrado}, D. and {Barros}, M. and
         {Barstow}, M.~A. and {Bartholom{\'e} Mu{\~n}oz}, S. and
         {Bassilana}, J. -L. and {Becciani}, U. and {Bellazzini}, M. and
         {Berihuete}, A. and {Bertone}, S. and {Bianchi}, L. and
         {Bienaym{\'e}}, O. and {Blanco-Cuaresma}, S. and {Boch}, T. and
         {Boeche}, C. and {Bombrun}, A. and {Borrachero}, R. and {Bossini}, D. and
         {Bouquillon}, S. and {Bourda}, G. and {Bragaglia}, A. and
         {Bramante}, L. and {Bressan}, A. and {Brouillet}, N. and
         {Br{\"u}semeister}, T. and {Brugaletta}, E. and {Bucciarelli}, B. and
         {Burlacu}, A. and {Busonero}, D. and {Butkevich}, A.~G. and
         {Buzzi}, R. and {Caffau}, E. and {Cancelliere}, R. and
         {Cannizzaro}, G. and {Cantat-Gaudin}, T. and {Carballo}, R. and
         {Carlucci}, T. and {Carrasco}, J.~M. and {Casamiquela}, L. and
         {Castellani}, M. and {Castro-Ginard}, A. and {Charlot}, P. and
         {Chemin}, L. and {Chiavassa}, A. and {Cocozza}, G. and {Costigan}, G. and
         {Cowell}, S. and {Crifo}, F. and {Crosta}, M. and {Crowley}, C. and
         {Cuypers}, J. and {Dafonte}, C. and {Damerdji}, Y. and
         {Dapergolas}, A. and {David}, P. and {David}, M. and {de Laverny}, P. and
         {De Luise}, F. and {De March}, R. and {de Martino}, D. and
         {de Souza}, R. and {de Torres}, A. and {Debosscher}, J. and
         {del Pozo}, E. and {Delbo}, M. and {Delgado}, A. and {Delgado}, H.~E. and
         {Di Matteo}, P. and {Diakite}, S. and {Diener}, C. and {Distefano}, E. and
         {Dolding}, C. and {Drazinos}, P. and {Dur{\'a}n}, J. and
         {Edvardsson}, B. and {Enke}, H. and {Eriksson}, K. and {Esquej}, P. and
         {Eynard Bontemps}, G. and {Fabre}, C. and {Fabrizio}, M. and
         {Faigler}, S. and {Falc{\~a}o}, A.~J. and {Farr{\`a}s Casas}, M. and
         {Federici}, L. and {Fedorets}, G. and {Fernique}, P. and
         {Figueras}, F. and {Filippi}, F. and {Findeisen}, K. and {Fonti}, A. and
         {Fraile}, E. and {Fraser}, M. and {Fr{\'e}zouls}, B. and {Gai}, M. and
         {Galleti}, S. and {Garabato}, D. and {Garc{\'\i}a-Sedano}, F. and
         {Garofalo}, A. and {Garralda}, N. and {Gavel}, A. and {Gavras}, P. and
         {Gerssen}, J. and {Geyer}, R. and {Giacobbe}, P. and {Gilmore}, G. and
         {Girona}, S. and {Giuffrida}, G. and {Glass}, F. and {Gomes}, M. and
         {Granvik}, M. and {Gueguen}, A. and {Guerrier}, A. and {Guiraud}, J. and
         {Guti{\'e}rrez-S{\'a}nchez}, R. and {Hofmann}, W. and {Holland}, G. and
         {Huckle}, H.~E. and {Hypki}, A. and {Icardi}, V. and {Jan{\ss}en}, K. and
         {Jevardat de Fombelle}, G. and {Jonker}, P.~G. and
         {Juh{\'a}sz}, {\'A}. L. and {Julbe}, F. and {Karampelas}, A. and
         {Kewley}, A. and {Klar}, J. and {Kochoska}, A. and {Kohley}, R. and
         {Kolenberg}, K. and {Kontizas}, M. and {Kontizas}, E. and
         {Koposov}, S.~E. and {Kordopatis}, G. and {Kostrzewa-Rutkowska}, Z. and
         {Koubsky}, P. and {Lambert}, S. and {Lanza}, A.~F. and {Lasne}, Y. and
         {Lavigne}, J. -B. and {Le Fustec}, Y. and {Le Poncin-Lafitte}, C. and
         {Lebreton}, Y. and {Leccia}, S. and {Leclerc}, N. and
         {Lecoeur-Taibi}, I. and {Lenhardt}, H. and {Leroux}, F. and {Liao}, S. and
         {Licata}, E. and {Lindstr{\o}m}, H.~E.~P. and {Lister}, T.~A. and
         {Livanou}, E. and {Lobel}, A. and {L{\'o}pez}, M. and {Managau}, S. and
         {Mann}, R.~G. and {Mantelet}, G. and {Marchal}, O. and
         {Marchant}, J.~M. and {Marconi}, M. and {Marinoni}, S. and
         {Marschalk{\'o}}, G. and {Marshall}, D.~J. and {Martino}, M. and
         {Marton}, G. and {Mary}, N. and {Matijevi{\v{c}}}, G. and {Mazeh}, T. and
         {Messina}, S. and {Michalik}, D. and {Millar}, N.~R. and {Molina}, D. and
         {Molinaro}, R. and {Moln{\'a}r}, L. and {Montegriffo}, P. and
         {Mor}, R. and {Morbidelli}, R. and {Morel}, T. and {Morris}, D. and
         {Mulone}, A.~F. and {Muraveva}, T. and {Musella}, I. and
         {Nelemans}, G. and {Nicastro}, L. and {Noval}, L. and {O'Mullane}, W. and
         {Ord{\'e}novic}, C. and {Ord{\'o}{\~n}ez-Blanco}, D. and {Osborne}, P. and
         {Pagani}, C. and {Pagano}, I. and {Pailler}, F. and {Palacin}, H. and
         {Palaversa}, L. and {Panahi}, A. and {Pawlak}, M. and
         {Piersimoni}, A.~M. and {Pineau}, F. -X. and {Plachy}, E. and
         {Plum}, G. and {Poggio}, E. and {Poujoulet}, E. and {Pr{\v{s}}a}, A. and
         {Pulone}, L. and {Racero}, E. and {Ragaini}, S. and {Rambaux}, N. and
         {Ramos-Lerate}, M. and {Regibo}, S. and {Riclet}, F. and {Ripepi}, V. and
         {Riva}, A. and {Rivard}, A. and {Rixon}, G. and {Roegiers}, T. and
         {Roelens}, M. and {Romero-G{\'o}mez}, M. and {Rowell}, N. and
         {Royer}, F. and {Ruiz-Dern}, L. and {Sadowski}, G. and
         {Sagrist{\`a} Sell{\'e}s}, T. and {Sahlmann}, J. and {Salgado}, J. and
         {Salguero}, E. and {Sanna}, N. and {Santana-Ros}, T. and {Sarasso}, M. and
         {Savietto}, H. and {Schultheis}, M. and {Sciacca}, E. and {Segol}, M. and
         {Segovia}, J.~C. and {S{\'e}gransan}, D. and {Shih}, I. -C. and
         {Siltala}, L. and {Silva}, A.~F. and {Smart}, R.~L. and {Smith}, K.~W. and
         {Solano}, E. and {Solitro}, F. and {Sordo}, R. and {Soria Nieto}, S. and
         {Souchay}, J. and {Spagna}, A. and {Spoto}, F. and {Stampa}, U. and
         {Steele}, I.~A. and {Steidelm{\"u}ller}, H. and {Stephenson}, C.~A. and
         {Stoev}, H. and {Suess}, F.~F. and {Surdej}, J. and {Szabados}, L. and
         {Szegedi-Elek}, E. and {Tapiador}, D. and {Taris}, F. and {Tauran}, G. and
         {Taylor}, M.~B. and {Teixeira}, R. and {Terrett}, D. and {Teyssand
        ier}, P. and {Thuillot}, W. and {Titarenko}, A. and {Torra Clotet}, F. and
         {Turon}, C. and {Ulla}, A. and {Utrilla}, E. and {Uzzi}, S. and
         {Vaillant}, M. and {Valentini}, G. and {Valette}, V. and
         {van Elteren}, A. and {Van Hemelryck}, E. and {van Leeuwen}, M. and
         {Vaschetto}, M. and {Vecchiato}, A. and {Viala}, Y. and {Vicente}, D. and
         {Vogt}, S. and {von Essen}, C. and {Voss}, H. and {Votruba}, V. and
         {Voutsinas}, S. and {Walmsley}, G. and {Weiler}, M. and {Wertz}, O. and
         {Wevems}, T. and {Wyrzykowski}, {\L}. and {Yoldas}, A. and
         {{\v{Z}}erjal}, M. and {Ziaeepour}, H. and {Zorec}, J. and
         {Zschocke}, S. and {Zucker}, S. and {Zurbach}, C. and {Zwitter}, T.},
        title = "{Gaia Data Release 2. Kinematics of globular clusters and dwarf galaxies around the Milky Way}",
      journal = {\aap},
         year = "2018",
        month = "Aug",
       volume = {616},
          eid = {A12},
        pages = {A12},
          doi = {10.1051/0004-6361/201832698},
archivePrefix = {arXiv},
       eprint = {1804.09381},
 primaryClass = {astro-ph.GA},
       adsurl = {https://ui.adsabs.harvard.edu/\#abs/2018A&A...616A..12G}
}

@ARTICLE{donghia2016,
   author = {{D'Onghia}, E. and {Fox}, A.~J.},
    title = "{The Magellanic Stream: Circumnavigating the Galaxy}",
  journal = {\araa},
archivePrefix = "arXiv",
   eprint = {1511.05853},
     year = 2016,
    month = sep,
   volume = 54,
    pages = {363-400},
      doi = {10.1146/annurev-astro-081915-023251},
   adsurl = {http://adsabs.harvard.edu/abs/2016ARA%26A..54..363D}
}

@ARTICLE{henize1956,
       author = {{Henize}, Karl G.},
        title = "{Catalogues of H{\ensuremath{\alpha}}-emission Stars and Nebulae in the Magellanic Clouds.}",
      journal = {\apjs},
         year = 1956,
        month = sep,
       volume = {2},
        pages = {315},
          doi = {10.1086/190025},
       adsurl = {https://ui.adsabs.harvard.edu/abs/1956ApJS....2..315H}
}

@ARTICLE{wong2022,
       author = {{Wong}, Tony and {Oudshoorn}, Luuk and {Sofovich}, Eliyahu and {Green}, Alex and {Shah}, Charmi and {Indebetouw}, R{\'e}my and {Meixner}, Margaret and {Hacar}, Alvaro and {Nayak}, Omnarayani and {Tokuda}, Kazuki and {Bolatto}, Alberto D. and {Chevance}, M{\'e}lanie and {De Marchi}, Guido and {Fukui}, Yasuo and {Hirschauer}, Alec S. and {Jameson}, K.~E. and {Kalari}, Venu and {Lebouteiller}, Vianney and {Looney}, Leslie W. and {Madden}, Suzanne C. and {Onishi}, Toshikazu and {Roman-Duval}, Julia and {Rubio}, M{\'o}nica and {Tielens}, A.~G.~G.~M.},
        title = "{The 30 Doradus Molecular Cloud at 0.4 pc Resolution with the Atacama Large Millimeter/submillimeter Array: Physical Properties and the Boundedness of CO-emitting Structures}",
      journal = {\apj},
         year = 2022,
        month = jun,
       volume = {932},
       number = {1},
          eid = {47},
        pages = {47},
          doi = {10.3847/1538-4357/ac723a},
archivePrefix = {arXiv},
       eprint = {2206.06528},
 primaryClass = {astro-ph.GA},
       adsurl = {https://ui.adsabs.harvard.edu/abs/2022ApJ...932...47W}
}

@ARTICLE{Chevance2020,
       author = {{Chevance}, M{\'e}lanie and {Madden}, Suzanne C. and {Fischer}, Christian and {Vacca}, William D. and {Lebouteiller}, Vianney and {Fadda}, Dario and {Galliano}, Fr{\'e}d{\'e}ric and {Indebetouw}, Remy and {Kruijssen}, J.~M. Diederik and {Lee}, Min-Young and {Poglitsch}, Albrecht and {Polles}, Fiorella L. and {Cormier}, Diane and {Hony}, Sacha and {Iserlohe}, Christof and {Krabbe}, Alfred and {Meixner}, Margaret and {Sabbi}, Elena and {Zinnecker}, Hans},
        title = "{The CO-dark molecular gas mass in 30 Doradus}",
      journal = {\mnras},
         year = 2020,
        month = jun,
       volume = {494},
       number = {4},
        pages = {5279-5292},
          doi = {10.1093/mnras/staa1106},
archivePrefix = {arXiv},
       eprint = {2004.09516},
 primaryClass = {astro-ph.GA},
       adsurl = {https://ui.adsabs.harvard.edu/abs/2020MNRAS.494.5279C}
}

@ARTICLE{kolmogorov,
       author = {{Kolmogorov}, A.},
        title = "{The Local Structure of Turbulence in Incompressible Viscous Fluid for Very Large Reynolds' Numbers}",
      journal = {Akademiia Nauk SSSR Doklady},
         year = 1941,
        month = jan,
       volume = {30},
        pages = {301-305},
       adsurl = {https://ui.adsabs.harvard.edu/abs/1941DoSSR..30..301K}
}

@ARTICLE{casey2012,
       author = {{Casey}, C.~M. and {Berta}, S. and {B{\'e}thermin}, M. and {Bock}, J. and {Bridge}, C. and {Budynkiewicz}, J. and {Burgarella}, D. and {Chapin}, E. and {Chapman}, S.~C. and {Clements}, D.~L. and {Conley}, A. and {Conselice}, C.~J. and {Cooray}, A. and {Farrah}, D. and {Hatziminaoglou}, E. and {Ivison}, R.~J. and {le Floc'h}, E. and {Lutz}, D. and {Magdis}, G. and {Magnelli}, B. and {Oliver}, S.~J. and {Page}, M.~J. and {Pozzi}, F. and {Rigopoulou}, D. and {Riguccini}, L. and {Roseboom}, I.~G. and {Sanders}, D.~B. and {Scott}, Douglas and {Seymour}, N. and {Valtchanov}, I. and {Vieira}, J.~D. and {Viero}, M. and {Wardlow}, J.},
        title = "{A Redshift Survey of Herschel Far-infrared Selected Starbursts and Implications for Obscured Star Formation}",
      journal = {\apj},
         year = 2012,
        month = dec,
       volume = {761},
       number = {2},
          eid = {140},
        pages = {140},
          doi = {10.1088/0004-637X/761/2/140},
archivePrefix = {arXiv},
       eprint = {1210.4928},
 primaryClass = {astro-ph.CO},
       adsurl = {https://ui.adsabs.harvard.edu/abs/2012ApJ...761..140C}
}

@ARTICLE{meixner2006,
       author = {{Meixner}, Margaret and {Gordon}, Karl D. and {Indebetouw}, Remy and {Hora}, Joseph L. and {Whitney}, Barbara and {Blum}, Robert and {Reach}, William and {Bernard}, Jean-Philippe and {Meade}, Marilyn and {Babler}, Brian and {Engelbracht}, Charles W. and {For}, Bi-Qing and {Misselt}, Karl and {Vijh}, Uma and {Leitherer}, Claus and {Cohen}, Martin and {Churchwell}, Ed B. and {Boulanger}, Francois and {Frogel}, Jay A. and {Fukui}, Yasuo and {Gallagher}, Jay and {Gorjian}, Varoujan and {Harris}, Jason and {Kelly}, Douglas and {Kawamura}, Akiko and {Kim}, SoYoung and {Latter}, William B. and {Madden}, Suzanne and {Markwick-Kemper}, Ciska and {Mizuno}, Akira and {Mizuno}, Norikazu and {Mould}, Jeremy and {Nota}, Antonella and {Oey}, M.~S. and {Olsen}, Knut and {Onishi}, Toshikazu and {Paladini}, Roberta and {Panagia}, Nino and {Perez-Gonzalez}, Pablo and {Shibai}, Hiroshi and {Sato}, Shuji and {Smith}, Linda and {Staveley-Smith}, Lister and {Tielens}, A.~G.~G.~M. and {Ueta}, Toshiya and {van Dyk}, Schuyler and {Volk}, Kevin and {Werner}, Michael and {Zaritsky}, Dennis},
        title = "{Spitzer Survey of the Large Magellanic Cloud: Surveying the Agents of a Galaxy's Evolution (SAGE). I. Overview and Initial Results}",
      journal = {\aj},
         year = 2006,
        month = dec,
       volume = {132},
       number = {6},
        pages = {2268-2288},
          doi = {10.1086/508185},
archivePrefix = {arXiv},
       eprint = {astro-ph/0606356},
 primaryClass = {astro-ph},
       adsurl = {https://ui.adsabs.harvard.edu/abs/2006AJ....132.2268M}
}

@ARTICLE{skowron2021,
       author = {{Skowron}, D.~M. and {Skowron}, J. and {Udalski}, A. and {Szyma{\'n}ski}, M.~K. and {Soszy{\'n}ski}, I. and {Wyrzykowski}, {\L}. and {Ulaczyk}, K. and {Poleski}, R. and {Koz{\l}owski}, S. and {Pietrukowicz}, P. and {Mr{\'o}z}, P. and {Rybicki}, K. and {Iwanek}, P. and {Wrona}, M. and {Gromadzki}, M.},
        title = "{OGLE-ing the Magellanic System: Optical Reddening Maps of the Large and Small Magellanic Clouds from Red Clump Stars}",
      journal = {\apjs},
         year = 2021,
        month = feb,
       volume = {252},
       number = {2},
          eid = {23},
        pages = {23},
          doi = {10.3847/1538-4365/abcb81},
archivePrefix = {arXiv},
       eprint = {2006.02448},
 primaryClass = {astro-ph.SR},
       adsurl = {https://ui.adsabs.harvard.edu/abs/2021ApJS..252...23S}
}

@ARTICLE{Padoan2001,
       author = {{Padoan}, Paolo and {Kim}, Sungeun and {Goodman}, Alyssa and {Staveley-Smith}, Lister},
        title = "{A New Method to Measure and Map the Gas Scale Height of Disk Galaxies}",
      journal = {\apjl},
         year = 2001,
        month = jul,
       volume = {555},
       number = {1},
        pages = {L33-L36},
          doi = {10.1086/321735},
archivePrefix = {arXiv},
       eprint = {astro-ph/0103251},
 primaryClass = {astro-ph},
       adsurl = {https://ui.adsabs.harvard.edu/abs/2001ApJ...555L..33P}
}

@INPROCEEDINGS{Subramaniam2010,
       author = {{Subramaniam}, Annapurni and {Subramanian}, Smitha},
        title = "{Internal and foreground reddening maps of the Magellanic Clouds}",
    booktitle = {Astronomical Society of India Conference Series},
         year = 2010,
       series = {Astronomical Society of India Conference Series},
       volume = {1},
        month = jan,
        pages = {107-110},
       adsurl = {https://ui.adsabs.harvard.edu/abs/2010ASInC...1..107S}
}

@ARTICLE{cressie1992,
       author = {{Cressie}, Noel},
        title = "{Statistics for Spatial Data}",
      journal = {Terra Nova},
         year = 1992,
        month = sep,
       volume = {4},
       number = {5},
        pages = {613-617},
          doi = {10.1111/j.1365-3121.1992.tb00605.x},
       adsurl = {https://ui.adsabs.harvard.edu/abs/1992TeNov...4..613C}
}

@ARTICLE{matheron1963,
       author = {{Matheron}, Georges},
        title = "{Principles of geostatistics}",
      journal = {Economic Geology},
         year = 1963,
        month = dec,
       volume = {58},
       number = {8},
        pages = {1246-1266},
          doi = {10.2113/gsecongeo.58.8.1246},
       adsurl = {https://ui.adsabs.harvard.edu/abs/1963EcGeo..58.1246M}
}

@ARTICLE{murray2019,
       author = {{Murray}, Claire E. and {Peek}, J.~E.~G. and {Di Teodoro}, Enrico M. and {McClure-Griffiths}, N.~M. and {Dickey}, John M. and {D{\'e}nes}, Helga},
        title = "{The 3D Kinematics of Gas in the Small Magellanic Cloud}",
      journal = {\apj},
         year = 2019,
        month = dec,
       volume = {887},
       number = {2},
          eid = {267},
        pages = {267},
          doi = {10.3847/1538-4357/ab510f},
archivePrefix = {arXiv},
       eprint = {1910.11283},
 primaryClass = {astro-ph.GA},
       adsurl = {https://ui.adsabs.harvard.edu/abs/2019ApJ...887..267M}
}

@ARTICLE{russell1992,
   author = {{Russell}, S.~C. and {Dopita}, M.~A.},
    title = "{Abundances of the heavy elements in the Magellanic Clouds. III - Interpretation of results}",
  journal = {\apj},
     year = 1992,
    month = jan,
   volume = 384,
    pages = {508-522},
      doi = {10.1086/170893},
   adsurl = {http://adsabs.harvard.edu/abs/1992ApJ...384..508R}
}

@ARTICLE{nidever2008,
   author = {{Nidever}, D.~L. and {Majewski}, S.~R. and {Butler Burton}, W.
	},
    title = "{The Origin of the Magellanic Stream and Its Leading Arm}",
  journal = {\apj},
     year = 2008,
    month = may,
   volume = 679,
      eid = {432-459},
    pages = {432-459},
      doi = {10.1086/587042},
   adsurl = {http://adsabs.harvard.edu/abs/2008ApJ...679..432N}
}

@ARTICLE{lindberg2025,
       author = {{Lindberg}, Christina W. and {Murray}, Claire E. and {Yanchulova Merica-Jones}, Petia and {Bot}, Caroline and {Burhenne}, Clare and {Choi}, Yumi and {Clark}, Christopher J.~R. and {Cohen}, Roger E. and {Gilbert}, Karoline M. and {Goldman}, Steven R. and {Gordon}, Karl D. and {Hirschauer}, Alec S. and {McQuinn}, Kristen B.~W. and {Roman-Duval}, Julia C. and {Sandstrom}, Karin M. and {Tarantino}, Elizabeth and {Williams}, Benjamin F.},
        title = "{Scylla. IV. Intrinsic Stellar Properties and Line-of-sight Dust Extinction Measurements toward 1.5 Million Stars in the SMC and LMC}",
      journal = {\apj},
         year = 2025,
        month = mar,
       volume = {982},
       number = {1},
          eid = {33},
        pages = {33},
          doi = {10.3847/1538-4357/adb4e8},
archivePrefix = {arXiv},
       eprint = {2410.19910},
 primaryClass = {astro-ph.GA},
       adsurl = {https://ui.adsabs.harvard.edu/abs/2025ApJ...982...33L}
}

@article{dalcanton2012,
	Adsurl = {http://adsabs.harvard.edu/abs/2012ApJS..200...18D},
	Archiveprefix = {arXiv},
	Author = {{Dalcanton}, J.~J. and {Williams}, B.~F. and {Lang}, D. and {Lauer}, T.~R. and
	 {Kalirai}, J.~S. and {Seth}, A.~C. and {Dolphin}, A. and {Rosenfield}, P. and {Weisz}, D.~R. and
	 {Bell}, E.~F. and {Bianchi}, L.~C. and {Boyer}, M.~L. and {Caldwell}, N. and {Dong}, H. and
	 {Dorman}, C.~E. and {Gilbert}, K.~M. and {Girardi}, L. and {Gogarten}, S.~M. and {Gordon}, K.~D. and
	 {Guhathakurta}, P. and {Hodge}, P.~W. and {Holtzman}, J.~A. and {Johnson}, L.~C. and
	 {Larsen}, S.~S. and {Lewis}, A. and {Melbourne}, J.~L. and {Olsen}, K.~A.~G. and {Rix}, H.-W. and
	 {Rosema}, K. and {Saha}, A. and {Sarajedini}, A. and {Skillman}, E.~D. and {Stanek}, K.~Z.},
	Doi = {10.1088/0067-0049/200/2/18},
	Eid = {18},
	Eprint = {1204.0010},
	Journal = {\apjs},
	Month = jun,
	Pages = {18},
	Primaryclass = {astro-ph.CO},
	Title = {{The Panchromatic Hubble Andromeda Treasury}},
	Volume = 200,
	Year = 2012}

@article{williams2014,
	Author = {{Williams}, B.~F. and {Lang}, D. and {Dalcanton}, J.~J. and {Dolphin}, A.~E. and
	{Weisz}, D.~R. and {Bell}, E.~F. and {Bianchi}, L. and {Byler}, N. and {Gilbert}, K.~M. and
	{Girardi}, L. and {Gordon}, K. and {Gregersen}, D. and {Johnson}, L.~C. and {Kalirai}, J. and
	{Lauer}, T.~R. and {Monachesi}, A. and {Rosenfield}, P. and {Seth}, A. and {Skillman}, E.},
	Journal = {\apjs},
	Month = nov,
	Pages = {9},
	Title = {{The Panchromatic Hubble Andromeda Treasury. X. Ultraviolet to Infrared Photometry of 117 Million Equidistant Stars}},
	Volume = 215,
	Year = 2014}

@ARTICLE{dolphin2002,
   author = {{Dolphin}, A.~E.},
    title = "{Numerical methods of star formation history measurement and applications to seven dwarf spheroidals}",
  journal = {\mnras},
   eprint = {astro-ph/0112331},
     year = 2002,
    month = may,
   volume = 332,
    pages = {91-108},
      doi = {10.1046/j.1365-8711.2002.05271.x},
   adsurl = {http://adsabs.harvard.edu/abs/2002MNRAS.332...91D}
}

@article{draine2003,
	Adsurl = {http://adsabs.harvard.edu/abs/2003ARA%26A..41..241D},
	Author = {{Draine}, B.~T.},
	Doi = {10.1146/annurev.astro.41.011802.094840},
	Eprint = {astro-ph/0304489},
	Journal = {\araa},
	Pages = {241-289},
	Title = {{Interstellar Dust Grains}},
	Volume = 41,
	Year = 2003}

@ARTICLE{leike2020,
       author = {{Leike}, R.~H. and {Glatzle}, M. and {En{\ss}lin}, T.~A.},
        title = "{Resolving nearby dust clouds}",
      journal = {\aap},
         year = 2020,
        month = jul,
       volume = {639},
          eid = {A138},
        pages = {A138},
          doi = {10.1051/0004-6361/202038169},
archivePrefix = {arXiv},
       eprint = {2004.06732},
 primaryClass = {astro-ph.GA},
       adsurl = {https://ui.adsabs.harvard.edu/abs/2020A&A...639A.138L}
}

@ARTICLE{miller2022,
       author = {{Miller}, Andrew C. and {Anderson}, Lauren and {Leistedt}, Boris and {Cunningham}, John P. and {Hogg}, David W. and {Blei}, David M.},
        title = "{Mapping Interstellar Dust with Gaussian Processes}",
      journal = {arXiv e-prints},
         year = 2022,
        month = feb,
          eid = {arXiv:2202.06797},
        pages = {arXiv:2202.06797},
          doi = {10.48550/arXiv.2202.06797},
archivePrefix = {arXiv},
       eprint = {2202.06797},
 primaryClass = {astro-ph.GA},
       adsurl = {https://ui.adsabs.harvard.edu/abs/2022arXiv220206797M}
}

@ARTICLE{soding2025,
       author = {{S{\"o}ding}, Laurin and {Edenhofer}, Gordian and {En{\ss}lin}, Torsten A. and {Frank}, Philipp and {Kissmann}, Ralf and {Phan}, Vo Hong Minh and {Ram{\'\i}rez}, Andr{\'e}s and {Zandinejad}, Hanieh and {Mertsch}, Philipp},
        title = "{Spatially coherent 3D distributions of HI and CO in the Milky Way}",
      journal = {\aap},
         year = 2025,
        month = jan,
       volume = {693},
          eid = {A139},
        pages = {A139},
          doi = {10.1051/0004-6361/202451361},
archivePrefix = {arXiv},
       eprint = {2407.02859},
 primaryClass = {astro-ph.GA},
       adsurl = {https://ui.adsabs.harvard.edu/abs/2025A&A...693A.139S}
}

@ARTICLE{thavisha2023,
       author = {{Dharmawardena}, T.~E. and {Bailer-Jones}, C.~A.~L. and {Fouesneau}, M. and {Foreman-Mackey}, D. and {Coronica}, P. and {Colnaghi}, T. and {M{\"u}ller}, T. and {Henshaw}, J.},
        title = "{The three-dimensional structure of galactic molecular cloud complexes out to 2.5 kpc}",
      journal = {\mnras},
         year = 2023,
        month = feb,
       volume = {519},
       number = {1},
        pages = {228-247},
          doi = {10.1093/mnras/stac2790},
archivePrefix = {arXiv},
       eprint = {2210.03615},
 primaryClass = {astro-ph.GA},
       adsurl = {https://ui.adsabs.harvard.edu/abs/2023MNRAS.519..228D}
}

@ARTICLE{Rezaei2018,
       author = {{Rezaei Kh.}, Sara and {Bailer-Jones}, Coryn A.~L. and {Hogg}, David W. and {Schultheis}, Mathias},
        title = "{Detection of the Milky Way spiral arms in dust from 3D mapping}",
      journal = {\aap},
         year = 2018,
        month = oct,
       volume = {618},
          eid = {A168},
        pages = {A168},
          doi = {10.1051/0004-6361/201833284},
archivePrefix = {arXiv},
       eprint = {1808.00015},
 primaryClass = {astro-ph.GA},
       adsurl = {https://ui.adsabs.harvard.edu/abs/2018A&A...618A.168R}
}

@article{choi_mapping_2020,
    title = {Mapping the {Escape} {Fraction} of {Ionizing} {Photons} {Using} {Resolved} {Stars}: {A} {Much} {Higher} {Escape} {Fraction} for {NGC} 4214},
    volume = {902},
    doi = {10.3847/1538-4357/abb467},
    number = {1},
    journal = {The Astrophysical Journal},
    author = {Choi, Yumi and Dalcanton, Julianne J and Williams, Benjamin F and Skillman, Evan D and Fouesneau, Morgan and Gordon, Karl D and Sandstrom, Karin M and Weisz, Daniel R and Gilbert, Karoline M},
    year = {2020},
    pages = {54},
}

@article{zucker_three-dimensional_2021,
    title = {On the {Three}-{Dimensional} {Structure} of {Local} {Molecular} {Clouds}},
    volume = {919},
    issn = {0004-637X, 1538-4357},
    url = {http://arxiv.org/abs/2109.09765},
    doi = {10.3847/1538-4357/ac1f96},
    language = {en},
    number = {1},
    urldate = {2023-04-14},
    journal = {The Astrophysical Journal},
    author = {Zucker, Catherine and Goodman, Alyssa and Alves, João and Bialy, Shmuel and Koch, Eric W. and Speagle, Joshua S. and Foley, Michael M. and Finkbeiner, Douglas and Leike, Reimar and Enßlin, Torsten and Peek, Joshua E. G. and Edenhofer, Gordian},
    month = sep,
    year = {2021},
    note = {arXiv:2109.09765 [astro-ph]},
    pages = {35},
}

@ARTICLE {xsede2014,
author = {J. Towns and T. Cockerill and M. Dahan and I. Foster and K. Gaither and A. Grimshaw and V. Hazlewood and S. Lathrop and D. Lifka and G. D. Peterson and R. Roskies and J. Scott and N. Wilkins-Diehr},
journal = {Computing in Science \&amp; Engineering},
title = {XSEDE: Accelerating Scientific Discovery},
year = {2014},
volume = {16},
number = {05},
issn = {1558-366X},
pages = {62-74},
doi = {10.1109/MCSE.2014.80},
publisher = {IEEE Computer Society},
address = {Los Alamitos, CA, USA},
month = {sep}
}

@ARTICLE{Clark2023,
       author = {{Clark}, Christopher J.~R. and {Roman-Duval}, Julia C. and {Gordon}, Karl D. and {Bot}, Caroline and {Smith}, Matthew W.~L. and {Hagen}, Lea M.~Z.},
        title = "{The Quest for the Missing Dust. II. Two Orders of Magnitude of Evolution in the Dust-to-gas Ratio Resolved within Local Group Galaxies}",
      journal = {\apj},
         year = 2023,
        month = mar,
       volume = {946},
       number = {1},
          eid = {42},
        pages = {42},
          doi = {10.3847/1538-4357/acbb66},
archivePrefix = {arXiv},
       eprint = {2302.07378},
 primaryClass = {astro-ph.GA},
       adsurl = {https://ui.adsabs.harvard.edu/abs/2023ApJ...946...42C}
}

@ARTICLE{choi2022,
       author = {{Choi}, Yumi and {Olsen}, Knut A.~G. and {Besla}, Gurtina and {van der Marel}, Roeland P. and {Zivick}, Paul and {Kallivayalil}, Nitya and {Nidever}, David L.},
        title = "{The Recent LMC-SMC Collision: Timing and Impact Parameter Constraints from Comparison of Gaia LMC Disk Kinematics and N-body Simulations}",
      journal = {\apj},
         year = 2022,
        month = mar,
       volume = {927},
       number = {2},
          eid = {153},
        pages = {153},
          doi = {10.3847/1538-4357/ac4e90},
archivePrefix = {arXiv},
       eprint = {2201.04648},
 primaryClass = {astro-ph.GA},
       adsurl = {https://ui.adsabs.harvard.edu/abs/2022ApJ...927..153C}
}

@ARTICLE{Cullinane2022,
       author = {{Cullinane}, L.~R. and {Mackey}, A.~D. and {Da Costa}, G.~S. and {Erkal}, D. and {Koposov}, S.~E. and {Belokurov}, V.},
        title = "{The Magellanic Edges Survey - II. Formation of the LMC's northern arm}",
      journal = {\mnras},
         year = 2022,
        month = feb,
       volume = {510},
       number = {1},
        pages = {445-468},
          doi = {10.1093/mnras/stab3350},
archivePrefix = {arXiv},
       eprint = {2106.03274},
 primaryClass = {astro-ph.GA},
       adsurl = {https://ui.adsabs.harvard.edu/abs/2022MNRAS.510..445C}
}

@ARTICLE{lindberg2024,
       author = {{Lindberg}, Christina Willecke and {Murray}, Claire E. and {Dalcanton}, Julianne J. and {Peek}, J.~E.~G. and {Gordon}, Karl D.},
        title = "{Dust around Massive Stars Is Agnostic to Galactic Environment: New Insights from PHAT/BEAST}",
      journal = {\apj},
         year = 2024,
        month = mar,
       volume = {963},
       number = {1},
          eid = {58},
        pages = {58},
          doi = {10.3847/1538-4357/ad18cc},
archivePrefix = {arXiv},
       eprint = {2401.10991},
 primaryClass = {astro-ph.GA},
       adsurl = {https://ui.adsabs.harvard.edu/abs/2024ApJ...963...58L}
}

@ARTICLE{dalcanton2015,
       author = {{Dalcanton}, Julianne J. and {Fouesneau}, Morgan and {Hogg}, David W. and {Lang}, Dustin and {Leroy}, Adam K. and {Gordon}, Karl D. and {Sandstrom}, Karin and {Weisz}, Daniel R. and {Williams}, Benjamin F. and {Bell}, Eric F. and {Dong}, Hui and {Gilbert}, Karoline M. and {Gouliermis}, Dimitrios A. and {Guhathakurta}, Puragra and {Lauer}, Tod R. and {Schruba}, Andreas and {Seth}, Anil C. and {Skillman}, Evan D.},
        title = "{The Panchromatic Hubble Andromeda Treasury. VIII. A Wide-area, High-resolution Map of Dust Extinction in M31}",
      journal = {\apj},
         year = 2015,
        month = nov,
       volume = {814},
       number = {1},
          eid = {3},
        pages = {3},
          doi = {10.1088/0004-637X/814/1/3},
archivePrefix = {arXiv},
       eprint = {1509.06988},
 primaryClass = {astro-ph.GA},
       adsurl = {https://ui.adsabs.harvard.edu/abs/2015ApJ...814....3D}
}

@ARTICLE{gordon2023,
       author = {{Gordon}, Karl D. and {Clayton}, Geoffrey C. and {Decleir}, Marjorie and {Fitzpatrick}, E.~L. and {Massa}, Derck and {Misselt}, Karl A. and {Tollerud}, Erik J.},
        title = "{One Relation for All Wavelengths: The Far-ultraviolet to Mid-infrared Milky Way Spectroscopic R(V)-dependent Dust Extinction Relationship}",
      journal = {\apj},
         year = 2023,
        month = jun,
       volume = {950},
       number = {2},
          eid = {86},
        pages = {86},
          doi = {10.3847/1538-4357/accb59},
archivePrefix = {arXiv},
       eprint = {2304.01991},
 primaryClass = {astro-ph.GA},
       adsurl = {https://ui.adsabs.harvard.edu/abs/2023ApJ...950...86G}
}

@ARTICLE{wolfire2010,
       author = {{Wolfire}, Mark G. and {Hollenbach}, David and {McKee}, Christopher F.},
        title = "{The Dark Molecular Gas}",
      journal = {\apj},
         year = 2010,
        month = jun,
       volume = {716},
       number = {2},
        pages = {1191-1207},
          doi = {10.1088/0004-637X/716/2/1191},
archivePrefix = {arXiv},
       eprint = {1004.5401},
 primaryClass = {astro-ph.GA},
       adsurl = {https://ui.adsabs.harvard.edu/abs/2010ApJ...716.1191W}
}

@ARTICLE{liszt2014,
       author = {{Liszt}, Harvey},
        title = "{E(B - V), N(H I), and N(H$_{2}$)}",
      journal = {\apj},
         year = 2014,
        month = mar,
       volume = {783},
       number = {1},
          eid = {17},
        pages = {17},
          doi = {10.1088/0004-637X/783/1/17},
archivePrefix = {arXiv},
       eprint = {1401.1837},
 primaryClass = {astro-ph.GA},
       adsurl = {https://ui.adsabs.harvard.edu/abs/2014ApJ...783...17L}
}

@ARTICLE{Subramanian2017,
       author = {{Subramanian}, Smitha and {Rubele}, Stefano and {Sun}, Ning-Chen and {Girardi}, L{\'e}o and {de Grijs}, Richard and {van Loon}, Jacco Th. and {Cioni}, Maria-Rosa L. and {Piatti}, Andr{\'e}s E. and {Bekki}, Kenji and {Emerson}, Jim and {Ivanov}, Valentin D. and {Kerber}, Leandro and {Marconi}, Marcella and {Ripepi}, Vincenzo and {Tatton}, Benjamin L.},
        title = "{The VMC Survey - XXIV. Signatures of tidally stripped stellar populations from the inner Small Magellanic Cloud}",
      journal = {\mnras},
         year = 2017,
        month = may,
       volume = {467},
       number = {3},
        pages = {2980-2995},
          doi = {10.1093/mnras/stx205},
archivePrefix = {arXiv},
       eprint = {1701.05722},
 primaryClass = {astro-ph.GA},
       adsurl = {https://ui.adsabs.harvard.edu/abs/2017MNRAS.467.2980S}
}

@ARTICLE{choi2018,
       author = {{Choi}, Yumi and {Nidever}, David L. and {Olsen}, Knut and {Blum}, Robert D. and {Besla}, Gurtina and {Zaritsky}, Dennis and {van der Marel}, Roeland P. and {Bell}, Eric F. and {Gallart}, Carme and {Cioni}, Maria-Rosa L. and {Johnson}, L. Clifton and {Vivas}, A. Katherina and {Saha}, Abhijit and {de Boer}, Thomas J.~L. and {No{\"e}l}, Noelia E.~D. and {Monachesi}, Antonela and {Massana}, Pol and {Conn}, Blair C. and {Martinez-Delgado}, David and {Mu{\~n}oz}, Ricardo R. and {Stringfellow}, Guy S.},
        title = "{SMASHing the LMC: A Tidally Induced Warp in the Outer LMC and a Large-scale Reddening Map}",
      journal = {\apj},
         year = 2018,
        month = oct,
       volume = {866},
       number = {2},
          eid = {90},
        pages = {90},
          doi = {10.3847/1538-4357/aae083},
archivePrefix = {arXiv},
       eprint = {1804.07765},
 primaryClass = {astro-ph.GA},
       adsurl = {https://ui.adsabs.harvard.edu/abs/2018ApJ...866...90C}
}

@ARTICLE{zaritsky2002,
       author = {{Zaritsky}, Dennis and {Harris}, Jason and {Thompson}, Ian B. and {Grebel}, Eva K. and {Massey}, Philip},
        title = "{The Magellanic Clouds Photometric Survey: The Small Magellanic Cloud Stellar Catalog and Extinction Map}",
      journal = {\aj},
         year = 2002,
        month = feb,
       volume = {123},
       number = {2},
        pages = {855-872},
          doi = {10.1086/338437},
archivePrefix = {arXiv},
       eprint = {astro-ph/0110665},
 primaryClass = {astro-ph},
       adsurl = {https://ui.adsabs.harvard.edu/abs/2002AJ....123..855Z}
}

@ARTICLE{zaritsky2004,
       author = {{Zaritsky}, Dennis and {Harris}, Jason and {Thompson}, Ian B. and {Grebel}, Eva K.},
        title = "{The Magellanic Clouds Photometric Survey: The Large Magellanic Cloud Stellar Catalog and Extinction Map}",
      journal = {\aj},
         year = 2004,
        month = oct,
       volume = {128},
       number = {4},
        pages = {1606-1614},
          doi = {10.1086/423910},
archivePrefix = {arXiv},
       eprint = {astro-ph/0407006},
 primaryClass = {astro-ph},
       adsurl = {https://ui.adsabs.harvard.edu/abs/2004AJ....128.1606Z}
}

@ARTICLE{planck2016,
       author = {{Planck Collaboration} and {Ade}, P.~A.~R. and {Aghanim}, N. and {Alves}, M.~I.~R. and {Aniano}, G. and {Arnaud}, M. and {Ashdown}, M. and {Aumont}, J. and {Baccigalupi}, C. and {Banday}, A.~J. and {Barreiro}, R.~B. and {Bartolo}, N. and {Battaner}, E. and {Benabed}, K. and {Benoit-L{\'e}vy}, A. and {Bernard}, J. -P. and {Bersanelli}, M. and {Bielewicz}, P. and {Bonaldi}, A. and {Bonavera}, L. and {Bond}, J.~R. and {Borrill}, J. and {Bouchet}, F.~R. and {Boulanger}, F. and {Burigana}, C. and {Butler}, R.~C. and {Calabrese}, E. and {Cardoso}, J. -F. and {Catalano}, A. and {Chamballu}, A. and {Chiang}, H.~C. and {Christensen}, P.~R. and {Clements}, D.~L. and {Colombi}, S. and {Colombo}, L.~P.~L. and {Couchot}, F. and {Crill}, B.~P. and {Curto}, A. and {Cuttaia}, F. and {Danese}, L. and {Davies}, R.~D. and {Davis}, R.~J. and {de Bernardis}, P. and {de Rosa}, A. and {de Zotti}, G. and {Delabrouille}, J. and {Dickinson}, C. and {Diego}, J.~M. and {Dole}, H. and {Donzelli}, S. and {Dor{\'e}}, O. and {Douspis}, M. and {Draine}, B.~T. and {Ducout}, A. and {Dupac}, X. and {Efstathiou}, G. and {Elsner}, F. and {En{\ss}lin}, T.~A. and {Eriksen}, H.~K. and {Falgarone}, E. and {Finelli}, F. and {Forni}, O. and {Frailis}, M. and {Fraisse}, A.~A. and {Franceschi}, E. and {Frejsel}, A. and {Galeotta}, S. and {Galli}, S. and {Ganga}, K. and {Ghosh}, T. and {Giard}, M. and {Gjerl{\o}w}, E. and {Gonz{\'a}lez-Nuevo}, J. and {G{\'o}rski}, K.~M. and {Gregorio}, A. and {Gruppuso}, A. and {Guillet}, V. and {Hansen}, F.~K. and {Hanson}, D. and {Harrison}, D.~L. and {Henrot-Versill{\'e}}, S. and {Hern{\'a}ndez-Monteagudo}, C. and {Herranz}, D. and {Hildebrandt}, S.~R. and {Hivon}, E. and {Holmes}, W.~A. and {Hovest}, W. and {Huffenberger}, K.~M. and {Hurier}, G. and {Jaffe}, A.~H. and {Jaffe}, T.~R. and {Jones}, W.~C. and {Keih{\"a}nen}, E. and {Keskitalo}, R. and {Kisner}, T.~S. and {Kneissl}, R. and {Knoche}, J. and {Kunz}, M. and {Kurki-Suonio}, H. and {Lagache}, G. and {Lamarre}, J. -M. and {Lasenby}, A. and {Lattanzi}, M. and {Lawrence}, C.~R. and {Leonardi}, R. and {Levrier}, F. and {Liguori}, M. and {Lilje}, P.~B. and {Linden-V{\o}rnle}, M. and {L{\'o}pez-Caniego}, M. and {Lubin}, P.~M. and {Mac{\'\i}as-P{\'e}rez}, J.~F. and {Maffei}, B. and {Maino}, D. and {Mandolesi}, N. and {Maris}, M. and {Marshall}, D.~J. and {Martin}, P.~G. and {Mart{\'\i}nez-Gonz{\'a}lez}, E. and {Masi}, S. and {Matarrese}, S. and {Mazzotta}, P. and {Melchiorri}, A. and {Mendes}, L. and {Mennella}, A. and {Migliaccio}, M. and {Miville-Desch{\^e}nes}, M. -A. and {Moneti}, A. and {Montier}, L. and {Morgante}, G. and {Mortlock}, D. and {Munshi}, D. and {Murphy}, J.~A. and {Naselsky}, P. and {Natoli}, P. and {N{\o}rgaard-Nielsen}, H.~U. and {Novikov}, D. and {Novikov}, I. and {Oxborrow}, C.~A. and {Pagano}, L. and {Pajot}, F. and {Paladini}, R. and {Paoletti}, D. and {Pasian}, F. and {Perdereau}, O. and {Perotto}, L. and {Perrotta}, F. and {Pettorino}, V. and {Piacentini}, F. and {Piat}, M. and {Plaszczynski}, S. and {Pointecouteau}, E. and {Polenta}, G. and {Ponthieu}, N. and {Popa}, L. and {Pratt}, G.~W. and {Prunet}, S. and {Puget}, J. -L. and {Rachen}, J.~P. and {Reach}, W.~T. and {Rebolo}, R. and {Reinecke}, M. and {Remazeilles}, M. and {Renault}, C. and {Ristorcelli}, I. and {Rocha}, G. and {Roudier}, G. and {Rubi{\~n}o-Mart{\'\i}n}, J.~A. and {Rusholme}, B. and {Sandri}, M. and {Santos}, D. and {Scott}, D. and {Spencer}, L.~D. and {Stolyarov}, V. and {Sudiwala}, R. and {Sunyaev}, R. and {Sutton}, D. and {Suur-Uski}, A. -S. and {Sygnet}, J. -F. and {Tauber}, J.~A. and {Terenzi}, L. and {Toffolatti}, L. and {Tomasi}, M. and {Tristram}, M. and {Tucci}, M. and {Umana}, G. and {Valenziano}, L. and {Valiviita}, J. and {Van Tent}, B. and {Vielva}, P. and {Villa}, F. and {Wade}, L.~A. and {Wandelt}, B.~D. and {Wehus}, I.~K. and {Ysard}, N. and {Yvon}, D. and {Zacchei}, A. and {Zonca}, A.},
        title = "{Planck intermediate results. XXIX. All-sky dust modelling with Planck, IRAS, and WISE observations}",
      journal = {\aap},
         year = 2016,
        month = feb,
       volume = {586},
          eid = {A132},
        pages = {A132},
          doi = {10.1051/0004-6361/201424945},
archivePrefix = {arXiv},
       eprint = {1409.2495},
 primaryClass = {astro-ph.GA},
       adsurl = {https://ui.adsabs.harvard.edu/abs/2016A&A...586A.132P}
}

@ARTICLE{vandeputte2020,
       author = {{Van De Putte}, Dries and {Gordon}, Karl D. and {Roman-Duval}, Julia and {Williams}, Benjamin F. and {Baes}, Maarten and {Tchernyshyov}, Kirill and {Lawton}, Brandon L. and {Arab}, Heddy},
        title = "{Evidence of Dust Grain Evolution from Extinction Mapping in the IC 63 Photodissociation Region}",
      journal = {\apj},
         year = 2020,
        month = jan,
       volume = {888},
       number = {1},
          eid = {22},
        pages = {22},
          doi = {10.3847/1538-4357/ab557f},
archivePrefix = {arXiv},
       eprint = {1911.03406},
 primaryClass = {astro-ph.GA},
       adsurl = {https://ui.adsabs.harvard.edu/abs/2020ApJ...888...22V}
}

@ARTICLE{Jacyszyn2016,
       author = {{Jacyszyn-Dobrzeniecka}, A.~M. and {Skowron}, D.~M. and {Mr{\'o}z}, P. and {Skowron}, J. and {Soszy{\'n}ski}, I. and {Udalski}, A. and {Pietrukowicz}, P. and {Koz{\l}owski}, S. and {Wyrzykowski}, {\L}. and {Poleski}, R. and {Pawlak}, M. and {Szyma{\'n}ski}, M.~K. and {Ulaczyk}, K.},
        title = "{OGLE-ing the Magellanic System: Three-Dimensional Structure of the Clouds and the Bridge Using Classical Cepheids}",
      journal = {\actaa},
         year = 2016,
        month = jun,
       volume = {66},
       number = {2},
        pages = {149-196},
          doi = {10.48550/arXiv.1602.09141},
archivePrefix = {arXiv},
       eprint = {1602.09141},
 primaryClass = {astro-ph.GA},
       adsurl = {https://ui.adsabs.harvard.edu/abs/2016AcA....66..149J}
}

@ARTICLE{massana2022,
       author = {{Massana}, P. and {Ruiz-Lara}, T. and {No{\"e}l}, N.~E.~D. and {Gallart}, C. and {Nidever}, D.~L. and {Choi}, Y. and {Sakowska}, J.~D. and {Besla}, G. and {Olsen}, K.~A.~G. and {Monelli}, M. and {Dorta}, A. and {Stringfellow}, G.~S. and {Cassisi}, S. and {Bernard}, E.~J. and {Zaritsky}, D. and {Cioni}, M.-R.~L. and {Monachesi}, A. and {van der Marel}, R.~P. and {de Boer}, T.~J.~L. and {Walker}, A.~R.},
        title = "{The synchronized dance of the magellanic clouds' star formation history}",
      journal = {\mnras},
         year = 2022,
        month = jun,
       volume = {513},
       number = {1},
        pages = {L40-L45},
          doi = {10.1093/mnrasl/slac030},
archivePrefix = {arXiv},
       eprint = {2203.09523},
 primaryClass = {astro-ph.GA},
       adsurl = {https://ui.adsabs.harvard.edu/abs/2022MNRAS.513L..40M}
}

@ARTICLE{chen2022,
       author = {{Chen}, B. -Q. and {Guo}, H. -L. and {Gao}, J. and {Yang}, M. and {Liu}, Y. -L. and {Jiang}, B. -W.},
        title = "{Dust distributions in the magellanic clouds}",
      journal = {\mnras},
         year = 2022,
        month = mar,
       volume = {511},
       number = {1},
        pages = {1317-1329},
          doi = {10.1093/mnras/stac072},
archivePrefix = {arXiv},
       eprint = {2201.03152},
 primaryClass = {astro-ph.GA},
       adsurl = {https://ui.adsabs.harvard.edu/abs/2022MNRAS.511.1317C}
}

@ARTICLE{mccluregriffiths2009,
       author = {{McClure-Griffiths}, N.~M. and {Pisano}, D.~J. and {Calabretta}, M.~R. and {Ford}, H. Alyson and {Lockman}, Felix J. and {Staveley-Smith}, L. and {Kalberla}, P.~M.~W. and {Bailin}, J. and {Dedes}, L. and {Janowiecki}, S. and {Gibson}, B.~K. and {Murphy}, T. and {Nakanishi}, H. and {Newton-McGee}, K.},
        title = "{Gass: The Parkes Galactic All-Sky Survey. I. Survey Description, Goals, and Initial Data Release}",
      journal = {\apjs},
         year = 2009,
        month = apr,
       volume = {181},
       number = {2},
        pages = {398-412},
          doi = {10.1088/0067-0049/181/2/398},
archivePrefix = {arXiv},
       eprint = {0901.1159},
 primaryClass = {astro-ph.GA},
       adsurl = {https://ui.adsabs.harvard.edu/abs/2009ApJS..181..398M}
}

@ARTICLE{Mastropietro2009,
       author = {{Mastropietro}, Chiara and {Burkert}, Andreas and {Moore}, Ben},
        title = "{Effects of ram pressure on the gas distribution and star formation in the Large Magellanic Cloud}",
      journal = {\mnras},
         year = 2009,
        month = nov,
       volume = {399},
       number = {4},
        pages = {2004-2020},
          doi = {10.1111/j.1365-2966.2009.15406.x},
archivePrefix = {arXiv},
       eprint = {0905.1126},
 primaryClass = {astro-ph.CO},
       adsurl = {https://ui.adsabs.harvard.edu/abs/2009MNRAS.399.2004M}
}

@ARTICLE{tsuge2019,
       author = {{Tsuge}, Kisetsu and {Sano}, Hidetoshi and {Tachihara}, Kengo and {Yozin}, Cameron and {Bekki}, Kenji and {Inoue}, Tsuyoshi and {Mizuno}, Norikazu and {Kawamura}, Akiko and {Onishi}, Toshikazu and {Fukui}, Yasuo},
        title = "{Formation of the Active Star-forming Region LHA 120-N 44 Triggered by Tidally Driven Colliding H I Flows}",
      journal = {\apj},
         year = 2019,
        month = jan,
       volume = {871},
       number = {1},
          eid = {44},
        pages = {44},
          doi = {10.3847/1538-4357/aaf4fb},
archivePrefix = {arXiv},
       eprint = {1803.00713},
 primaryClass = {astro-ph.GA},
       adsurl = {https://ui.adsabs.harvard.edu/abs/2019ApJ...871...44T}
}

@ARTICLE{romanduval2021,
       author = {{Roman-Duval}, Julia and {Jenkins}, Edward B. and {Tchernyshyov}, Kirill and {Williams}, Benjamin and {Clark}, Christopher J.~R. and {Gordon}, Karl D. and {Meixner}, Margaret and {Hagen}, Lea and {Peek}, Joshua and {Sandstrom}, Karin and {Werk}, Jessica and {Yanchulova Merica-Jones}, Petia},
        title = "{METAL: The Metal Evolution, Transport, and Abundance in the Large Magellanic Cloud Hubble Program. II. Variations of Interstellar Depletions and Dust-to-gas Ratio within the LMC}",
      journal = {\apj},
         year = 2021,
        month = apr,
       volume = {910},
       number = {2},
          eid = {95},
        pages = {95},
          doi = {10.3847/1538-4357/abdeb6},
archivePrefix = {arXiv},
       eprint = {2101.09399},
 primaryClass = {astro-ph.GA},
       adsurl = {https://ui.adsabs.harvard.edu/abs/2021ApJ...910...95R}
}

@ARTICLE{fox2018,
       author = {{Fox}, Andrew J. and {Barger}, Kathleen A. and {Wakker}, Bart P. and {Richter}, Philipp and {Antwi-Danso}, Jacqueline and {Casetti-Dinescu}, Dana I. and {Howk}, J. Christopher and {Lehner}, Nicolas and {D'Onghia}, Elena and {Crowther}, Paul A. and {Lockman}, Felix J.},
        title = "{Chemical Abundances in the Leading Arm of the Magellanic Stream}",
      journal = {\apj},
         year = 2018,
        month = feb,
       volume = {854},
       number = {2},
          eid = {142},
        pages = {142},
          doi = {10.3847/1538-4357/aaa9bb},
archivePrefix = {arXiv},
       eprint = {1801.06446},
 primaryClass = {astro-ph.GA},
       adsurl = {https://ui.adsabs.harvard.edu/abs/2018ApJ...854..142F}
}

@ARTICLE{clark2021,
       author = {{Clark}, Christopher J.~R. and {Roman-Duval}, Julia C. and {Gordon}, Karl D. and {Bot}, Caroline and {Smith}, Matthew W.~L.},
        title = "{The Quest for the Missing Dust. I. Restoring Large-scale Emission in Herschel Maps of Local Group Galaxies}",
      journal = {\apj},
         year = 2021,
        month = nov,
       volume = {921},
       number = {1},
          eid = {35},
        pages = {35},
          doi = {10.3847/1538-4357/ac16d4},
archivePrefix = {arXiv},
       eprint = {2107.14302},
 primaryClass = {astro-ph.GA},
       adsurl = {https://ui.adsabs.harvard.edu/abs/2021ApJ...921...35C}
}

@ARTICLE{draine2014,
       author = {{Draine}, B.~T. and {Aniano}, G. and {Krause}, Oliver and {Groves}, Brent and {Sandstrom}, Karin and {Braun}, Robert and {Leroy}, Adam and {Klaas}, Ulrich and {Linz}, Hendrik and {Rix}, Hans-Walter and {Schinnerer}, Eva and {Schmiedeke}, Anika and {Walter}, Fabian},
        title = "{Andromeda's Dust}",
      journal = {\apj},
         year = 2014,
        month = jan,
       volume = {780},
       number = {2},
          eid = {172},
        pages = {172},
          doi = {10.1088/0004-637X/780/2/172},
archivePrefix = {arXiv},
       eprint = {1306.2304},
 primaryClass = {astro-ph.CO},
       adsurl = {https://ui.adsabs.harvard.edu/abs/2014ApJ...780..172D}
}

@ARTICLE{Rachford2009,
       author = {{Rachford}, Brian L. and {Snow}, Theodore P. and {Destree}, Joshua D. and {Ross}, Teresa L. and {Ferlet}, Roger and {Friedman}, Scott D. and {Gry}, Cecile and {Jenkins}, Edward B. and {Morton}, Donald C. and {Savage}, Blair D. and {Shull}, J. Michael and {Sonnentrucker}, Paule and {Tumlinson}, Jason and {Vidal-Madjar}, Alfred and {Welty}, Daniel E. and {York}, Donald G.},
        title = "{Molecular Hydrogen in the Far Ultraviolet Spectroscopic Explorer Translucent Lines of Sight: The Full Sample}",
      journal = {\apjs},
         year = 2009,
        month = jan,
       volume = {180},
       number = {1},
        pages = {125-137},
          doi = {10.1088/0067-0049/180/1/125},
archivePrefix = {arXiv},
       eprint = {0809.3831},
 primaryClass = {astro-ph},
       adsurl = {https://ui.adsabs.harvard.edu/abs/2009ApJS..180..125R}
}

@ARTICLE{Boulanger1996,
       author = {{Boulanger}, F. and {Abergel}, A. and {Bernard}, J. -P. and {Burton}, W.~B. and {Desert}, F. -X. and {Hartmann}, D. and {Lagache}, G. and {Puget}, J. -L.},
        title = "{The dust/gas correlation at high Galactic latitude.}",
      journal = {\aap},
         year = 1996,
        month = aug,
       volume = {312},
        pages = {256-262},
       adsurl = {https://ui.adsabs.harvard.edu/abs/1996A&A...312..256B}
}

@ARTICLE{akeson2019,
       author = {{Akeson}, Rachel and {Armus}, Lee and {Bachelet}, Etienne and {Bailey}, Vanessa and {Bartusek}, Lisa and {Bellini}, Andrea and {Benford}, Dominic and {Bennett}, David and {Bhattacharya}, Aparna and {Bohlin}, Ralph and {Boyer}, Martha and {Bozza}, Valerio and {Bryden}, Geoffrey and {Calchi Novati}, Sebastiano and {Carpenter}, Kenneth and {Casertano}, Stefano and {Choi}, Ami and {Content}, David and {Dayal}, Pratika and {Dressler}, Alan and {Dor{\'e}}, Olivier and {Fall}, S. Michael and {Fan}, Xiaohui and {Fang}, Xiao and {Filippenko}, Alexei and {Finkelstein}, Steven and {Foley}, Ryan and {Furlanetto}, Steven and {Kalirai}, Jason and {Gaudi}, B. Scott and {Gilbert}, Karoline and {Girard}, Julien and {Grady}, Kevin and {Greene}, Jenny and {Guhathakurta}, Puragra and {Heinrich}, Chen and {Hemmati}, Shoubaneh and {Hendel}, David and {Henderson}, Calen and {Henning}, Thomas and {Hirata}, Christopher and {Ho}, Shirley and {Huff}, Eric and {Hutter}, Anne and {Jansen}, Rolf and {Jha}, Saurabh and {Johnson}, Samson and {Jones}, David and {Kasdin}, Jeremy and {Kelly}, Patrick and {Kirshner}, Robert and {Koekemoer}, Anton and {Kruk}, Jeffrey and {Lewis}, Nikole and {Macintosh}, Bruce and {Madau}, Piero and {Malhotra}, Sangeeta and {Mandel}, Kaisey and {Massara}, Elena and {Masters}, Daniel and {McEnery}, Julie and {McQuinn}, Kristen and {Melchior}, Peter and {Melton}, Mark and {Mennesson}, Bertrand and {Peeples}, Molly and {Penny}, Matthew and {Perlmutter}, Saul and {Pisani}, Alice and {Plazas}, Andr{\'e}s and {Poleski}, Radek and {Postman}, Marc and {Ranc}, Cl{\'e}ment and {Rauscher}, Bernard and {Rest}, Armin and {Roberge}, Aki and {Robertson}, Brant and {Rodney}, Steven and {Rhoads}, James and {Rhodes}, Jason and {Ryan}, Jr., Russell and {Sahu}, Kailash and {Sand}, David and {Scolnic}, Dan and {Seth}, Anil and {Shvartzvald}, Yossi and {Siellez}, Karelle and {Smith}, Arfon and {Spergel}, David and {Stassun}, Keivan and {Street}, Rachel and {Strolger}, Louis-Gregory and {Szalay}, Alexander and {Trauger}, John and {Troxel}, M.~A. and {Turnbull}, Margaret and {van der Marel}, Roeland and {von der Linden}, Anja and {Wang}, Yun and {Weinberg}, David and {Williams}, Benjamin and {Windhorst}, Rogier and {Wollack}, Edward and {Wu}, Hao-Yi and {Yee}, Jennifer and {Zimmerman}, Neil},
        title = "{The Wide Field Infrared Survey Telescope: 100 Hubbles for the 2020s}",
      journal = {arXiv e-prints},
         year = 2019,
        month = feb,
          eid = {arXiv:1902.05569},
        pages = {arXiv:1902.05569},
          doi = {10.48550/arXiv.1902.05569},
archivePrefix = {arXiv},
       eprint = {1902.05569},
 primaryClass = {astro-ph.IM},
       adsurl = {https://ui.adsabs.harvard.edu/abs/2019arXiv190205569A}
}

@ARTICLE{ivezic2019,
       author = {{Ivezi{\'c}}, {\v{Z}}eljko and {Kahn}, Steven M. and {Tyson}, J. Anthony and {Abel}, Bob and {Acosta}, Emily and {Allsman}, Robyn and {Alonso}, David and {AlSayyad}, Yusra and {Anderson}, Scott F. and {Andrew}, John and {Angel}, James Roger P. and {Angeli}, George Z. and {Ansari}, Reza and {Antilogus}, Pierre and {Araujo}, Constanza and {Armstrong}, Robert and {Arndt}, Kirk T. and {Astier}, Pierre and {Aubourg}, {\'E}ric and {Auza}, Nicole and {Axelrod}, Tim S. and {Bard}, Deborah J. and {Barr}, Jeff D. and {Barrau}, Aurelian and {Bartlett}, James G. and {Bauer}, Amanda E. and {Bauman}, Brian J. and {Baumont}, Sylvain and {Bechtol}, Ellen and {Bechtol}, Keith and {Becker}, Andrew C. and {Becla}, Jacek and {Beldica}, Cristina and {Bellavia}, Steve and {Bianco}, Federica B. and {Biswas}, Rahul and {Blanc}, Guillaume and {Blazek}, Jonathan and {Blandford}, Roger D. and {Bloom}, Josh S. and {Bogart}, Joanne and {Bond}, Tim W. and {Booth}, Michael T. and {Borgland}, Anders W. and {Borne}, Kirk and {Bosch}, James F. and {Boutigny}, Dominique and {Brackett}, Craig A. and {Bradshaw}, Andrew and {Brandt}, William Nielsen and {Brown}, Michael E. and {Bullock}, James S. and {Burchat}, Patricia and {Burke}, David L. and {Cagnoli}, Gianpietro and {Calabrese}, Daniel and {Callahan}, Shawn and {Callen}, Alice L. and {Carlin}, Jeffrey L. and {Carlson}, Erin L. and {Chandrasekharan}, Srinivasan and {Charles-Emerson}, Glenaver and {Chesley}, Steve and {Cheu}, Elliott C. and {Chiang}, Hsin-Fang and {Chiang}, James and {Chirino}, Carol and {Chow}, Derek and {Ciardi}, David R. and {Claver}, Charles F. and {Cohen-Tanugi}, Johann and {Cockrum}, Joseph J. and {Coles}, Rebecca and {Connolly}, Andrew J. and {Cook}, Kem H. and {Cooray}, Asantha and {Covey}, Kevin R. and {Cribbs}, Chris and {Cui}, Wei and {Cutri}, Roc and {Daly}, Philip N. and {Daniel}, Scott F. and {Daruich}, Felipe and {Daubard}, Guillaume and {Daues}, Greg and {Dawson}, William and {Delgado}, Francisco and {Dellapenna}, Alfred and {de Peyster}, Robert and {de Val-Borro}, Miguel and {Digel}, Seth W. and {Doherty}, Peter and {Dubois}, Richard and {Dubois-Felsmann}, Gregory P. and {Durech}, Josef and {Economou}, Frossie and {Eifler}, Tim and {Eracleous}, Michael and {Emmons}, Benjamin L. and {Fausti Neto}, Angelo and {Ferguson}, Henry and {Figueroa}, Enrique and {Fisher-Levine}, Merlin and {Focke}, Warren and {Foss}, Michael D. and {Frank}, James and {Freemon}, Michael D. and {Gangler}, Emmanuel and {Gawiser}, Eric and {Geary}, John C. and {Gee}, Perry and {Geha}, Marla and {Gessner}, Charles J.~B. and {Gibson}, Robert R. and {Gilmore}, D. Kirk and {Glanzman}, Thomas and {Glick}, William and {Goldina}, Tatiana and {Goldstein}, Daniel A. and {Goodenow}, Iain and {Graham}, Melissa L. and {Gressler}, William J. and {Gris}, Philippe and {Guy}, Leanne P. and {Guyonnet}, Augustin and {Haller}, Gunther and {Harris}, Ron and {Hascall}, Patrick A. and {Haupt}, Justine and {Hernandez}, Fabio and {Herrmann}, Sven and {Hileman}, Edward and {Hoblitt}, Joshua and {Hodgson}, John A. and {Hogan}, Craig and {Howard}, James D. and {Huang}, Dajun and {Huffer}, Michael E. and {Ingraham}, Patrick and {Innes}, Walter R. and {Jacoby}, Suzanne H. and {Jain}, Bhuvnesh and {Jammes}, Fabrice and {Jee}, M. James and {Jenness}, Tim and {Jernigan}, Garrett and {Jevremovi{\'c}}, Darko and {Johns}, Kenneth and {Johnson}, Anthony S. and {Johnson}, Margaret W.~G. and {Jones}, R. Lynne and {Juramy-Gilles}, Claire and {Juri{\'c}}, Mario and {Kalirai}, Jason S. and {Kallivayalil}, Nitya J. and {Kalmbach}, Bryce and {Kantor}, Jeffrey P. and {Karst}, Pierre and {Kasliwal}, Mansi M. and {Kelly}, Heather and {Kessler}, Richard and {Kinnison}, Veronica and {Kirkby}, David and {Knox}, Lloyd and {Kotov}, Ivan V. and {Krabbendam}, Victor L. and {Krughoff}, K. Simon and {Kub{\'a}nek}, Petr and {Kuczewski}, John and {Kulkarni}, Shri and {Ku}, John and {Kurita}, Nadine R. and {Lage}, Craig S. and {Lambert}, Ron and {Lange}, Travis and {Langton}, J. Brian and {Le Guillou}, Laurent and {Levine}, Deborah and {Liang}, Ming and {Lim}, Kian-Tat and {Lintott}, Chris J. and {Long}, Kevin E. and {Lopez}, Margaux and {Lotz}, Paul J. and {Lupton}, Robert H. and {Lust}, Nate B. and {MacArthur}, Lauren A. and {Mahabal}, Ashish and {Mandelbaum}, Rachel and {Markiewicz}, Thomas W. and {Marsh}, Darren S. and {Marshall}, Philip J. and {Marshall}, Stuart and {May}, Morgan and {McKercher}, Robert and {McQueen}, Michelle and {Meyers}, Joshua and {Migliore}, Myriam and {Miller}, Michelle and {Mills}, David J.},
        title = "{LSST: From Science Drivers to Reference Design and Anticipated Data Products}",
      journal = {\apj},
         year = 2019,
        month = mar,
       volume = {873},
       number = {2},
          eid = {111},
        pages = {111},
          doi = {10.3847/1538-4357/ab042c},
archivePrefix = {arXiv},
       eprint = {0805.2366},
 primaryClass = {astro-ph},
       adsurl = {https://ui.adsabs.harvard.edu/abs/2019ApJ...873..111I}
}

@ARTICLE{hawcroft2024,
       author = {{Hawcroft}, C. and {Sana}, H. and {Mahy}, L. and {Sundqvist}, J.~O. and {de Koter}, A. and {Crowther}, P.~A. and {Bestenlehner}, J.~M. and {Brands}, S.~A. and {David-Uraz}, A. and {Decin}, L. and {Erba}, C. and {Garcia}, M. and {Hamann}, W. -R. and {Herrero}, A. and {Ignace}, R. and {Kee}, N.~D. and {Kub{\'a}tov{\'a}}, B. and {Lefever}, R. and {Moffat}, A. and {Najarro}, F. and {Oskinova}, L. and {Pauli}, D. and {Prinja}, R. and {Puls}, J. and {Sander}, A.~A.~C. and {Shenar}, T. and {St-Louis}, N. and {ud-Doula}, A. and {Vink}, J.~S.},
        title = "{X-Shooting ULLYSES: Massive stars at low metallicity. III. Terminal wind speeds of ULLYSES massive stars}",
      journal = {\aap},
         year = 2024,
        month = aug,
       volume = {688},
          eid = {A105},
        pages = {A105},
          doi = {10.1051/0004-6361/202245588},
archivePrefix = {arXiv},
       eprint = {2303.12165},
 primaryClass = {astro-ph.SR},
       adsurl = {https://ui.adsabs.harvard.edu/abs/2024A&A...688A.105H}
}

@ARTICLE{bekki2009,
   author = {{Bekki}, K. and {Stanimirovi{\'c}}, S.},
    title = "{The total mass and dark halo properties of the Small Magellanic Cloud}",
  journal = {\mnras},
archivePrefix = "arXiv",
   eprint = {0807.2102},
     year = 2009,
    month = may,
   volume = 395,
    pages = {342-350},
      doi = {10.1111/j.1365-2966.2009.14514.x},
   adsurl = {http://adsabs.harvard.edu/abs/2009MNRAS.395..342B}
}

@ARTICLE{subramanian2012,
   author = {{Subramanian}, S. and {Subramaniam}, A.},
    title = "{The Three-dimensional Structure of the Small Magellanic Cloud}",
  journal = {\apj},
archivePrefix = "arXiv",
   eprint = {1109.3980},
     year = 2012,
    month = jan,
   volume = 744,
      eid = {128},
    pages = {128},
      doi = {10.1088/0004-637X/744/2/128},
   adsurl = {http://adsabs.harvard.edu/abs/2012ApJ...744..128S}
}

@ARTICLE{draine2007,
       author = {{Draine}, B.~T. and {Dale}, D.~A. and {Bendo}, G. and {Gordon}, K.~D. and {Smith}, J.~D.~T. and {Armus}, L. and {Engelbracht}, C.~W. and {Helou}, G. and {Kennicutt}, R.~C., Jr. and {Li}, A. and {Roussel}, H. and {Walter}, F. and {Calzetti}, D. and {Moustakas}, J. and {Murphy}, E.~J. and {Rieke}, G.~H. and {Bot}, C. and {Hollenbach}, D.~J. and {Sheth}, K. and {Teplitz}, H.~I.},
        title = "{Dust Masses, PAH Abundances, and Starlight Intensities in the SINGS Galaxy Sample}",
      journal = {\apj},
         year = 2007,
        month = jul,
       volume = {663},
       number = {2},
        pages = {866-894},
          doi = {10.1086/518306},
archivePrefix = {arXiv},
       eprint = {astro-ph/0703213},
 primaryClass = {astro-ph},
       adsurl = {https://ui.adsabs.harvard.edu/abs/2007ApJ...663..866D}
}

@ARTICLE{Virtanen2020,
       author = {{Virtanen}, Pauli and {Gommers}, Ralf and {Oliphant}, Travis E. and {Haberland}, Matt and {Reddy}, Tyler and {Cournapeau}, David and {Burovski}, Evgeni and {Peterson}, Pearu and {Weckesser}, Warren and {Bright}, Jonathan and {van der Walt}, St{\'e}fan J. and {Brett}, Matthew and {Wilson}, Joshua and {Millman}, K. Jarrod and {Mayorov}, Nikolay and {Nelson}, Andrew R.~J. and {Jones}, Eric and {Kern}, Robert and {Larson}, Eric and {Carey}, C.~J. and {Polat}, {\.I}lhan and {Feng}, Yu and {Moore}, Eric W. and {VanderPlas}, Jake and {Laxalde}, Denis and {Perktold}, Josef and {Cimrman}, Robert and {Henriksen}, Ian and {Quintero}, E.~A. and {Harris}, Charles R. and {Archibald}, Anne M. and {Ribeiro}, Ant{\^o}nio H. and {Pedregosa}, Fabian and {van Mulbregt}, Paul and {SciPy 1. 0 Contributors}},
        title = "{SciPy 1.0: fundamental algorithms for scientific computing in Python}",
      journal = {Nature Methods},
         year = 2020,
        month = feb,
       volume = {17},
        pages = {261-272},
          doi = {10.1038/s41592-019-0686-2},
archivePrefix = {arXiv},
       eprint = {1907.10121},
 primaryClass = {cs.MS},
       adsurl = {https://ui.adsabs.harvard.edu/abs/2020NatMe..17..261V}
}

@ARTICLE{Hunter2007,
       author = {{Hunter}, John D.},
        title = "{Matplotlib: A 2D Graphics Environment}",
      journal = {Computing in Science and Engineering},
         year = 2007,
        month = may,
       volume = {9},
       number = {3},
        pages = {90-95},
          doi = {10.1109/MCSE.2007.55},
       adsurl = {https://ui.adsabs.harvard.edu/abs/2007CSE.....9...90H}
}

@Article{         harris2020array,
 title         = {Array programming with {NumPy}},
 author        = {Charles R. Harris and K. Jarrod Millman and St{\'{e}}fan J.
                 van der Walt and Ralf Gommers and Pauli Virtanen and David
                 Cournapeau and Eric Wieser and Julian Taylor and Sebastian
                 Berg and Nathaniel J. Smith and Robert Kern and Matti Picus
                 and Stephan Hoyer and Marten H. van Kerkwijk and Matthew
                 Brett and Allan Haldane and Jaime Fern{\'{a}}ndez del
                 R{\'{i}}o and Mark Wiebe and Pearu Peterson and Pierre
                 G{\'{e}}rard-Marchant and Kevin Sheppard and Tyler Reddy and
                 Warren Weckesser and Hameer Abbasi and Christoph Gohlke and
                 Travis E. Oliphant},
 year          = {2020},
 month         = sep,
 journal       = {Nature},
 volume        = {585},
 number        = {7825},
 pages         = {357--362},
 doi           = {10.1038/s41586-020-2649-2},
 publisher     = {Springer Science and Business Media {LLC}},
 url           = {https://doi.org/10.1038/s41586-020-2649-2}
}

@ARTICLE{scowcroft2016,
   author = {{Scowcroft}, V. and {Freedman}, W.~L. and {Madore}, B.~F. and 
	{Monson}, A. and {Persson}, S.~E. and {Rich}, J. and {Seibert}, M. and 
	{Rigby}, J.~R.},
    title = "{The Carnegie Hubble Program: The Distance and Structure of the SMC as Revealed by Mid-infrared Observations of Cepheids}",
  journal = {\apj},
archivePrefix = "arXiv",
   eprint = {1502.06995},
     year = 2016,
    month = jan,
   volume = 816,
      eid = {49},
    pages = {49},
      doi = {10.3847/0004-637X/816/2/49},
   adsurl = {http://adsabs.harvard.edu/abs/2016ApJ...816...49S}
}

@ARTICLE{fitzpatrick1999,
   author = {{Fitzpatrick}, E.~L.},
    title = "{Correcting for the Effects of Interstellar Extinction}",
  journal = {\pasp},
   eprint = {astro-ph/9809387},
     year = 1999,
    month = jan,
   volume = 111,
    pages = {63-75},
      doi = {10.1086/316293},
   adsurl = {http://adsabs.harvard.edu/abs/1999PASP..111...63F}
}

@ARTICLE{gordon2016,
   author = {{Gordon}, K.~D. and {Fouesneau}, M. and {Arab}, H. and {Tchernyshyov}, K. and 
	{Weisz}, D.~R. and {Dalcanton}, J.~J. and {Williams}, B.~F. and 
	{Bell}, E.~F. and {Bianchi}, L. and {Boyer}, M. and {Choi}, Y. and 
	{Dolphin}, A. and {Girardi}, L. and {Hogg}, D.~W. and {Kalirai}, J.~S. and 
	{Kapala}, M. and {Lewis}, A.~R. and {Rix}, H.-W. and {Sandstrom}, K. and 
	{Skillman}, E.~D.},
    title = "{The Panchromatic Hubble Andromeda Treasury. XV. The BEAST: Bayesian Extinction and Stellar Tool}",
  journal = {\apj},
archivePrefix = "arXiv",
   eprint = {1606.06182},
     year = 2016,
    month = aug,
   volume = 826,
      eid = {104},
    pages = {104},
      doi = {10.3847/0004-637X/826/2/104},
   adsurl = {http://adsabs.harvard.edu/abs/2016ApJ...826..104G}
}

@ARTICLE{gilbert2024,
       author = {{Gilbert}, Karoline M. and {Choi}, Yumi and {Boyer}, Martha L. and {Williams}, Benjamin F. and {Weisz}, Daniel R. and {Bell}, Eric F. and {Dalcanton}, Julianne J. and {McQuinn}, Kristen B.~W. and {Skillman}, Evan D. and {Costa}, Guglielmo and {Dolphin}, Andrew E. and {Fouesneau}, Morgan and {Girardi}, L{\'e}o and {Goldman}, Steven R. and {Gordon}, Karl D. and {Guhathakurta}, Puragra and {Gull}, Maude and {Hagen}, Lea and {Huynh}, Ky and {Lindberg}, Christina W. and {Marigo}, Paola and {Murray}, Claire E. and {Pastorelli}, Giada and {Yanchulova Merica-Jones}, Petia},
        title = "{The Local Ultraviolet to Infrared Treasury. I. Survey Overview of the Broadband Imaging}",
      journal = {\apjs},
         year = 2025,
        month = jan,
       volume = {276},
       number = {1},
          eid = {8},
        pages = {8},
          doi = {10.3847/1538-4365/ad76af},
archivePrefix = {arXiv},
       eprint = {2410.20454},
 primaryClass = {astro-ph.GA},
       adsurl = {https://ui.adsabs.harvard.edu/abs/2025ApJS..276....8G}
}

@ARTICLE{murray2024b,
       author = {{Murray}, Claire E. and {Lindberg}, Christina W. and {Yanchulova Merica-Jones}, Petia and {Williams}, Benjamin F. and {Cohen}, Roger E. and {Gordon}, Karl D. and {McQuinn}, Kristen B.~W. and {Choi}, Yumi and {Burhenne}, Clare and {Sandstrom}, Karin M. and {Bot}, Caroline and {Johnson}, L. Clifton and {Goldman}, Steven R. and {Clark}, Christopher J.~R. and {Roman-Duval}, Julia C. and {Gilbert}, Karoline M. and {Peek}, J.~E.~G. and {Hirschauer}, Alec S. and {Boyer}, Martha L. and {Dolphin}, Andrew E.},
        title = "{Scylla. I. A Pure-parallel, Multiwavelength Imaging Survey of the ULLYSES Fields in the LMC and SMC}",
      journal = {\apjs},
         year = 2024,
        month = nov,
       volume = {275},
       number = {1},
          eid = {5},
        pages = {5},
          doi = {10.3847/1538-4365/ad6de2},
archivePrefix = {arXiv},
       eprint = {2410.11695},
 primaryClass = {astro-ph.GA},
       adsurl = {https://ui.adsabs.harvard.edu/abs/2024ApJS..275....5M}
}

@ARTICLE{boggess1992,
       author = {{Boggess}, N.~W. and {Mather}, J.~C. and {Weiss}, R. and {Bennett}, C.~L. and {Cheng}, E.~S. and {Dwek}, E. and {Gulkis}, S. and {Hauser}, M.~G. and {Janssen}, M.~A. and {Kelsall}, T. and {Meyer}, S.~S. and {Moseley}, S.~H. and {Murdock}, T.~L. and {Shafer}, R.~A. and {Silverberg}, R.~F. and {Smoot}, G.~F. and {Wilkinson}, D.~T. and {Wright}, E.~L.},
        title = "{The COBE Mission: Its Design and Performance Two Years after Launch}",
      journal = {\apj},
         year = 1992,
        month = oct,
       volume = {397},
        pages = {420},
          doi = {10.1086/171797},
       adsurl = {https://ui.adsabs.harvard.edu/abs/1992ApJ...397..420B}
}

@INPROCEEDINGS{silverberg1993,
       author = {{Silverberg}, Robert F. and {Hauser}, Michael G. and {Boggess}, Nancy W. and {Kelsall}, Thomas J. and {Moseley}, S.~H. and {Murdock}, Tom L.},
        title = "{Design of the diffuse infrared background experiment (DIRBE) on COBE}",
    booktitle = {Infrared Spaceborne Remote Sensing},
         year = 1993,
       editor = {{Scholl}, Marija S.},
       series = {Society of Photo-Optical Instrumentation Engineers (SPIE) Conference Series},
       volume = {2019},
        month = oct,
        pages = {180-189},
          doi = {10.1117/12.157825},
       adsurl = {https://ui.adsabs.harvard.edu/abs/1993SPIE.2019..180S}
}

@ARTICLE{astropy,
       author = {{Astropy Collaboration} and {Price-Whelan}, A.~M. and {Sip{\H{o}}cz}, B.~M. and {G{\"u}nther}, H.~M. and {Lim}, P.~L. and {Crawford}, S.~M. and {Conseil}, S. and {Shupe}, D.~L. and {Craig}, M.~W. and {Dencheva}, N. and {Ginsburg}, A. and {VanderPlas}, J.~T. and {Bradley}, L.~D. and {P{\'e}rez-Su{\'a}rez}, D. and {de Val-Borro}, M. and {Aldcroft}, T.~L. and {Cruz}, K.~L. and {Robitaille}, T.~P. and {Tollerud}, E.~J. and {Ardelean}, C. and {Babej}, T. and {Bach}, Y.~P. and {Bachetti}, M. and {Bakanov}, A.~V. and {Bamford}, S.~P. and {Barentsen}, G. and {Barmby}, P. and {Baumbach}, A. and {Berry}, K.~L. and {Biscani}, F. and {Boquien}, M. and {Bostroem}, K.~A. and {Bouma}, L.~G. and {Brammer}, G.~B. and {Bray}, E.~M. and {Breytenbach}, H. and {Buddelmeijer}, H. and {Burke}, D.~J. and {Calderone}, G. and {Cano Rodr{\'\i}guez}, J.~L. and {Cara}, M. and {Cardoso}, J.~V.~M. and {Cheedella}, S. and {Copin}, Y. and {Corrales}, L. and {Crichton}, D. and {D'Avella}, D. and {Deil}, C. and {Depagne}, {\'E}. and {Dietrich}, J.~P. and {Donath}, A. and {Droettboom}, M. and {Earl}, N. and {Erben}, T. and {Fabbro}, S. and {Ferreira}, L.~A. and {Finethy}, T. and {Fox}, R.~T. and {Garrison}, L.~H. and {Gibbons}, S.~L.~J. and {Goldstein}, D.~A. and {Gommers}, R. and {Greco}, J.~P. and {Greenfield}, P. and {Groener}, A.~M. and {Grollier}, F. and {Hagen}, A. and {Hirst}, P. and {Homeier}, D. and {Horton}, A.~J. and {Hosseinzadeh}, G. and {Hu}, L. and {Hunkeler}, J.~S. and {Ivezi{\'c}}, {\v{Z}}. and {Jain}, A. and {Jenness}, T. and {Kanarek}, G. and {Kendrew}, S. and {Kern}, N.~S. and {Kerzendorf}, W.~E. and {Khvalko}, A. and {King}, J. and {Kirkby}, D. and {Kulkarni}, A.~M. and {Kumar}, A. and {Lee}, A. and {Lenz}, D. and {Littlefair}, S.~P. and {Ma}, Z. and {Macleod}, D.~M. and {Mastropietro}, M. and {McCully}, C. and {Montagnac}, S. and {Morris}, B.~M. and {Mueller}, M. and {Mumford}, S.~J. and {Muna}, D. and {Murphy}, N.~A. and {Nelson}, S. and {Nguyen}, G.~H. and {Ninan}, J.~P. and {N{\"o}the}, M. and {Ogaz}, S. and {Oh}, S. and {Parejko}, J.~K. and {Parley}, N. and {Pascual}, S. and {Patil}, R. and {Patil}, A.~A. and {Plunkett}, A.~L. and {Prochaska}, J.~X. and {Rastogi}, T. and {Reddy Janga}, V. and {Sabater}, J. and {Sakurikar}, P. and {Seifert}, M. and {Sherbert}, L.~E. and {Sherwood-Taylor}, H. and {Shih}, A.~Y. and {Sick}, J. and {Silbiger}, M.~T. and {Singanamalla}, S. and {Singer}, L.~P. and {Sladen}, P.~H. and {Sooley}, K.~A. and {Sornarajah}, S. and {Streicher}, O. and {Teuben}, P. and {Thomas}, S.~W. and {Tremblay}, G.~R. and {Turner}, J.~E.~H. and {Terr{\'o}n}, V. and {van Kerkwijk}, M.~H. and {de la Vega}, A. and {Watkins}, L.~L. and {Weaver}, B.~A. and {Whitmore}, J.~B. and {Woillez}, J. and {Zabalza}, V. and {Astropy Contributors}},
        title = "{The Astropy Project: Building an Open-science Project and Status of the v2.0 Core Package}",
      journal = {\aj},
         year = 2018,
        month = sep,
       volume = {156},
       number = {3},
          eid = {123},
        pages = {123},
          doi = {10.3847/1538-3881/aabc4f},
archivePrefix = {arXiv},
       eprint = {1801.02634},
 primaryClass = {astro-ph.IM},
       adsurl = {https://ui.adsabs.harvard.edu/abs/2018AJ....156..123A}
}

@ARTICLE{neugebauer1984,
       author = {{Neugebauer}, G. and {Habing}, H.~J. and {van Duinen}, R. and {Aumann}, H.~H. and {Baud}, B. and {Beichman}, C.~A. and {Beintema}, D.~A. and {Boggess}, N. and {Clegg}, P.~E. and {de Jong}, T. and {Emerson}, J.~P. and {Gautier}, T.~N. and {Gillett}, F.~C. and {Harris}, S. and {Hauser}, M.~G. and {Houck}, J.~R. and {Jennings}, R.~E. and {Low}, F.~J. and {Marsden}, P.~L. and {Miley}, G. and {Olnon}, F.~M. and {Pottasch}, S.~R. and {Raimond}, E. and {Rowan-Robinson}, M. and {Soifer}, B.~T. and {Walker}, R.~G. and {Wesselius}, P.~R. and {Young}, E.},
        title = "{The Infrared Astronomical Satellite (IRAS) mission.}",
      journal = {\apjl},
         year = 1984,
        month = mar,
       volume = {278},
        pages = {L1-L6},
          doi = {10.1086/184209},
       adsurl = {https://ui.adsabs.harvard.edu/abs/1984ApJ...278L...1N}
}

@ARTICLE{meixner2013,
       author = {{Meixner}, M. and {Panuzzo}, P. and {Roman-Duval}, J. and {Engelbracht}, C. and {Babler}, B. and {Seale}, J. and {Hony}, S. and {Montiel}, E. and {Sauvage}, M. and {Gordon}, K. and {Misselt}, K. and {Okumura}, K. and {Chanial}, P. and {Beck}, T. and {Bernard}, J. -P. and {Bolatto}, A. and {Bot}, C. and {Boyer}, M.~L. and {Carlson}, L.~R. and {Clayton}, G.~C. and {Chen}, C. -H.~R. and {Cormier}, D. and {Fukui}, Y. and {Galametz}, M. and {Galliano}, F. and {Hora}, J.~L. and {Hughes}, A. and {Indebetouw}, R. and {Israel}, F.~P. and {Kawamura}, A. and {Kemper}, F. and {Kim}, S. and {Kwon}, E. and {Lebouteiller}, V. and {Li}, A. and {Long}, K.~S. and {Madden}, S.~C. and {Matsuura}, M. and {Muller}, E. and {Oliveira}, J.~M. and {Onishi}, T. and {Otsuka}, M. and {Paradis}, D. and {Poglitsch}, A. and {Reach}, W.~T. and {Robitaille}, T.~P. and {Rubio}, M. and {Sargent}, B. and {Sewi{\l}o}, M. and {Skibba}, R. and {Smith}, L.~J. and {Srinivasan}, S. and {Tielens}, A.~G.~G.~M. and {van Loon}, J. Th. and {Whitney}, B.},
        title = "{The HERSCHEL Inventory of The Agents of Galaxy Evolution in the Magellanic Clouds, a Herschel Open Time Key Program}",
      journal = {\aj},
         year = 2013,
        month = sep,
       volume = {146},
       number = {3},
          eid = {62},
        pages = {62},
          doi = {10.1088/0004-6256/146/3/62},
       adsurl = {https://ui.adsabs.harvard.edu/abs/2013AJ....146...62M}
}

@ARTICLE{degrijs2015,
       author = {{de Grijs}, Richard and {Bono}, Giuseppe},
        title = "{Clustering of Local Group Distances: Publication Bias or Correlated Measurements? III. The Small Magellanic Cloud}",
      journal = {\aj},
         year = 2015,
        month = jun,
       volume = {149},
       number = {6},
          eid = {179},
        pages = {179},
          doi = {10.1088/0004-6256/149/6/179},
archivePrefix = {arXiv},
       eprint = {1504.00417},
 primaryClass = {astro-ph.SR},
       adsurl = {https://ui.adsabs.harvard.edu/abs/2015AJ....149..179D}
}
